\documentclass[apj]{emulateapj}

\usepackage{graphicx}
\usepackage{rotating}
\usepackage{afterpage}
\usepackage{txfonts}
\usepackage[sort&compress]{natbib}
\bibpunct{(}{)}{;}{a}{}{,}

\newcommand{\um}{$\mu$m}

\newcommand{\kms}{km\thinspace s$^{-1}$}

\def\arcmin{\hbox{$^\prime$}}
\def\arcsec{\hbox{$^{\prime\prime}$}}
\def\utw{\smash{\rlap{\lower5pt\hbox{$\sim$}}}}
\def\udtw{\smash{\rlap{\lower6pt\hbox{$\approx$}}}}

\def\Teff{\hbox{\it T$_{\rm eff}$}}

\def\Msun{\hbox{\it M$_\odot$}}

\def\Mbol{\hbox{\it M$_{bol}$}}

\def\Teff{\hbox{\it T$_{\rm eff}$}}

\def\Mk{\hbox{\it M$_{\rm K}$}}
\newcommand{\Ks}{{\it K$_{\rm s}$}}
\newcommand{\Kso}{{\it K$_{\rm so}$}}

\newcommand{\Aks}{{\it A$_{\it K_{\rm s}}$}}

\def\BCK{\hbox{\it BC$_{\it K}$}}

\def\simgr{\mathrel{\hbox{\rlap{\hbox{\lower4pt\hbox{$\sim$}}}\hbox{$>$}}}}
\def\HH{H{\sc ii}}	

\def\brg{\hbox{\it Br$_\gamma$}}

\def\Vlsr{\hbox{\it V$_{\rm lsr}$}}
\def\nodata{\hbox{ $..$}}

\DeclareMathAlphabet{\mathsc}{OT1}{cmr}{m}{sc}
\def\testbx{bx}%
\DeclareRobustCommand{\ion}[2]{%
\relax\ifmmode
\ifx\testbx\f@series
{\mathbf{#1\,\mathsc{#2}}}\else
{\mathrm{#1\,\mathsc{#2}}}\fi
\else\textup{#1\,{\mdseries\textsc{#2}}}%
\fi}

\shorttitle{Massive stars in the inner Disk.}
\shortauthors{Messineo et al.}

\begin{document}


\title{Detections of massive stars in the cluster MCM2005b77,  
   in the star-forming regions GRS G331.34$-$00.36 (S62) and  GRS G337.92$-$00.48 (S36).
   \thanks{Based on observations collected at the European Southern Observatory 
   (ESO Programme 089.D-876(A)).}
    }
      
\author{Maria~Messineo\altaffilmark{1,2}, 
        Karl M. Menten\altaffilmark{2},
        Donald F. Figer\altaffilmark{3},
        C.-H. Rosie Chen\altaffilmark{2},
        R. Michael Rich\altaffilmark{4},
	  }
 	  
\altaffiltext{1}{Key Laboratory for Researches in Galaxies and Cosmology, University of Science and Technology of China, 
Chinese Academy of Sciences, Hefei, Anhui, 230026, China
\email{messineo@ustc.edu.cn}
}
\altaffiltext{2}{Max-Planck-Institut f\"ur Radioastronomie, Auf dem H\"ugel 69, D-53121 Bonn, Germany	}
\altaffiltext{3}{Center for Detectors, Rochester Institute of Technology, 54 Memorial Drive, Rochester, NY 14623, USA	}
\altaffiltext{4}{Physics and Astronomy Building, 430 Portola Plaza, Box 951547, Department of Physics 
and Astronomy, University of California, Los Angeles, CA 90095-1547.}

\begin{abstract}
Large infrared and millimeter wavelength
surveys of the Galactic plane have unveiled more than 600 
new bubble \HH\ regions and more than 3000 candidate star clusters.
We present a study of the candidate clusters  MCM2005b72,
DBS2003-157, DBS2003-172, and MCM2005b77, based on
near-infrared spectroscopy taken with SofI on the NTT and infrared photometry 
from the 2MASS, VVV,  and GLIMPSE surveys.
We find that (1) MCM2005b72 and DBS2003-157 are subregions of the same  
star-forming region, \HH\ GRS G331.34$-$00.36 (bubble S62). 
MCM2005b72 coincides with the central 
part of this \HH\ region,  while DBS2003-157 is  a bright mid-infrared  knot of 
the S62 shell. 
We detected two O-type stars at extinction \Aks=1.0-1.3 mag. Their spectrophotometric properties 
are consistent with the near-kinematic distance to GRS G331.34$-$00.36 of $3.9\pm0.3$ kpc.
(2) DBS2003-172 coincides with a bright mid-infrared knot in the S36 shell 
(GRS G337.92$-$00.48), where we detected a pair of candidate  He I stars 
embedded in a small cometary nebula. 
(3) The stellar cluster MCM2005b77  is   rich in B-type stars, has an average 
\Aks\ of 0.91 mag,  and is adjacent to the \HH\ region IRAS 16137$-$5025.
The average spectrophotometric distance of $\sim 5.0$ kpc matches the near-kinematic
distance to IRAS 16137$-$5025 of $5.2\pm0.1$ kpc.
\end{abstract}



%

\section{Introduction}

Massive stars form continuously in the Galactic disk, chemically enriching the 
interstellar medium through by losing mass at high rates and by exploding. 
They are usually detected in groups or clusters of stars and make good tracers 
of Galactic structure \citep[e.g.][]{georgelin76,russeil03}.
More than 3000  candidate stellar clusters  
have been detected in large near-infrared  and mid-infrared 
surveys of the Galactic plane 
\citep[e.g.][]{mercer05,dutra03,froebrich07,glushkova10,borissova11,solin12,camargo14}.
The most concentrated and populous candidate clusters are usually studied first --
with about 20  young  clusters known to be more massive than $\ga 10\,000$ \Msun-- 
but there is a plethora of sparse groups of stars or extended regions 
of increased stellar counts in the direction of nebulosities.

In the course of the Galactic Legacy 
Infrared Mid-plane Survey Extraordinaire (GLIMPSE), marvelous nebulae were detected   
along the  Galactic plane,  making it a  ``bubbling" environment
\citep[e.g. ][]{watson08, watson09, churchwell09}.
A typical bubble is marked by a shell of mid-infrared emission from dust and 
polycyclic aromatic hydrocarbon (PAH) bands 
(e.g.\ at  5.8 \um\ and 8.0 \um) surrounding
a \HH\ region. Indeed, GLIMPSE bubbles are filled 
with radio continuum emission \citep[e.g.][]{watson08,richards12,deharveng10}.
The bubbles   are created by winds, explosions, and UV radiation 
from massive stars (O-type and early B-type stars).
To investigate their origin, it is of primary importance to study   loose candidate clusters
found in the direction of \HH\ regions, to detect the ionizing stars and to  characterize
the continuum photons and the associated bubbles.
The types of the dominant stars determine the number of 
Lyman photons ($N_{\rm lyc}$) of a system.
About 90\% of bubbles are expected to contain massive OB stars. 
Undoubtedly, the Westerlund 2 cluster sustains the  RCW49 nebula \citep[e.g.][]{povich08},
 but, so far, massive stars have remained undetected in most bubbles.
For several bubbles associated with HII regions, candidate ionizing stars  (photometrically identified) 
are listed in the works  of, for example, \citet[e.g.][]{watson08}, \citet{watson09}, and \citet{sidorin14}.
Interstellar extinction toward bubbles is highly patchy and 
some candidate clusters may simply be  zones of lower interstellar extinction 
\citep[e.g.][]{messineo15, dutra01, gonzalez12}.
Typically, more than 50\% of the detected candidate clusters are found to be spurious
 when followed up with spectroscopic studies 
of the  brightest stars \citep[e.g.][]{messineo14,froebrich07,borissova05}.

In this paper,  we  report on low-resolution infrared spectroscopy of several stellar overdensities 
that are projected toward Galactic bubbles. 
The candidates MCM2005b72 \citep[][]{mercer05} and DBS2003-157  
\citep[][]{dutra03} are both located in the direction of GRS G331.34$-$00.36
\citep[S62;][]{culverhouse11,churchwell06,simpson12}. 
The candidate cluster DBS2003-172 \citep[][]{dutra03} is projected onto GRS G337.92$-$00.48
\citep[S36;][]{culverhouse11,churchwell06}.
The candidate cluster MCM2005b77 \citep{mercer05} is located at  
$(l,b)=(332.\!\!^{\circ}780,+00.\!\!^{\circ}022)$  near the \HH\ region 
 (and far-infrared source) IRAS 16137-5025. 
The spectroscopic observations are described in Section \ref{observations} and
the data analysis  in Section \ref{analysis}. In Sections \ref{clusteranalysis} and 5,
we review the properties of the confirmed stellar clusters and massive stars.
A summary is provided in Section \ref{summary}.

\begin{table*}
\caption{\label{cand.clusters} Positions of candidate clusters projected toward the analyzed regions.}
\begin{tabular}{lllrrrlll}
\hline
\hline
Name         &  $\alpha$(J2000)&     $\delta$(J2000)                &   Longitude      & Latitude        &   Diameter    &  \HH\ region & Cluster references   \\
             & [hh~mm~ss]& [$^\circ ~ ^\prime ~ ^{\prime \prime}] $ & [$^\circ$] & [$^\circ$] &  [\arcsec] \\                            \\
\hline
DBS2003-157            & 16 12 20  & $-$51 46 12  & 331.330 & $-$0.334 &126 &     GRS G331.34$-$00.36$^a$  & 1, 2 \\
MCM2005b72             & 16 12 30  & $-$51 46 59  & 331.340 & $-$0.361 & 72 &     GRS G331.34$-$00.36$^a$  &3 \\
MCM2005b77$^b$         & 16 17 27.1& $-$50 30 41  & 332.780 &  0.022   & 98 &     IRAS 16137$-$5025        &3 \\
DBS2003-172$^b$         & 16 41 10.1& $-$47 07 24  & 337.923 & $-$0.469 &125 &     GRS G337.92$-$00.48$^c$ & 1,4 \\
\hline
\end{tabular}
\begin{list}{}{}{}
\item[]{\bf Notes.} 
($^a$) The \HH\ GRS G331.34$-$00.36 \citep{culverhouse11} includes the GLIMPSE 
bubble S62 \citep{churchwell06,simpson12}.
($^b$) Centers of the clusters DBS2003-172 and MCM2005b77 were determined using centroids of the flux overdensities
detected in the  \Ks\ images. The diameters are half-light diameters with respect to the 
peak positions.~  
($^c$) The \HH\  GRS G337.92$-$00.48 \citep{culverhouse11} coincides with the brightest part of the GLIMPSE 
bubble S36  \citep{churchwell06}.~
\item[]{\bf Cluster References.} 1 =\citet{dutra03};  2=\citet{pinheiro12}; 3=\citet{mercer05}; 4=\citet{borissova06}.
\end{list}
\end{table*}

\section{Observations and data reduction}
\label{observations}

\begin{figure*}
\begin{center}
\resizebox{0.33\hsize}{!}{\includegraphics[angle=0]{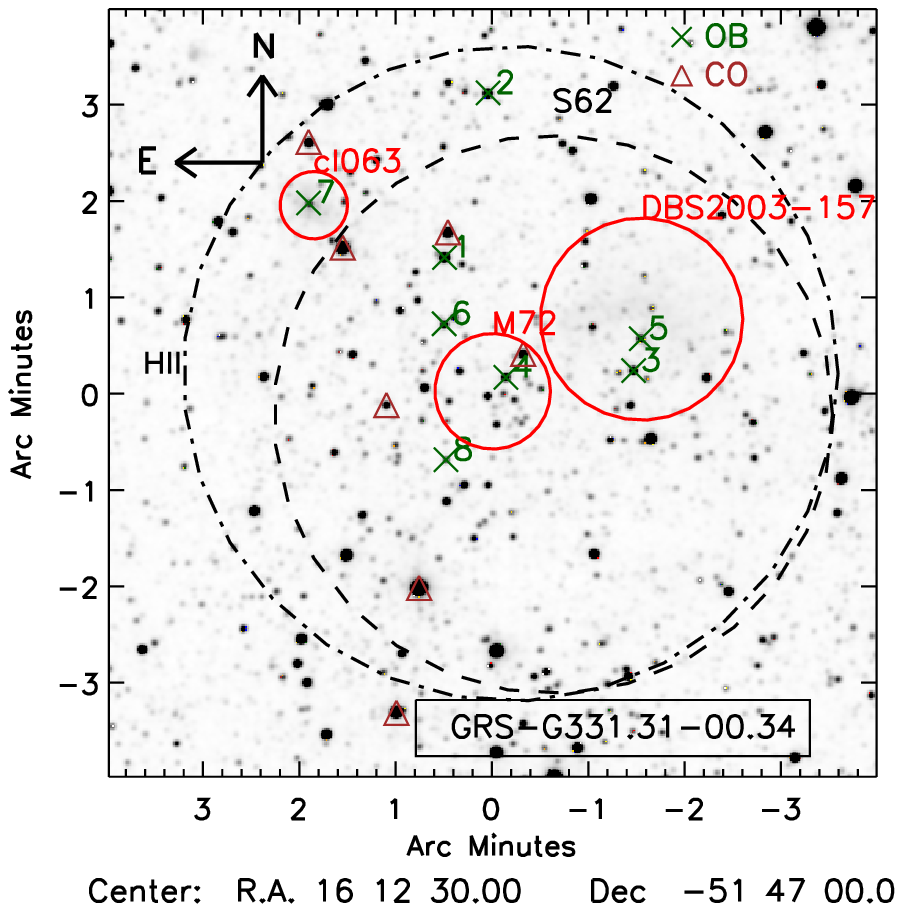}}
\resizebox{0.33\hsize}{!}{\includegraphics[angle=0]{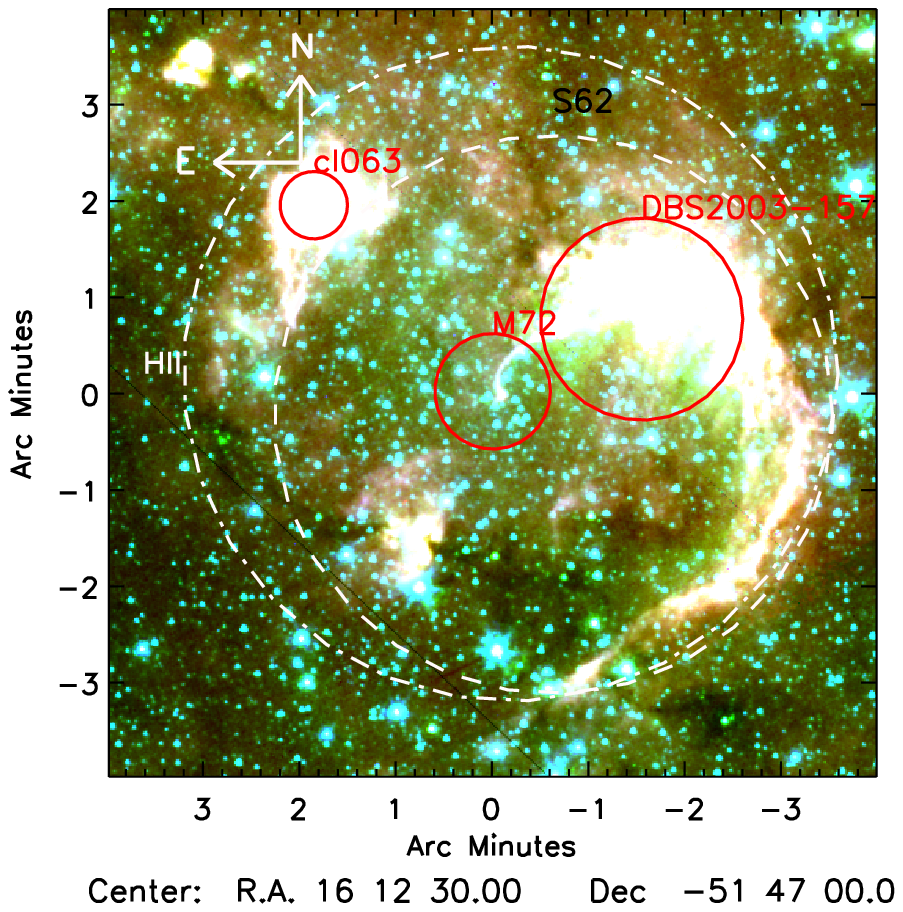}}
\resizebox{0.33\hsize}{!}{\includegraphics[angle=0]{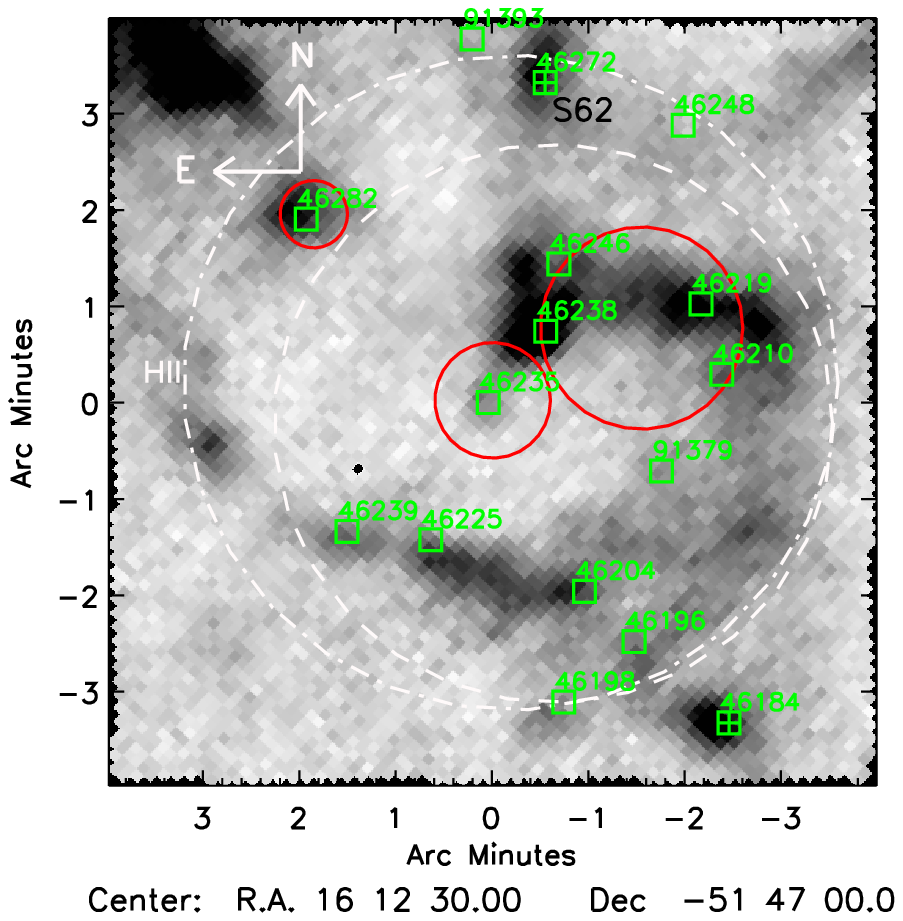}}
\end{center}
\begin{center}
\resizebox{0.33\hsize}{!}{\includegraphics[angle=0]{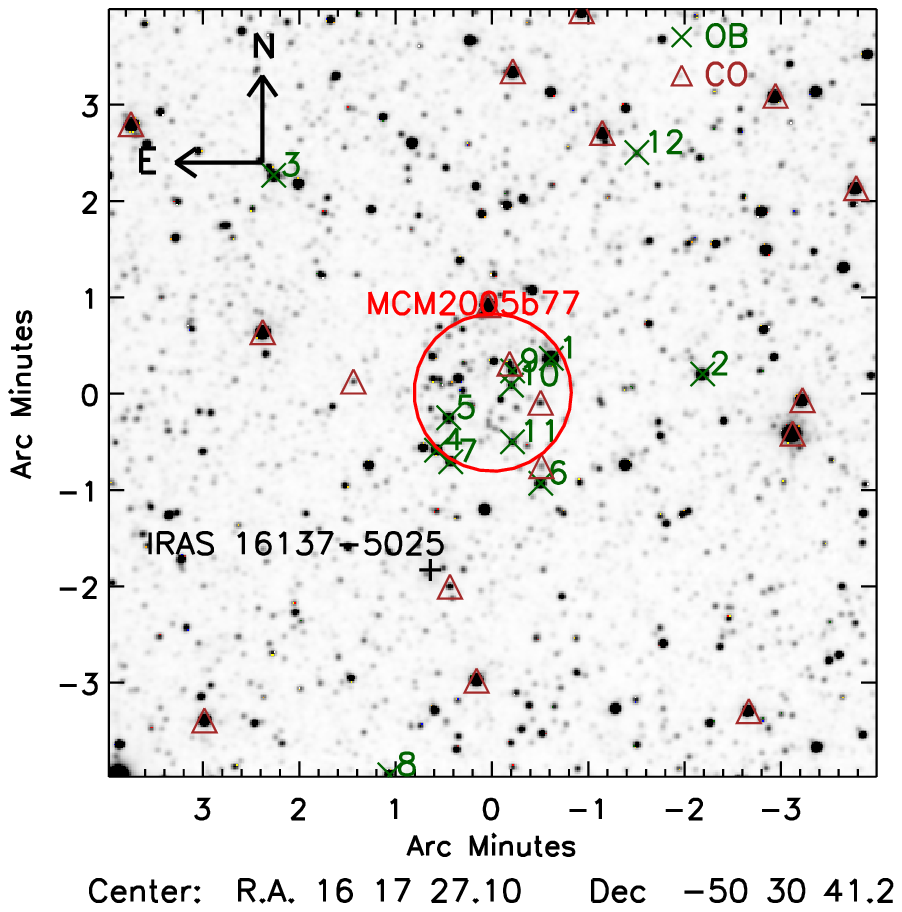}}
\resizebox{0.33\hsize}{!}{\includegraphics[angle=0]{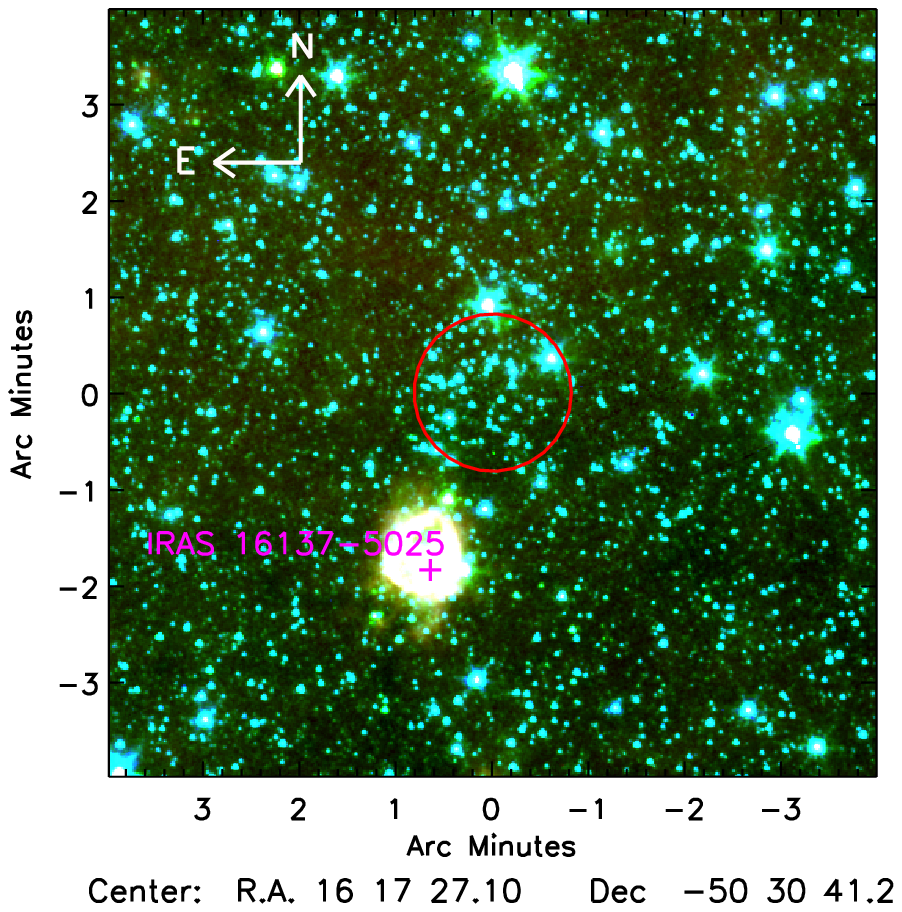}}
\resizebox{0.33\hsize}{!}{\includegraphics[angle=0]{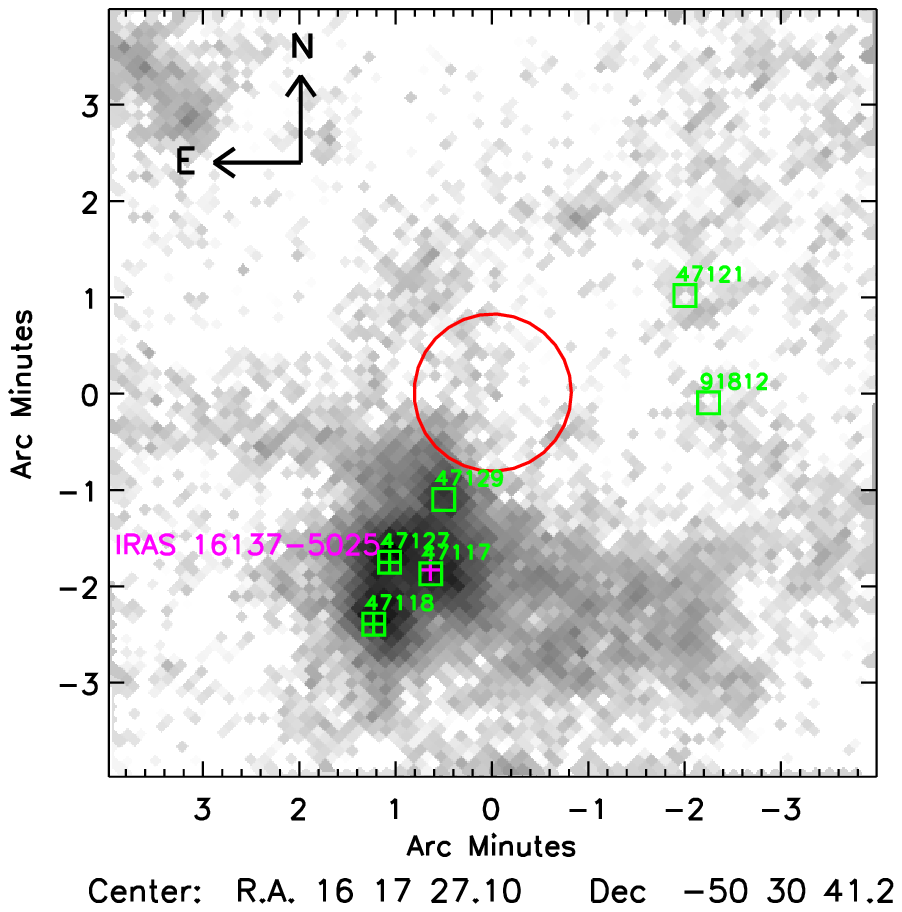}}
\end{center}
\begin{center}
\resizebox{0.33\hsize}{!}{\includegraphics[angle=0]{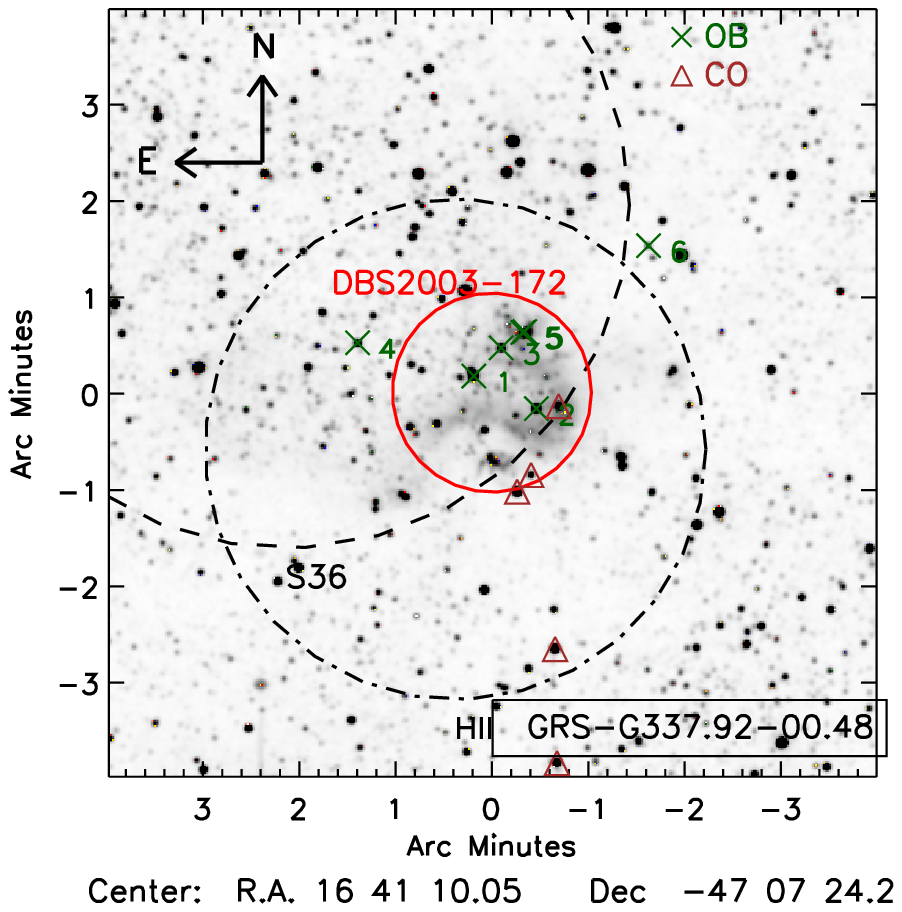}}
\resizebox{0.33\hsize}{!}{\includegraphics[angle=0]{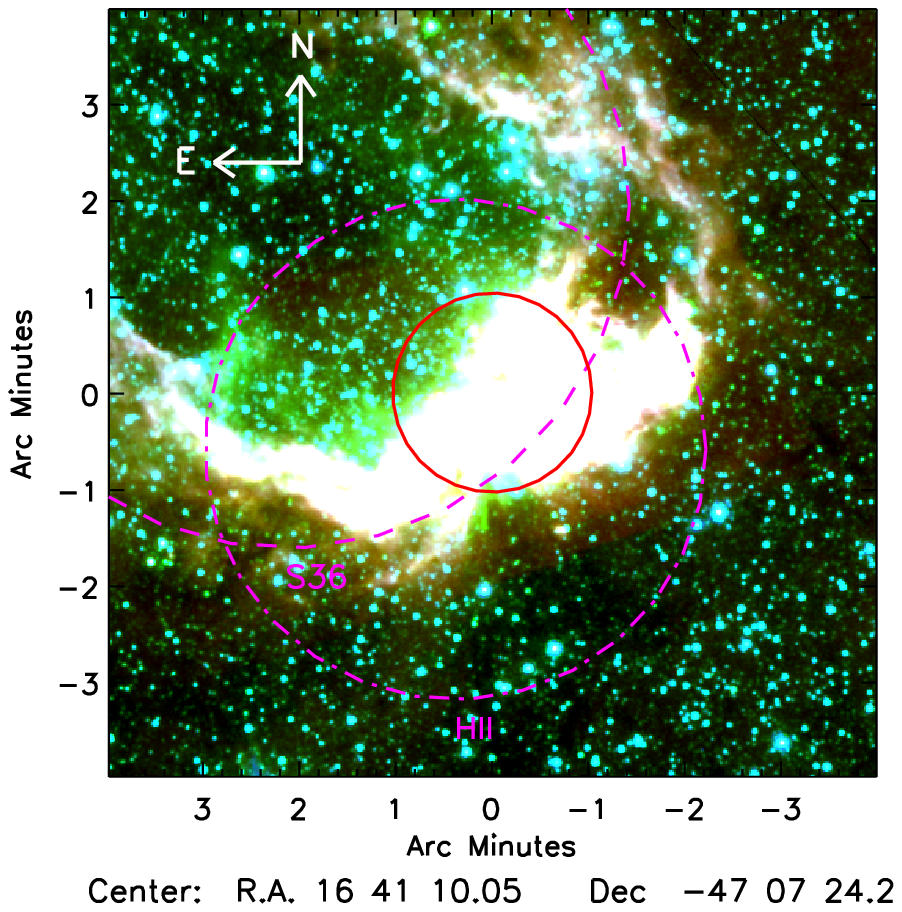}}
\resizebox{0.33\hsize}{!}{\includegraphics[angle=0]{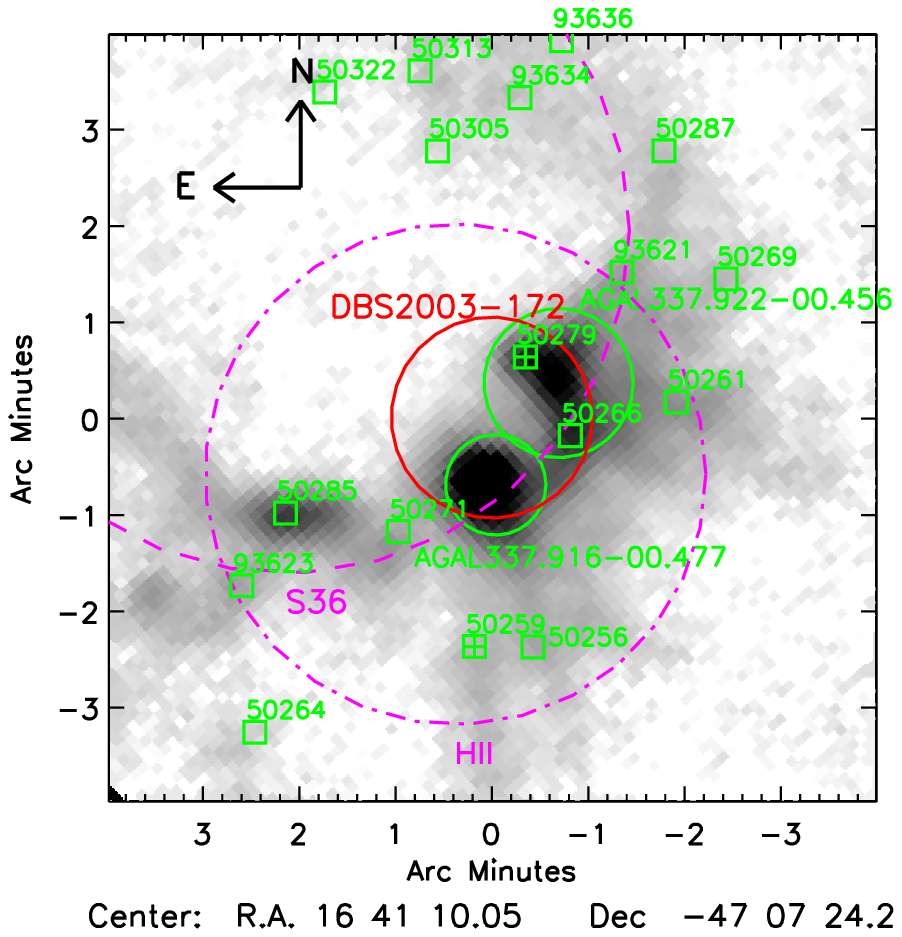}}
\end{center}
\caption{ \label{maps} {\it Left column:} 2MASS \Ks\ charts of late-type and early-type stars detected   
in the direction of  candidate clusters MCM2005b72/DBS2003-157, 
MCM2005b77, and DBS2003-172 (continuous circles).  
\HH\ regions identified by \citet{culverhouse11} and bubble sizes taken from \citet[S36,][]{churchwell06}  and   \citet[S62,][]{simpson12}
are marked with dotted-dashed and dashed circles (see Table \ref{cand.clusters}). 
{\it Middle column:} GLIMPSE images of the three regions. Composite images  
with GLIMPSE 3.6 \um-band in the blue channel, 4.5 \um-band in the green, 
and 8.0 \um-band in the red. {\it Right column:} 
 ATLASGAL 870 \um-band images \citep{schuller09}. Herschel/Hi-GAL compact
molecular clumps identified by \citet{elia17} are marked with green squares. Those
compact Hi-GAL clumps with estimated masses larger than 1000 \Msun\ are marked 
with green plus signs.  The two massive  (and more extended) ATLASGAL 
protostellar  condensations in S62 analyzed by \citet{koenig17},  \citet{urquhart14}, and \citet{urquhart18}
are marked with green circles.
}
\end{figure*}

Targets were selected among candidate clusters identified 
in GLIMPSE and 2MASS data, with bright near-infrared stars 
\citep[e.g.][]{mercer05,dutra03}, observable with 
the Son of ISAAC (SofI) spectrograph (see Table \ref{cand.clusters} and Fig.\  \ref{maps}),
and possibly located in the direction of bubble HII regions.

Observations  were taken with the SofI spectrograph on the ESO NTT/La Silla 
Telescope, under program  089.D-876(A) (P.I. Messineo),
over three nights, from 2012 May 31 to June 2.

The spectroscopic observations were carried out  using a 
$1^{\prime\prime} \times 290^{\prime\prime}$ slit. 
Typically, the slit orientation was adjusted to observe  two targets per slit.
For every target, a spectrum was taken with the high-resolution grism and the  
\Ks\ filter, at a resolving power $R \sim 2200$.
At least four exposures per target were taken in an ABBA cycle, nodding  along the slit. 

Data reduction was performed with IDL scripts and with the
Image Reduction and Analysis Facility (IRAF\footnote{IRAF is distributed by the 
National Optical Astronomy Observatories, which is operated by the Association 
of Universities.}) software.
Exposures close in time
were subtracted  from one another and
flat-fielded with spectroscopic lamp flats.
Up to five stellar traces were extracted in a single exposure.
Atmospheric transmission and instrumental response curves were
obtained with spectra of B-type stars observed in the same manner 
as the science targets, typically within a variation of 0.2 in airmass.  Stellar lines
(\brg\ and \ion{He}{I}) were removed from the standard spectra 
with a linear interpolation and the resulting spectra were
multiplied by  blackbody curves at the stellar effective temperatures.

Positions and spectral types of stars observed in 
 GRS G331.34$-$00.36 (S62), MCM2005b77, and GRS G337.92$-$00.48 (S36)
are  listed  in Tables \ref{spectra.early}, \ref{table5ab}, and \ref{spectra.late}.
Spectra are shown in Fig. \ref{spectra.early} and 11.

\begin{figure}[h]
\resizebox{0.85\hsize}{!}{\includegraphics[angle=0]{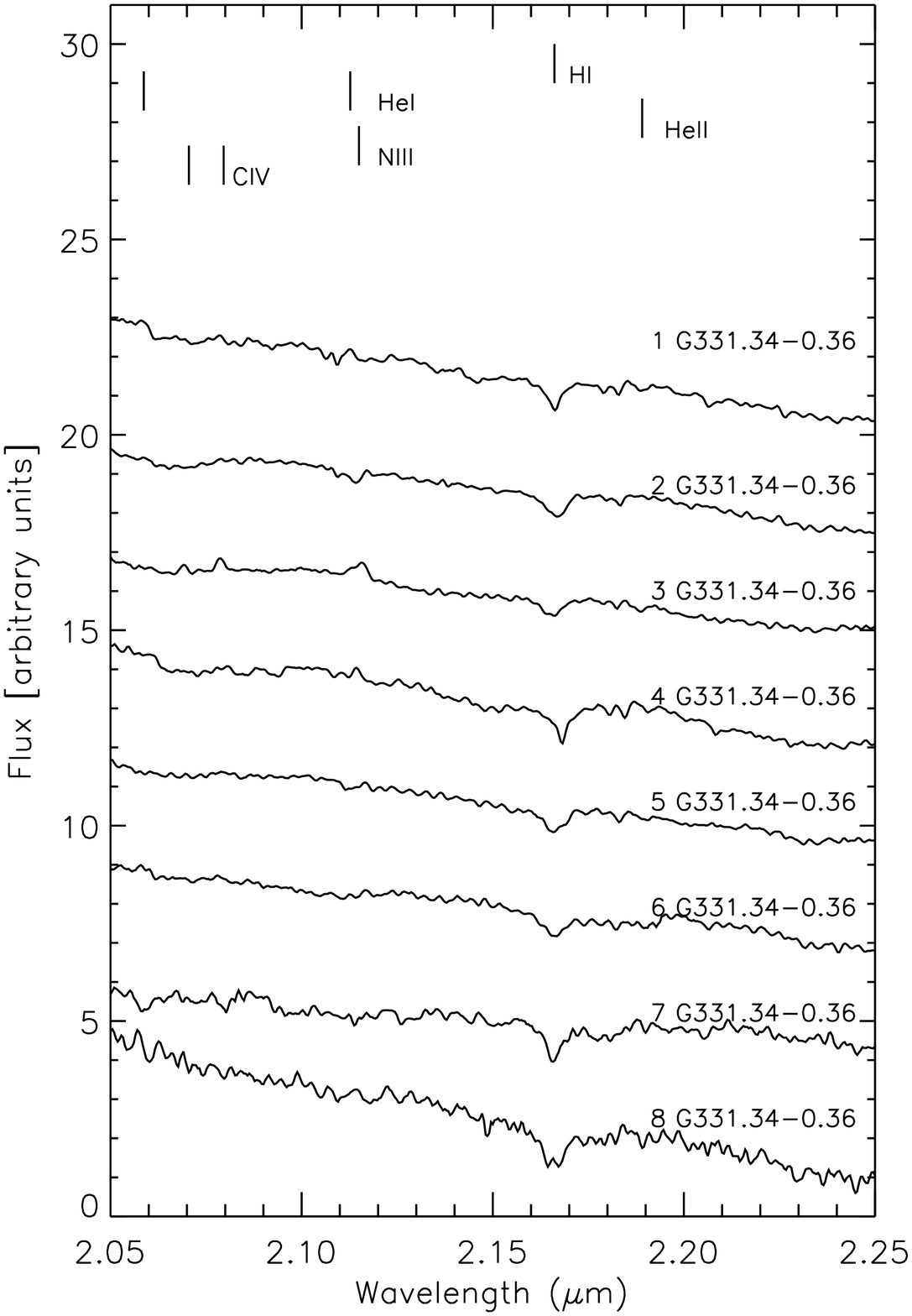}}
\resizebox{0.85\hsize}{!}{\includegraphics[angle=0]{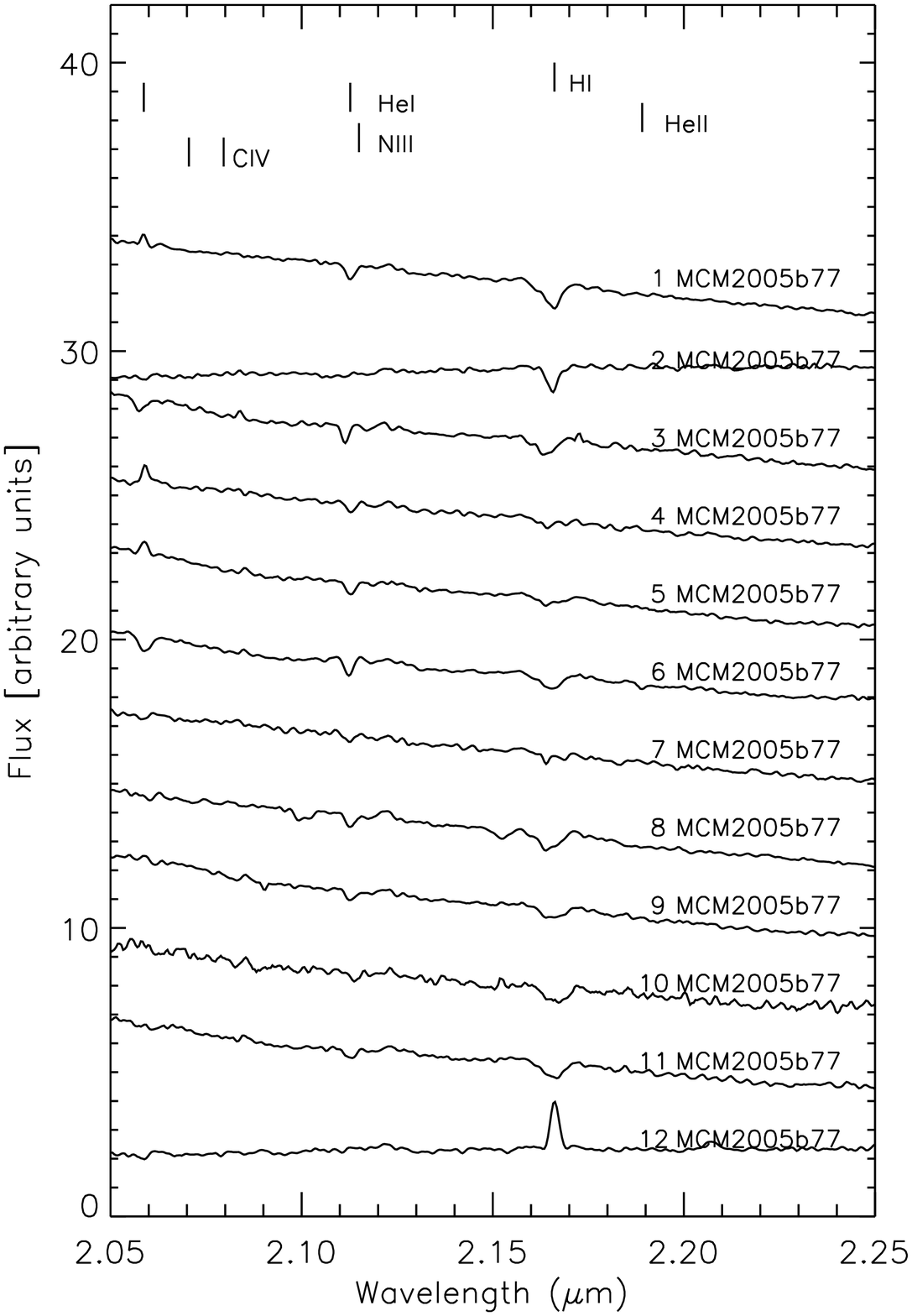}}
\caption{\label{spectra.early} Spectra of detected early-type stars.} 
\end{figure}

\addtocounter{figure}{-1}
\begin{figure}[h]
\resizebox{0.85\hsize}{!}{\includegraphics[angle=0]{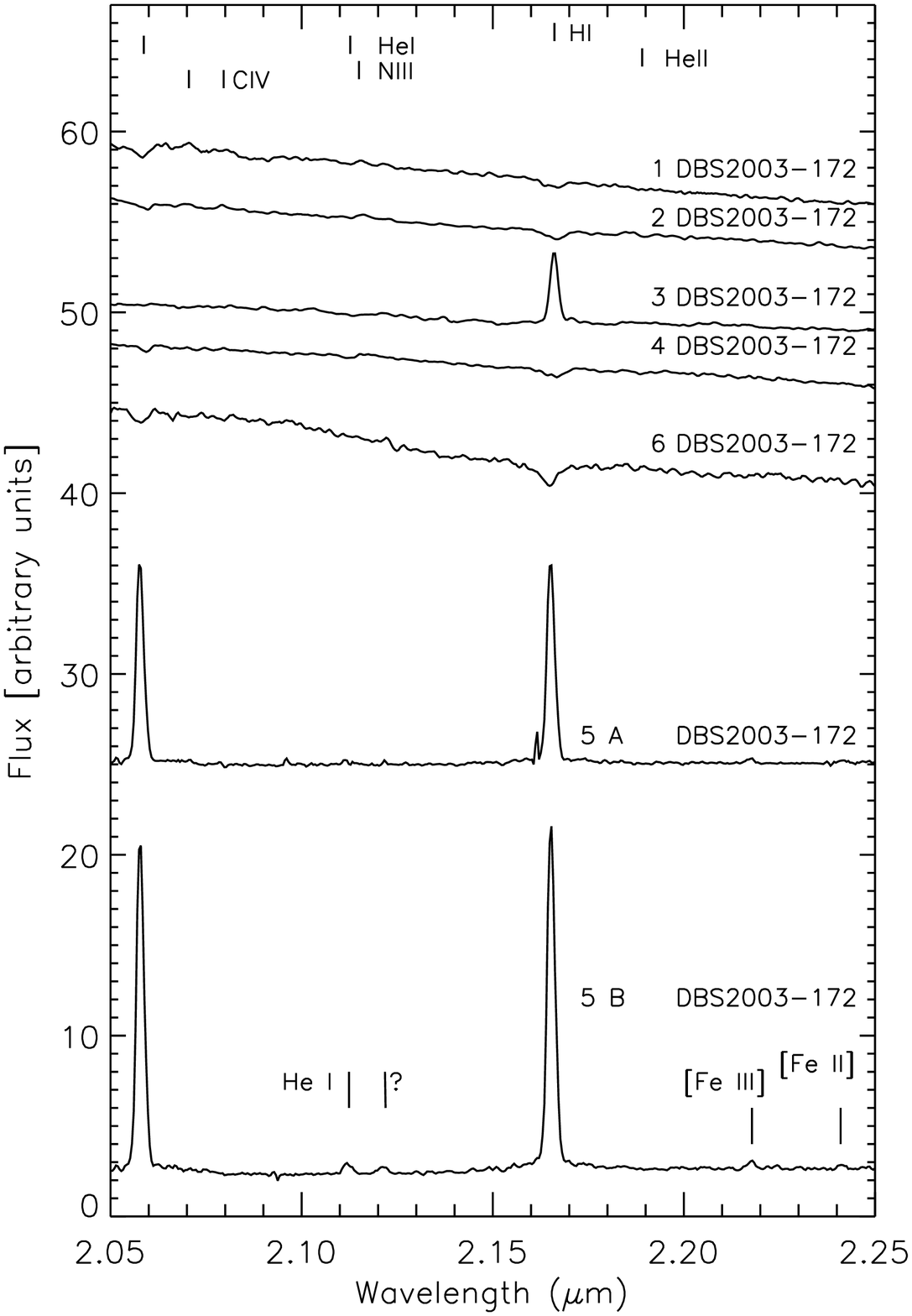}}
\caption{ Continuation of Fig.\ \ref{spectra.early}.} 
\end{figure}

\begin{figure}
\resizebox{\hsize}{!}{\includegraphics[angle=0]{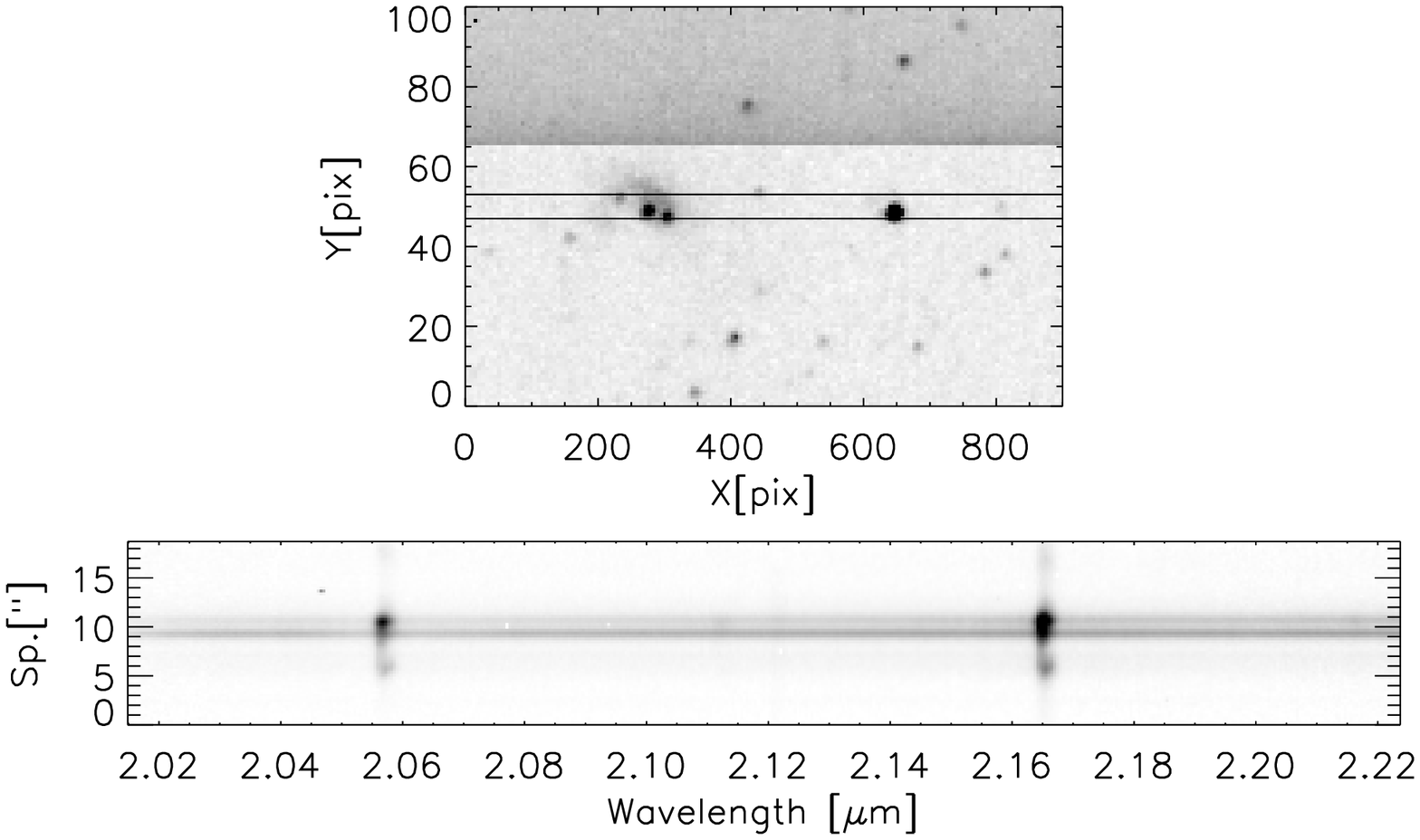}}
\caption{\label{acqn3} {\it Top:} 
acquisition of objects 5A (left) and 5B (right) in the field DBS2003-172 
(at position X= $\sim 250$ pix in the extracted image).  
The slit aperture is indicated with two black lines.
{\it Bottom:} stellar traces of stars 5A (bottom) and 5B (top). 
Knots of \ion{He}{I} at 2.058 \um\ 
and \brg\ are centered on the stellar traces, and are stronger in the
 weaker continuum component (B).
} 
\end{figure}

\section{Analysis}
\label{analysis}

\begin{table*}
\caption{ \label{spectra.early} List of detected early-type stars in the direction of the candidate stellar clusters 
MCM2005b72/DBS2003-157, MCM2005b77, and DBS2003-172. }
\begin{tabular}{lrrllll}
\hline
  Cluster    & ID          &  $\alpha$(J2000)   & $\delta$(J2000)  &           Sp. Type    & $T_{eff}$ & Alias/Comments  \\
             &             & [hh~mm~ss]   & [$^\circ ~ ^\prime ~ ^{\prime \prime}] $ &                        \\
\hline

                  G331.34$-$0.36      &    1     & 16 12 33.23 & $-$51~45~35.6     &            OBAF     &      $..$      &                                 \\
                  G331.34$-$0.36      &    2     & 16 12 30.31 & $-$51~43~53.1     &            B0-5     &  25000   $\pm$  6000      &                                 \\
                  G331.34$-$0.36      &    3     & 16 12 20.54 & $-$51~46~46.0     &            O4-6     &  41000   $\pm$  3000      &   PAC2012-IRS~298$^a$                                \\
                  G331.34$-$0.36      &    4     & 16 12 29.12 & $-$51~46~50.3     &            OBAF     &                 $..$      &   PAC2012-IRS~287                               \\
                  G331.34$-$0.36      &    5     & 16 12 20.03 & $-$51~46~26.2     &          O9-B0     &  31000   $\pm$  2000      &   PAC2012-IRS~339$^b$                              \\
                  G331.34$-$0.36      &    6     & 16 12 33.26 & $-$51~46~16.9     &            B0-3     &  25000   $\pm$  6000      &        2MASS~J16123324-5146173/ PAC2012-IRS~355$^c$    \\
                  G331.34$-$0.36      &    7     & 16 12 42.35 & $-$51~45~01.6     &            OBAF     &      $..$      &                                 \\
                  G331.34$-$0.36      &    8     & 16 12 33.14 & $-$51~47~41.6     &            OBAF     &      $..$      &                                 \\
  
                    MCM2005b77      &    1     & 16 17 23.27 & $-$50~30~19.7     &            B0-3     &  22000   $\pm$  6000      &                                 \\
                    MCM2005b77      &    2     & 16 17 13.39 & $-$50~30~29.3     &            G-F     &      $..$      &                                 \\
                    MCM2005b77      &    3     & 16 17 41.39 & $-$50~28~25.6     &            B0-5     &  21000   $\pm$  7000      &                                 \\
                    MCM2005b77      &    4     & 16 17 30.77 & $-$50~31~16.9     &            B0-3     &  22000   $\pm$  6000      &                                 \\
                    MCM2005b77      &    5     & 16 17 29.97 & $-$50~30~56.7     &            B0-3     &  22000   $\pm$  6000      &                                 \\
                    MCM2005b77      &    6     & 16 17 23.96 & $-$50~31~37.5     &            B0-5     &  21000   $\pm$  7000      &                                 \\
                    MCM2005b77      &    7     & 16 17 29.84 & $-$50~31~24.1     &            B0-5     &  21000   $\pm$  7000      &                                 \\
                    MCM2005b77      &    8     & 16 17 33.83 & $-$50~34~39.1     &            B0-5     &  21000   $\pm$  7000      &                                 \\
                    MCM2005b77      &    9     & 16 17 25.80 & $-$50~30~27.5     &            B0-5     &  21000   $\pm$  7000      &                                 \\
                    MCM2005b77      &   10     & 16 17 25.87 & $-$50~30~36.3     &            B0-5     &  21000   $\pm$  7000      &                      \\
                    MCM2005b77      &   11     & 16 17 25.81 & $-$50~31~11.7     &            B0-5     &  21000   $\pm$  7000      &                                 \\
                    MCM2005b77      &   12     & 16 17 17.70 & $-$50~28~11.5     &              OB     &   $..$      &                                 \\
                
                   DBS2003-172      &    1     & 16 41 11.20 & $-$47~07~13.7     &            OB     &      $..$      &                                 \\
                   DBS2003-172      &    2     & 16 41 07.38 & $-$47~07~33.9     &            OB     &      $..$      &                                 \\
                   DBS2003-172      &    3     & 16 41 09.53 & $-$47~06~56.2     &              OB     &  $..$      &                      \\
                   DBS2003-172      &    4     & 16 41 18.30 & $-$47~06~52.9     &            OB     &      $..$      &                                 \\
                   DBS2003-172      &    5A     & 16 41 08.10 & $-$47~06~46.0     &              OB     &  $..$      &     2MASS J16410805$-$4706466                  \\
                   DBS2003-172      &    5B     & 16 41 08.22 & $-$47~06~46.5     &              OB     &  $..$      &     2MASS J16410805$-$4706466                  \\
                   DBS2003-172      &    6     & 16 41 00.55 & $-$47~05~52.4     &            OBAF     &      $..$      &                                 \\
\hline
\end{tabular}
\begin{list}{}
\item $^a$ PAC2012-IRS~298 is classified as an O6V by  \citet{pinheiro12};
$^b$ PAC2012-IRS~339 is classified as an O9V by  \citet{pinheiro12};
$^c$ PAC2012-IRS~355 is classified as a possible B0 by  \citet{pinheiro12}.
\end{list}
\end{table*}
          
\subsection{Spectral classification of early-type stars}

Spectral classification was performed by comparing  the 
shapes of detected spectral lines from H, He and other 
heavier atoms (e.g. C, Na, Mg, and Fe) with those  in atlases of 
known early-type stars \citep[e.g.][]{hanson96,hanson05,morris96}.

\vspace{+0.8cm}
\noindent
{\it  G331.34$-$0.36 (S62)}\\
In the direction of the \HH\  region G331.34$-$0.36, we detected two B0-5 stars, 
one massive O4-6 star and five other early types with less well determined spectral types.
In Fig. \ref{spectra.early},  stars  \#2 and \#5 show the 
\ion{He}{I} line at 2.1126 \um\  and  \brg\  in absorption, as seen in the spectra of  B0-5 stars. 
It could be argued that there is a possible \ion{He}{II} line 
at 2.1891 \um\ is present in the spectrum of star
\#5  (O9-B0).  The spectrum of the O4-6 star \#3 shows  
two \ion{C}{IV} lines (2.0705 \um\ and 2.0796 \um),  
\ion{N}{III}/\ion{C}{III} emission at $\sim 2.115$ \um,
the \brg\ line in absorption, and the \ion{He}{II} line at 2.1891 \um\ in absorption.
The spectra of stars  \#1, \#4, \#6, \#7, and \#8 show 
only \brg\ in absorption; their spectral types may therefore range from O to F types.
For stars \#3 (O6V), \#5 (O9V), and \#6 (B0) spectra in the bands $J$,$H$, and $K$ were acquired and analyzed
by \citet{pinheiro12}. By having added the detection of CIV lines we confirm that
star \#3 is an early O-type star.

The feature at 2.183 \um\ that appears   in some spectra is 
a glitch,  at the juncture  of the lower and upper quadrants 
(pixels 510-512) of the SofI $1024 \times 1024$ array.

\vspace{+0.8cm}
\noindent
{\it MCM2005b77}\\
In the cluster MCM2005b77, 
the spectra of stars \#1, \#3, \#4, \#5, \#6, \#7, \#8, \#9, and \#11
are characterized by \brg\ lines in absorption, \ion{He}{I} at 2.1126 
\um\ in absorption, and possible \ion{He}{I} at 2.0587 \um.
The latter \ion{He}{I} line appears in emission in the spectra
of stars \#1, \#4, and \#5, indicating a supergiant luminosity class
\citep{davies12,hanson96}.\\
The spectra of stars \#2 and \#10 show only \brg\ in absorption, which is 
compatible with spectral types from B to F.
In Sect. \ref{clMCM2005b77}, we refine their spectral types photometrically, 
to a G-F and a B0-5 star, respectively. 
The spectrum of star \#12 has a strong \brg\ line in emission.

\vspace{+0.8cm}
\noindent
{\it  DBS2003-172 in GRS G337.92$-$00.48 (S36)}\\
We detected seven early-type stars in the direction of DBS2003-172.
The spectra of stars \#1,  \#2, \#4, and \#6  display only the \brg\ 
line in absorption. Their spectral types are therefore poorly constrained and
could range from O to F.
The spectrum of star \#3 shows a strong \brg\ line in emission.

On the spectroscopic exposure, 
two stellar traces are seen at the location of 2MASS J16410805$-$4706466 
(stars \#5A and \#5B).  
Indeed, in the SofI acquisition image taken
with the filter NB\_2.248 (0.030 \um\ wide) and shown in Fig. \ref{acqn3},
there are two stars, separated by about 1\farcs4 (5 pixels), 
which were unresolved in 2MASS images.
We extracted two spectra from the rectified traces, using 
narrow apertures of 2 pixels, centered on the two maxima.
Knots of the \ion{He}{I} line at 2.058 \um\ 
and \brg\ are clearly seen in the traces of both components,
but appear to be stronger in the B component (the source with weaker continuum). 
Diffuse nebular emission was also detected.
For the B component, the ratio between the equivalent width (EW) of the \ion{He}{I} line at 2.058 \um\ and that of the \brg\ 
line is found to be equal to $\sim 1.2$ (with no extinction correction). For the A component, this
ratio is  $\sim 0.9$.
The B component displays  an additional  weak \ion{He}{I} line  at 2.112 \um, 
[FeIII] at 2.2178 \um, and possible [FeII] at 2.241 \um\ (see Table \ref{table5ab}).

\begin{table*}
\caption{ \label{table5ab} Line parameters of stars 5A and 5B in DBS2003-172.}
\begin{tabular}{lllrrl}
\hline
B component\\
\hline
Atom                                        & $\lambda_{rest}$ & $\lambda_{obs}$ & EW  & FWHM &   Line Refs. \\
                                            & [$\mu$m]         & [$\mu$m]        & [$\AA$] & [${\rm km~ s^{-1}}$] & \\ 
\hline
\ion{He}{I}                                    &   2.05869     & 2.05824 & $185\pm3$   &  $332\pm05$    &                   \\ 
\ion{He}{I}                                    &   2.11238     & 2.11250 &   $7\pm1$   &  $424\pm35$    &                      \\
\ion{C}{III} or                                &   2.1217      & 2.12165 &   $4\pm1$   &  $318\pm32$    &             \citet{najarro97}             \\
H$_2$ (ISM)                                    &               &         &             &                &             \citet{tanner05} \\
 \brg                                          &   2.16613     & 2.16573 & $160\pm2$   &  $326\pm05$    &                      \\
$[{\rm Fe~ III}]$   $a^3 G_5 a^3 H_6$          &   2.2184      & 2.21808 &   $6\pm1$   &  $274\pm40$    &             \citet{nahar96} \\ 
$[{\rm Fe~ II}]$   $Z^4 D_{3/2} - C^4 P_{3/2}$ &   2.241       & 2.24167 &   $2\pm2$   &  $273\pm50$    &             \citet{geballe00}              \\
\hline
A component\\
\hline
\ion{He}{I}                                    &   2.05869     & 2.05818 & $109\pm1$    &  $331\pm6$    &                   \\ 
\brg                                           &   2.16613     & 2.16565 & $103\pm1$    &  $320\pm8$    &                      \\
\end{tabular}
\end{table*}

\subsection{Spectral classification of late-type stars}

$K$-band spectra of late-type stars are easily recognizable by  
\ion{CO}{} absorption bands with a band head at 2.2935 \um.  
The strongest  overtone band ($^{12}{\rm CO}$, $\nu$=2-0) was used 
to estimate stellar effective temperatures \citep[e.g.][]{blum03,figer06, messineo10, ramirez00}.
The EW of the CO bands, EW(CO), linearly increases with decreasing stellar temperature
and increases with increasing luminosity class, so red giants and 
supergiants follow  two different linear relations.
We  classified the late-type stars by quantitatively comparing the CO bands with those
of stars in the atlas by \citet{kleinmann86}.
The spectra were initially dereddened with an  average color of $J-$\Ks= 1.05 mag 
and $H-$\Ks = 0.24 mag. The average color implies an uncertainty in \Aks\ of 0.1 mag, 
which has a negligible effect on the EW(CO) measurements. 
EW(CO)s were measured from 2.290 \um\ to 2.307 \um.
Uncertainties were estimated by measuring the EW(CO) twice, with two different assumptions about
the continuum, i.e. by assuming a constant continuum 
\citep[from 2.285 \um\ to 2.290 \um;][]{figer06,kleinmann86} and by using  a first-order 
linear fit to  several continuum regions \citep[from  2.250 \um\ to 2.257 \um,  from 
2.270 \um\ to 2.277 \um, and from 2.285 \um\ to  2.290 \um,][]{ramirez00}. Average
EW(CO)s  are listed in Table \ref{spectra.late}.
Due to a  loss of instrumental sensitivity longward of 2.3 \um,
the high-resolution mode of SofI is not optimal for spectral classification of 
late-type stars. Spectral types were inferred  by comparing
EW(CO) measurements of  \citet{kleinmann86} atlas stars with those of 
stars observed with both the low-resolution mode, which covers to 2.4 \um, and 
high-resolution mode of SofI, as described in \citet{messineo14, messineo17}.
The resulting spectral types are within $\pm2$ subclasses.

\begin{table*}
\caption{\label{spectra.late} List of late-type stars in the direction of the candidate stellar clusters
MCM2005b72/DBS2003-157 (GRS G331.34$-$0.36), MCM2005b77, and DBS2003-172 (GRS G337.92$-$00.48).
}
\begin{tabular}{llllllrrr}       
\hline
\hline
Field                & ID  &   $\alpha$(J2000)  &  $\delta$(J2000)  & EW(CO) &     Sp(RGB) &   $K_S$& $A_{K_S}$ &BCk\\
                     &     &   [hh~mm~ss] & [$^\circ ~ ^\prime ~ ^{\prime\prime}]$ & \AA &  & [mag]& [mag]&[mag]\\
\hline
        G331.34$-$0.36 &   9 &  16 12 34.94 & $-$51 49 01.5 &20.54 $\pm$ 1.39 &         M6 & 5.79$\pm$ 0.02& 0.67$\pm$ 0.02& 3.04    \\
        G331.34$-$0.36 &  10 &  16 12 25.17 & $-$51 55 56.4 &15.20 $\pm$ 3.18 &         M1 & 6.60$\pm$ 0.02& 0.95$\pm$ 0.01& 2.73    \\
        G331.34$-$0.36 &  11 &  16 12 40.06 & $-$51 45 29.5 &20.69 $\pm$ 1.65 &         M6 & 7.21$\pm$ 0.03& 0.62$\pm$ 0.02& 3.04    \\
        G331.34$-$0.36 &  12 &  16 12 36.47 & $-$51 50 19.3 &19.76 $\pm$ 2.12 &         M5 & 7.44$\pm$ 0.03& 0.83$\pm$ 0.02& 2.96    \\
        G331.34$-$0.36 &  13 &  16 12 33.00 & $-$51 45 19.9 &22.90 $\pm$ 2.17 &         M7 & 8.65$\pm$ 0.02& 2.10$\pm$ 0.02& 3.13    \\
        G331.34$-$0.36 &  14 &  16 12 02.22 & $-$51 49 25.1 &10.02 $\pm$ 1.62 &         K1 & 8.91$\pm$ 0.03& 0.17$\pm$ 0.02& 2.45    \\
        G331.34$-$0.36 &  15 &  16 12 42.35 & $-$51 44 23.6 &04.82 $\pm$ 1.00 &         $\la$K0 & 9.00$\pm$ 0.02& 0.14$\pm$ 0.02& 2.40    \\
        G331.34$-$0.36 &  16 &  16 12 27.95 & $-$51 46 35.7 &13.77 $\pm$ 1.36 &         K5 & 9.21$\pm$ 0.03& 0.62$\pm$ 0.02& 2.64    \\
        G331.34$-$0.36 &  17 &  16 12 37.14 & $-$51 47 07.6 &14.14 $\pm$ 0.37 &         K5 &10.00$\pm$ 0.03& 0.22$\pm$ 0.02& 2.64    \\
          MCM2005b77 &  13 &  16 17 07.51 & $-$50 31 06.4 &27.17 $\pm$ 0.88 & $>$M7 Mira & 5.32$\pm$ 0.03& 0.65$\pm$ 0.02& 3.17    \\
          MCM2005b77 &  14 &  16 17 27.40 & $-$50 29 46.5 &31.52 $\pm$ 1.25 & $>$M7 Mira & 6.61$\pm$ 0.03& 1.26$\pm$ 0.02& 3.17    \\
          MCM2005b77 &  15 &  16 17 42.12 & $-$50 30 03.3 &20.03 $\pm$ 1.24 &         M5 & 6.79$\pm$ 0.02& 0.69$\pm$ 0.02& 2.96    \\
          MCM2005b77 &  16 &  16 17 50.70 & $-$50 27 53.6 &20.51 $\pm$ 1.22 &         M6 & 6.96$\pm$ 0.02& 0.35$\pm$ 0.02& 3.04    \\
          MCM2005b77 &  17 &  16 17 08.65 & $-$50 27 36.0 &20.38 $\pm$ 1.19 &         M6 & 7.04$\pm$ 0.02& 0.82$\pm$ 0.02& 3.04    \\
          MCM2005b77 &  18($^a$) &  16 17 06.87 & $-$50 30 45.6 &11.06 $\pm$ 0.47 &         K2 & 7.21$\pm$ 0.03& 0.18$\pm$ 0.02& 2.50    \\
          MCM2005b77 &  19 &  16 17 28.15 & $-$50 33 40.4 &20.31 $\pm$ 1.60 &         M6 & 7.29$\pm$ 0.02& 1.12$\pm$ 0.02& 3.04    \\
          MCM2005b77 &  20 &  16 17 03.38 & $-$50 28 33.4 &13.22 $\pm$ 1.05 &         K5 & 7.34$\pm$ 0.02& 0.26$\pm$ 0.01& 2.64    \\
          MCM2005b77 &  21 &  16 17 19.95 & $-$50 27 59.2 &21.32 $\pm$ 1.58 &         M7 & 7.48$\pm$ 0.02& 0.83$\pm$ 0.02& 3.13    \\
          MCM2005b77 &  22($^{b}$) &  16 17 25.79 & $-$50 27 20.8 &16.48 $\pm$ 0.37 &    M2 Mira & 7.59$\pm$ 0.03& 3.60$\pm$ 0.05& 2.80    \\
          MCM2005b77 &  23 &  16 17 45.95 & $-$50 34 05.1 &19.78 $\pm$ 1.41 &         M5 & 7.61$\pm$ 0.03& 0.88$\pm$ 0.02& 2.96    \\
          MCM2005b77 &  24 &  16 17 10.36 & $-$50 33 59.3 &21.67 $\pm$ 1.53 &         M7 & 7.83$\pm$ 0.02& 1.26$\pm$ 0.02& 3.13    \\
          MCM2005b77 &  25 &  16 17 21.30 & $-$50 26 43.4 &16.54 $\pm$ 1.43 &         M2 & 7.99$\pm$ 0.02& 1.10$\pm$ 0.02& 2.80    \\
          MCM2005b77 &  26 &  16 17 26.00 & $-$50 30 23.0 &09.77 $\pm$ 1.03 &         K1 &10.51$\pm$ 0.03& 0.17$\pm$ 0.04& 2.45    \\
          MCM2005b77 &  27 &  16 17 29.89 & $-$50 32 41.9 &10.86 $\pm$ 1.87 &         K2 &10.70$\pm$ 0.03& 0.60$\pm$ 0.02& 2.50    \\
          MCM2005b77 &  28 &  16 17 23.97 & $-$50 30 47.3 &10.15 $\pm$ 1.03 &         K2 &11.00$\pm$ 0.03& 0.31$\pm$ 0.02& 2.50    \\
          MCM2005b77 &  29 &  16 17 36.20 & $-$50 30 34.0 &15.84 $\pm$ 0.27 &         M1 &11.39$\pm$ 0.03& 1.28$\pm$ 0.03& 2.73    \\
          MCM2005b77 &  30 &  16 17 23.92 & $-$50 31 26.8 &20.99 $\pm$ 1.13 &         M6 &11.63$\pm$ 0.05& 1.30$\pm$ 0.05& 3.04    \\
         DBS2003-172 &   7 &  16 41 08.56 & $-$47 08 25.7 &04.32 $\pm$ 2.36 &         $\la$K0 & 8.51$\pm$ 0.02& 0.08$\pm$ 0.02& 2.40    \\
         DBS2003-172 &   8 &  16 41 06.24 & $-$47 10 03.7 &21.36 $\pm$ 0.50 &         M7 & 8.63$\pm$ 0.02& 2.54$\pm$ 0.02& 3.13    \\
         DBS2003-172 &   9 &  16 41 06.13 & $-$47 11 14.8 &07.49 $\pm$ 1.18 &         $\la$K0 & 9.07$\pm$ 0.03& 0.18$\pm$ 0.02& 2.40    \\
         DBS2003-172 &  10 &  16 41 06.02 & $-$47 07 32.4 &21.76 $\pm$ 0.20 &         M7 & 9.49$\pm$ 0.80& 2.16$\pm$ 0.44& 3.13    \\
         DBS2003-172 &  11 &  16 41 07.70 & $-$47 08 15.3 &23.63 $\pm$ 3.99 &         M7 &10.45$\pm$ 0.02& 3.13$\pm$ 0.04& 3.13    \\
\hline
\end{tabular}
\begin{list}{}{}
\item[{\bf Notes.}]  
($^a$) MCM2005b77-18=CD-50 10319.~ ($^b$) MCM2005b77-22=IRAS 16136-5020.~
\end{list}
\end{table*}

\subsection{Available photometry}
We searched for counterparts in the near-infrared catalogs  from 
the Two Micron all Sky Survey  \citep[2MASS,][]{skrutskie06} and the Deep Near-infrared Survey
\citep[DENIS,][]{epchtein99}, using a search radius of 2\farcs0.
Matches with 2MASS datapoints were visually checked by superimposing
2MASS datapoints on the acquisition charts (astrometrically aligned to 2MASS).

We also searched  the mid-infrared catalogs of the 
Galactic Legacy Infrared Mid-Plane Survey Extraordinaire  \citep[GLIMPSE,][]{benjamin03}, the
Midcourse Space Experiment   \citep[MSX,][]{egan03,price01}, and the Wide-field Infrared Survey 
Explorer    \citep[WISE,][]{wright10}, using search radii of 2\farcs0, 
5\farcs0, and 2\farcs0, respectively.
We searched for $BVR$ measurements in the Naval Observatory Merged Astrometric Data set
(NOMAD) by \citet{zacharias05} and for  data 
 in the  second release (DR2) of the GAIA catalog  \citep{gaia18}.

For all but three detected stars, infrared charts (provided in Appendix)
were  available from the
VISTA Variables in the Via Lactea survey (VVV) \citep[e.g.][]{soto13}.  
VVV \Ks-band sources are in the regime of linearity up to 
$\approx$ 11.5 mag. Detected targets have \Ks\ from 5.57 mag to 11.63 mag, 
so 2MASS measurements  are adopted  for most of the 
targets. 

In Tables  \ref{table:phot}  and \ref{table:photlate}, we show
the collected photometric measurements of the early-type and late-type 
stars detected in GRS G331.34$-$0.36, MCM2005b77, and GRS G337.92$-$00.48.

\subsubsection{PSF photometry from VVV images}
\begin{figure}
\begin{center}
\resizebox{1.0\hsize}{!}{\includegraphics[angle=0]{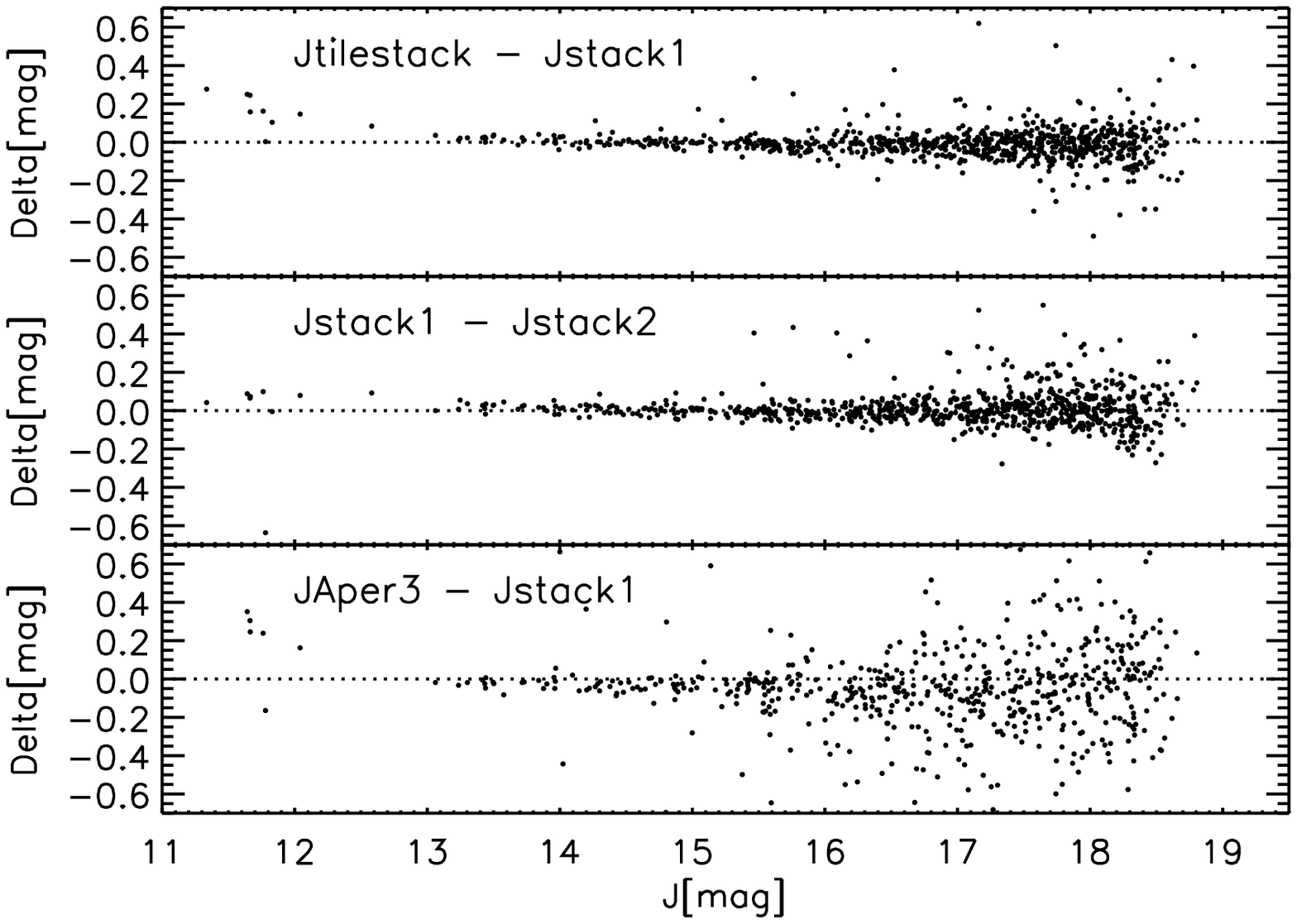}}
\caption{\label{psf} {\it Upper panel:} Difference between VVV $J$ magnitudes 
extracted from one tile stack and those from one stack  versus $J$ magnitudes, 
for sources detected in the direction of the star-forming region  
DBS2003-172. {\it Middle panel:} Difference between VVV $J$ magnitudes 
extracted from two  stacks  versus $J$ magnitudes. 
{\it Lower panel:}  Difference between $J$ magnitudes 
from the CASU pipeline (Aper3) and those from one stack  versus $J$ magnitudes. }
\end{center}
\end{figure}

For fields GRS G331.34$-$0.36 (MCM2005b72/DBS2003-157), MCM2005b77, 
and GRS G337.92$-$00.48 (DBS2003-172), photometric catalogs 
of fainter stars were extracted from VVV images using the point-spread function (PSF) fitting 
routines by \citet{stetson87}. 
Images (stacks and tile stacks) were retrieved from the VISTA Science Archive (VSA)\footnote{http://horus.roe.ac.uk/vsa/index.html} and reduced with the
the Cambridge Astronomy Survey Unit (CASU) pipeline\footnote{http://casu.ast.cam.ac.uk}.
Stacks are  created from two dithered exposures of the same chip.
Tile stacks are  obtained by combining six offset exposures per observation to yield 
 contiguous coverage. 
For  DBS2003-172, a comparison of measurements extracted from the tile stack
and from  individual stacks of  a $ 3^\prime \times 3^\prime$ field is given in  Fig.\ \ref{psf}.
For the DBS2003-172 field and for the GRS G331.34$-$0.36 field ($ 8^\prime \times 8^\prime$), 
target measurements were analyzed in four stacks 
(EXPTIME =10s, 10s, and 4s), but eventually, for the surrounding field, the magnitudes 
measured on the tile stack were retained.
For MCM2005b77, one stack that covered the whole cluster was available 
($ 6^\prime \times 6^\prime$, EXPTIME = 10 s in $J$-band, $H$-band, and  in \Ks\ bands).

Catalogs extracted from VVV data were combined with 2MASS datapoints.
Absolute calibration was performed  with 2MASS datapoints 
from  12.5 to 15 in $J$ mag, from 12.5 to 14.5 in $H$ band,
and from 11.5 to 13.0 mag in \Ks\ band.
2MASS magnitudes were retained for bright stars that were saturated in the VVV images.

For GRS G331.34$-$0.36 $JH$\Ks\ photometry of the central 
$2^\prime \times 2^\prime$ taken with a spatial 
scale of 0\farcs48  pix$^{-1}$ is also available from \citet{pinheiro12}.

\subsubsection{Photometry of DBS2003-172 \#5A \& \#5B}
Stars \#5 A and B are blended in 2MASS.
The acquisition image taken with the narrowband filter NB2.248 
(2.248 \um) on SofI shows two stars, DBS2003-172 \#5A and \#5B, surrounded by a nebular arc.
By measuring their counts, and those of nearby stars, on the acquisition image, using a PSF model
and scaling them by the 2MASS magnitudes,
we estimated   \Ks $ \approx 10.82$ mag for component A
and \Ks $\approx 11.24$ mag for component B. 
These \Ks\ magnitude estimates are consistent with values from the VVV survey.
In VVV images, component A  is clearly detected in all bands, but component B
is detected  only in the \Ks\ band. 
\Ks\ magnitudes are above the linear regime (within 0.6 mag): 
 we measured 
$(J,H,$\Ks)=$(14.257\pm0.010,11.750\pm0.400,10.785\pm0.035$ mag) 
for component A  and $(H,$\Ks)=$(13.495\pm0.140,11.115\pm0.125$ mag) for component B.

\begin{sidewaystable*} 
\vspace*{-5.5cm}
\caption{ \label{table:phot} Photometric data of early-type stars in  MCM2005b72/DBS2003-157, MCM2005b77, and DBS2003-172.  } 
\begin{tabular}{@{\extracolsep{-.06in}}lr|rrr|rrr|rrrr|r|rr|rrr|rrrr|rr}
\hline 
\hline 
    &  &   \multicolumn{3}{c}{\rm 2MASS}   &\multicolumn{3}{c}{\rm DENIS} &   \multicolumn{4}{c}{\rm GLIMPSE}   &  \multicolumn{1}{c}{\rm MSX}& \multicolumn{2}{c}{\rm WISE} & \multicolumn{3}{c}{\rm NOMAD} & \multicolumn{4}{c}{{\rm { GAIA}}} &&\\ 
\hline 
Field & {\rm ID} &  $J$ & $H$ & $K_S$  &
 $I$ & $J$ & $K_S$ & $[3.6]$ & $[4.5]$ & $[5.8]$ & $[8.0]$ &
 $A$  &$W1$ &$W2$ &
 $B$  &$V$ &$R$ &${ pmRa}$&${ pmDec}$ &${ G}$ &${ plx}$ & var\\  
\hline 
&    &{\rm [mag]}   &	{\rm [mag]}    & {\rm [mag]}     & {\rm [mag]} &{\rm [mag]}  & {\rm [mag]}  & {\rm [mag]}  & {\rm [mag]}  & {\rm [mag]}&{\rm [mag]}&{\rm [mag]}& {\rm [mag]}&{\rm [mag]}&{\rm [mag]}& {\rm [mag]}&{\rm [mag]}&{\rm [mas/yr]}&{\rm [mas/yr]}& {\rm [mag]}&{\rm [mas]}\\ 
\hline

                  G331.34$-$0.36  &    1  &  10.02 &  9.08 &  8.67 & 12.68 & 10.02 &  8.61 &  8.38 &  8.32 &  8.18 &  8.25 & \nodata &  8.36 &  8.25 & \nodata &  16.28 & \nodata &   $-$4.15$\pm$   0.16 &   $-$3.84$\pm$   0.10 &  14.26 &  $-$0.11$\pm$  0.10 &  0 &  \\
                  G331.34$-$0.36  &    2  &   9.74 &  9.18 &  8.89 & 11.70 &  9.71 &  8.87 &  8.72 &  8.63 &  8.60 &  8.83 & \nodata &  8.63 &  8.48 &  15.88 &  14.30 &  12.82 &   $-$3.76$\pm$   0.61 &   $-$1.39$\pm$   0.42 &  13.13 &   0.07$\pm$  0.36 &  0 &  \\
                  G331.34$-$0.36  &    3  &  11.70 & 10.27 &  9.57 & 16.67 & 11.62 &  9.47 &  9.01 &  8.88 &  8.80 &  8.87 & \nodata & \nodata & \nodata & \nodata & \nodata & \nodata &   $-$2.44$\pm$   0.42 &   $-$2.93$\pm$   0.28 &  17.87 &  $-$0.15$\pm$  0.28 &  0 &  \\
                  G331.34$-$0.36  &    4  &  11.20 & 10.14 &  9.72 & 14.14 & 11.24 &  9.65 &  9.36 &  9.32 &  9.28 & \nodata & \nodata &  9.29 &  9.03 &  15.74 &  15.17 & \nodata &   $-$3.94$\pm$   0.19 &   $-$4.45$\pm$   0.13 &  15.77 &  $-$0.06$\pm$  0.13 &  0 &  \\
                  G331.34$-$0.36  &    5$^+$  &  11.78 &  10.60 &  9.99 & 15.84 & 11.75 &  9.98 &  9.53 &  9.48 &  9.48 & \nodata & \nodata &  9.35 &  8.91 &  17.53 & \nodata &  17.35 &   $-$3.98$\pm$   0.36 &   $-$3.50$\pm$   0.21 &  17.45 &  $-$0.12$\pm$  0.21 &  0 &  \\
                  G331.34$-$0.36  &    6$^+$  &  11.73 & 10.69 & 10.15 & 15.36 & 11.75 & 10.07 &  9.76 &  9.51 &  9.17 &  7.46 & \nodata &  9.45 &  9.29 & \nodata & \nodata &  19.10 &   $-$4.19$\pm$   0.26 &   $-$3.68$\pm$   0.16 &  16.84 &   0.08$\pm$  0.17 &  0 &  \\
                  G331.34$-$0.36  &    7$^{*+}$  &  13.64 &  11.68 & 10.61 & \nodata & 13.47 & 10.65 & \nodata & \nodata & \nodata & \nodata &  2.35 & 7.89 & 6.90  & \nodata & \nodata & \nodata &   $-$1.21$\pm$   4.14 &   $-$4.24$\pm$   3.75 &  20.90 &  $-$0.44$\pm$  2.29 &  0 &  \\
                  G331.34$-$0.36  &    8  &  11.61 & 11.26 & 11.08 & 12.62 & 11.59 & 11.05 & 10.96 & 10.79 & 10.57 & \nodata & \nodata & 10.93 & 10.57 &  14.34 &  13.72 &  13.83 &   $-$2.88$\pm$   0.05 &   $-$3.84$\pm$   0.03 &  13.51 &   0.51$\pm$  0.03 &  0 &  \\
                    MCM2005b77  &    1  &   7.99 &  7.03 &  6.49 & 11.39 &  7.91 &  6.46 &  6.74 & \nodata &  5.96 &  5.93 &  5.86 &  6.20 &  5.92 &  18.63 &  16.22 &  13.14 &   $-$3.08$\pm$   0.20 &   $-$5.18$\pm$   0.14 &  12.86 &   1.02$\pm$  0.15 &  0 &  \\
                    MCM2005b77  &    2  &  12.89 &  9.51 &  7.80 & \nodata & 12.83 &  7.78 &  6.82 & \nodata &  5.91 &  6.04 &  5.92 &  6.69 &  6.07 & \nodata & \nodata & \nodata &    $..$  &    $..$  &  \nodata &   $..$  &  0 &  \\
                    MCM2005b77  &    3  &  10.24 &  8.91 &  8.22 & 14.85 & 10.15 &  8.14 &  7.61 &  7.52 &  7.42 &  7.39 & \nodata &  7.41 &  7.20 & \nodata & \nodata &  17.30 &   $-$3.49$\pm$   0.30 &   $-$4.34$\pm$   0.20 &  16.30 &   0.84$\pm$  0.19 &  0 &  \\
                    MCM2005b77  &    4  &   9.81 &  8.90 &  8.41 & 13.12 &  9.73 &  8.30 &  8.03 &  7.94 &  7.82 &  7.86 & \nodata &  8.00 &  7.85 & \nodata & \nodata & \nodata &   $-$3.94$\pm$   0.19 &   $-$3.95$\pm$   0.13 &  14.67 &   0.51$\pm$  0.12 &  0 &  \\
                    MCM2005b77  &    5  &   9.99 &  9.09 &  8.59 & 13.19 & 10.01 &  8.55 &  8.17 &  8.03 &  7.89 &  7.96 & \nodata &  8.11 &  7.94 &  16.20 &  15.78 &  14.92 &   $-$4.25$\pm$   0.20 &   $-$4.16$\pm$   0.13 &  14.85 &   0.39$\pm$  0.13 &  0 &  \\
                    MCM2005b77  &    6  &  10.40 &  9.26 &  8.71 & 14.49 & 10.40 &  8.70 &  8.33 &  8.13 &  7.99 &  8.05 & \nodata &  8.24 &  7.98 & \nodata & \nodata &  17.71 &   $-$4.47$\pm$   0.24 &   $-$4.03$\pm$   0.16 &  15.89 &   0.25$\pm$  0.16 &  0 &  \\
                    MCM2005b77  &    7  &  10.52 &  9.49 &  8.99 & 14.09 & 10.54 &  8.95 &  8.59 &  8.54 &  8.36 &  8.43 & \nodata &  8.62 &  8.46 & \nodata & \nodata &  16.79 &   $-$4.40$\pm$   0.20 &   $-$4.05$\pm$   0.13 &  15.57 &   0.40$\pm$  0.14 &  0 &  \\
                    MCM2005b77  &    8  &  10.57 &  9.71 &  9.30 & 13.44 & 10.49 &  9.33 &  8.99 &  8.92 &  8.85 &  8.97 & \nodata &  9.02 &  8.95 &  18.08 &  17.71 &  15.37 &   $-$4.51$\pm$   0.16 &   $-$3.76$\pm$   0.10 &  14.97 &   0.54$\pm$  0.10 &  0 &  \\
                    MCM2005b77  &    9  &  10.80 &  9.87 &  9.43 & 13.91 & 10.81 &  9.36 &  9.04 &  8.94 &  8.88 &  8.80 & \nodata &  8.72 &  8.49 &  16.11 &  15.40 & \nodata &   $-$4.38$\pm$   0.20 &   $-$3.89$\pm$   0.13 &  15.74 &   0.39$\pm$  0.13 &  0 &  \\
                    MCM2005b77  &   10  &  11.69 & 10.68 & 10.13 & 15.39 & 11.78 & 10.14 &  9.85 &  9.73 &  9.69 &  9.78 & \nodata & \nodata & \nodata &  17.83 & \nodata &  16.90 &   $-$7.73$\pm$   0.54 &   $-$7.14$\pm$   0.32 &  17.06 &  $-$0.54$\pm$  0.33 &  0 &  \\
                    MCM2005b77  &   11  &  11.94 & 10.82 & 10.28 & \nodata & 11.96 & 10.22 &9.89 &  9.74 &  9.66 &  9.80 & \nodata & \nodata & \nodata & \nodata & \nodata & \nodata &   $-$4.53$\pm$   0.33 &   $-$4.09$\pm$   0.20 &  17.54 &   0.29$\pm$  0.21 &  0 &  \\
                    MCM2005b77  &   12  &  14.34 & 12.31 & 10.97 & \nodata & 14.36 & 10.94 &  9.41 &  8.93 &  8.48 &  8.27 & \nodata &  9.56 &  8.83 & \nodata & \nodata & \nodata &    $..$  &    $..$  &  \nodata &   $..$  &  0 &  \\
                   DBS2003-172  &    1  &  10.45 &  9.25 &  8.55 & 14.59 & 10.25 &  8.58 &    8.33 &  8.15 &  8.09 & \nodata & \nodata & 7.77 & 7.14 & \nodata & \nodata &  17.97 &   $-$1.00$\pm$   0.46 &   $-$2.86$\pm$   0.34 &  16.40 &   0.13$\pm$  0.27 &  0 &  \\
                   DBS2003-172  &    2  &  10.45 &  9.43 &  8.95 & 13.90 & 10.44 &  8.91 &  8.60 &  8.41 & \nodata & \nodata & \nodata & \nodata & \nodata & \nodata & \nodata &  16.58 &   $-$1.30$\pm$   0.21 &   $-$3.59$\pm$   0.18 &  15.32 &  $-$0.43$\pm$  0.14 &  0 &  \\
                   DBS2003-172  &    3  &  11.63 & 10.38 &  9.59 & 15.42 & 11.50 &  9.64 &  8.89 &  8.53 &  8.08 & \nodata & \nodata & \nodata & \nodata & \nodata & \nodata &  19.65 &   $-$1.79$\pm$   0.34 &   $-$2.97$\pm$   0.24 &  17.05 &   0.01$\pm$  0.20 &  0 &  \\
                   DBS2003-172  &    4  &  11.46 & 10.45 &  9.98 & 14.99 & 11.50 &  9.95 &  9.70 &  9.55 &  9.51 &  9.47 & \nodata & \nodata & \nodata & \nodata & \nodata &  18.74 &   $-$1.61$\pm$   0.27 &   $-$3.15$\pm$   0.21 &  16.45 &   0.36$\pm$  0.18 &  0 &  \\
                   DBS2003-172  &    5A$^*$  &  14.26 & 11.75 & 10.78 & \nodata & 12.99 &  9.42 & \nodata & \nodata & \nodata & \nodata & \nodata & \nodata & \nodata & \nodata & \nodata & \nodata &    $..$  &    $..$  &  \nodata &   $..$  &  1 &  \\
                   DBS2003-172  &    5B$^*$  &  \nodata & 13.49 & 11.11 & \nodata & 12.99 &  9.42 & \nodata & \nodata & \nodata & \nodata & \nodata & \nodata & \nodata & \nodata & \nodata & \nodata &    $..$  &    $..$  &  \nodata &   $..$  &  1 &  \\
                   DBS2003-172  &    6  &  11.65 & 11.25 & 11.05 & 12.84 & 11.76 & 11.06 & 10.98 & 10.88 & 10.82 & \nodata & \nodata & \nodata & \nodata &  14.99 &  13.82 &  13.79 &   $-$3.34$\pm$   0.05 &   $-$2.40$\pm$   0.04 &  13.84 &   0.41$\pm$  0.03 &  0 &  \\
\hline
\end{tabular}
\begin{list}{}{}
\item[{\bf Notes.}] ($^+$) 2MASS upper limits were replaced with photometric measurements from \citet{pinheiro12}. 
(VVV magnitudes were  above  linearity.)
($^*$) Magnitudes are estimated from  VVV images.~ 
 The $H$ magnitude of 5 A is 
not in the linear regime (within 0.4 mag). The \Ks\ magnitudes are also above linearity.
However, estimates from the SofI acquisition image yielded values within 0.12 mag.~
(a) Star \#6  in GRS G331.34$-$0.36 (2MASS J16123324$-$5146173) was previously 
photometrically classified as a 
young stellar object \citep{robitalle08}. The detection of the  \brg\ 
line in absorption indicates a later stage (a dwarf).
\end{list}
\end{sidewaystable*}

\begin{sidewaystable*} 
\vspace*{+10.5cm}
\caption{ \label{table:photlate} Photometric data of late-type stars in  MCM2005b72/DBS2003-157, MCM2005b77, and DBS2003-172.} 
\begin{tabular}{@{\extracolsep{-.06in}}lr|rrr|rrr|rrrr|r|rr|rrr|rrrr|rr}
\hline 
\hline 
    &  &   \multicolumn{3}{c}{\rm 2MASS}   &\multicolumn{3}{c}{\rm DENIS} &   \multicolumn{4}{c}{\rm GLIMPSE}   &  \multicolumn{1}{c}{\rm MSX}& \multicolumn{2}{c}{\rm WISE} & \multicolumn{3}{c}{\rm NOMAD} & \multicolumn{4}{c}{\rm  GAIA} &&\\ 
\hline 
Field & {\rm ID} &  $J$ & $H$ & $K_S$  &
 $I$ & $J$ & $K_S$ & $[3.6]$ & $[4.5]$ & $[5.8]$ & $[8.0]$ &
 $A$  &$W1$ &$W2$ &
 $B$  &$V$ &$R$ &$ pmRa$&$ pmDec$ &$ G$ &$ plx$ & var\\  
\hline 
&    &{\rm [mag]}   &	{\rm [mag]}    & {\rm [mag]}     & {\rm [mag]} &{\rm [mag]}  & {\rm [mag]}  & {\rm [mag]}  & {\rm [mag]}  & {\rm [mag]}&{\rm [mag]}&{\rm [mag]}& {\rm [mag]}&{\rm [mag]}&{\rm [mag]}& {\rm [mag]}&{\rm [mag]}&{\rm [mas/yr]}&{\rm [mas/yr]}& {\rm [mag]}&{\rm [mas]}\\ 
\hline
                  G331.34$-$0.36  &    9  &   7.82 &  6.39 &  5.79 & 11.49 &  7.12 &  4.92 &  5.51 &  6.13 &  5.34 &  5.37 & \nodata &  5.41 &  5.36 &  19.56 & \nodata &  14.72 &   $-$1.38$\pm$   0.19 &   $-$4.23$\pm$   0.13 &  12.98 &   0.29$\pm$  0.13 &  0 &  \\
                  G331.34$-$0.36  &   10  &   8.99 &  7.28 &  6.60 & 13.43 &  9.02 &  6.69 &  6.66 &  6.35 &  6.05 &  6.11 & \nodata &  6.18 &  6.21 & \nodata & \nodata & \nodata &   $-$3.04$\pm$   0.21 &   $-$2.69$\pm$   0.14 &  14.88 &  $-$0.09$\pm$  0.13 &  0 &  \\
                  G331.34$-$0.36  &   11  &   9.32 &  7.83 &  7.21 & 13.02 &  9.31 &  7.22 &  6.92 &  7.01 &  6.76 &  6.83 & \nodata &  6.82 &  6.97 & \nodata &  14.99 & \nodata &   $-$3.25$\pm$   0.21 &   $-$1.80$\pm$   0.14 &  14.65 &  $-$0.15$\pm$  0.14 &  0 &  \\
                  G331.34$-$0.36  &   12  &   9.77 &  8.15 &  7.44 & 14.07 &  9.76 &  7.45 &  7.07 &  7.18 &  6.95 &  6.97 & \nodata &  7.04 &  7.13 & \nodata & \nodata &  18.92 &   $-$4.88$\pm$   0.23 &   $-$5.33$\pm$   0.16 &  15.46 &  $-$0.24$\pm$  0.15 &  0 &  \\
                  G331.34$-$0.36  &   13  &  13.56 & 10.19 &  8.65 & \nodata & 13.50 &  8.61 &  7.57 &  7.59 &  7.21 &  7.26 & \nodata &  7.68 &  7.53 & \nodata & \nodata & \nodata &    $..$  &    $..$  &  \nodata &   $..$  &  0 &  \\
                  G331.34$-$0.36  &   14  &   9.80 &  9.16 &  8.91 & 10.98 &  9.77 &  8.81 &  8.77 &  8.90 &  8.75 &  8.76 & \nodata &  8.54 &  8.66 &  13.54 &  12.55 &  11.96 &   $-$6.83$\pm$   0.06 &   $-$2.96$\pm$   0.04 &  12.02 &   0.81$\pm$  0.04 &  0 &  \\
                  G331.34$-$0.36  &   15  &   9.84 &  9.19 &  9.00 & 10.99 &  9.80 &  8.94 &  8.88 &  8.97 &  8.89 & \nodata & \nodata &  8.73 &  8.70 &  13.57 &  12.50 &  11.87 &   $-$2.85$\pm$   0.07 &   $-$5.64$\pm$   0.05 &  12.02 &   0.70$\pm$  0.04 &  0 &  \\
                  G331.34$-$0.36  &   16  &  10.98 &  9.72 &  9.21 & 13.98 & 10.95 &  9.19 &  8.98 &  9.02 &  8.75 & \nodata & \nodata &  8.74 &  8.72 &  19.91 & \nodata &  16.21 &   $-$3.16$\pm$   0.17 &   $-$3.91$\pm$   0.12 &  15.55 &   0.07$\pm$  0.12 &  0 &  \\
                  G331.34$-$0.36  &   17  &  11.02 & 10.27 & 10.00 & 12.52 & 11.07 &  9.98 &  9.88 &  9.94 &  9.79 &  9.74 & \nodata &  9.94 &  9.94 &  15.53 &  14.11 &  13.54 &   $-$1.27$\pm$   0.05 &   $-$4.55$\pm$   0.03 &  13.70 &   0.55$\pm$  0.03 &  0 &  \\
                    MCM2005b77  &   13  &   7.60 &  6.17 &  5.32 & 12.80 &  8.09 &  4.63 & \nodata & \nodata &  4.07 &  4.02 &  3.32 &  4.46 &  3.52 &  14.97 &  14.99 &  14.56 &   $-$3.68$\pm$   0.28 &   $-$3.40$\pm$   0.19 &  14.23 &   0.35$\pm$  0.19 &  1 &  \\
                    MCM2005b77  &   14  &  10.15 &  7.83 &  6.61 & \nodata & 10.18 &  6.65 & \nodata &  5.82 &  5.17 &  4.96 &  5.17 &  5.95 &  5.18 & \nodata & \nodata & \nodata &   $-$4.93$\pm$   0.76 &   $-$5.47$\pm$   0.47 &  19.08 &   0.27$\pm$  0.50 &  0 &  \\
                    MCM2005b77  &   15  &   8.85 &  7.43 &  6.79 & 12.69 &  8.77 &  6.72 &  6.67 &  6.62 &  6.36 &  6.29 & \nodata &  6.45 &  6.55 & \nodata & \nodata &  16.15 &   $-$1.84$\pm$   0.22 &   $-$0.19$\pm$   0.15 &  14.11 &   0.53$\pm$  0.14 &  0 &  \\
                    MCM2005b77  &   16  &   8.39 &  7.38 &  6.96 & 10.77 &  8.39 &  6.94 &  6.87 &  7.02 &  6.76 &  6.71 & \nodata &  6.60 &  6.83 &  14.87 &  13.85 & \nodata &   $-$5.50$\pm$   0.15 &   $-$2.46$\pm$   0.11 &  12.31 &   0.72$\pm$  0.10 &  0 &  \\
                    MCM2005b77  &   17  &   9.37 &  7.76 &  7.05 & 13.66 &  9.26 &  7.14 &  6.90 &  6.93 &  6.64 &  6.57 &  6.02 &  6.74 &  6.85 &  16.33 & \nodata &  16.82 &   $-$8.96$\pm$   0.23 &   $-$7.50$\pm$   0.15 &  14.99 &   0.47$\pm$  0.15 &  0 &  \\
                    MCM2005b77  &   18  &   8.13 &  7.44 &  7.21 &  9.49 &  8.03 &  7.21 &  7.18 &  7.24 &  7.13 &  7.12 & \nodata &  7.12 &  7.28 &  12.99 &  11.22 &  10.34 &   $-$2.75$\pm$   0.07 &   $-$0.19$\pm$   0.04 &  10.36 &   1.03$\pm$  0.04 &  0 &  \\
                    MCM2005b77  &   19  &  10.16 &  8.13 &  7.29 & 15.63 & 10.13 &  7.35 &  6.69 &  6.95 &  6.62 &  6.64 & \nodata &  6.72 &  6.80 & \nodata & \nodata & \nodata &   $-$5.22$\pm$   0.35 &   $-$5.04$\pm$   0.20 &  16.85 &   0.44$\pm$  0.21 &  0 &  \\
                    MCM2005b77  &   20  &   8.46 &  7.64 &  7.34 & 10.07 &  8.50 &  7.38 &  7.22 &  7.34 &  7.19 &  7.11 & \nodata &  7.09 &  7.19 &  13.29 &  12.45 &  11.20 &   $-$2.33$\pm$   0.10 &   $-$1.82$\pm$   0.07 &  11.18 &   0.39$\pm$  0.07 &  0 &  \\
                    MCM2005b77  &   21  &   9.99 &  8.24 &  7.48 & 14.71 &  9.99 &  7.43 &  7.04 &  7.26 &  6.93 &  6.89 & \nodata &  6.92 &  6.95 & \nodata & \nodata & \nodata &   $-$3.75$\pm$   0.27 &   $-$3.96$\pm$   0.17 &  16.01 &   0.61$\pm$  0.17 &  0 &  \\
                    MCM2005b77  &   22  &  14.92 & 10.45 &  7.59 & \nodata & 12.02 &  5.70 &  4.88 & \nodata & \nodata & \nodata &  2.54 &  5.32 &  3.40 & \nodata & \nodata & \nodata &    $..$  &    $..$  &  \nodata &   $..$  &  1 &  \\
                    MCM2005b77  &   23  &  10.04 &  8.31 &  7.61 & 14.43 & 10.02 &  7.58 &  7.08 &  7.33 &  7.04 &  7.02 & \nodata &  7.07 &  7.14 & \nodata& \nodata& \nodata&   $-$1.81$\pm$   0.25 &   $-$3.26$\pm$   0.16 &  15.90 &   0.52$\pm$  0.15 &  0 &  \\
                    MCM2005b77  &   24  &  11.14 &  8.84 &  7.83 & \nodata & 10.98 &  7.80 &  7.12 &  7.27 &  7.02 &  6.97 & \nodata &  7.28 &  7.28 & \nodata & \nodata & \nodata &   $-$4.39$\pm$   0.69 &   $-$5.30$\pm$   0.42 &  18.67 &   0.97$\pm$  0.48 &  0 &  \\
                    MCM2005b77  &   25  &  10.65 &  8.79 &  7.99 & 15.96 & 10.63 &  7.94 &  7.51 &  7.57 &  7.34 &  7.35 & \nodata &  7.48 &  7.52 & \nodata & \nodata & \nodata &   $-$3.32$\pm$   0.32 &   $-$3.10$\pm$   0.21 &  17.16 &   0.45$\pm$  0.21 &  0 &  \\
                    MCM2005b77  &   26  &  11.40 & 10.77 & 10.51 & 12.92 & 11.43 & 10.43 & 10.38 & 10.42 & 10.28 & 10.05 & \nodata & \nodata & \nodata &  14.99 &  14.36 &  13.19 &   $-$1.28$\pm$   0.04 &   $-$3.14$\pm$   0.03 &  14.07 &   0.39$\pm$  0.03 &  0 &  \\
                    MCM2005b77  &   27  &  12.40 & 11.19 & 10.70 & 15.73 & 12.45 & 10.67 & 10.29 & 10.31 & \nodata & \nodata & \nodata & \nodata & \nodata & \nodata & \nodata &  18.21 &   $-$4.40$\pm$   0.24 &   $-$3.89$\pm$   0.16 &  17.25 &   0.18$\pm$  0.16 &  0 &  \\
                    MCM2005b77  &   28  &  12.16 & 11.29 & 11.00 & 14.00 & 12.15 & 11.01 & 10.82 & 10.86 & 10.88 & 10.93 & \nodata & 10.68 & 10.59 & \nodata &  15.61 & \nodata &   $-$0.45$\pm$   0.08 &   $-$1.09$\pm$   0.05 &  15.33 &   0.29$\pm$  0.05 &  0 &  \\
                    MCM2005b77  &   29  &  14.40 & 12.29 & 11.39 & \nodata & 14.41 & 11.32 & 10.72 & 10.70 & 10.50 & 10.64 & \nodata & 10.66 & 10.57 & \nodata & \nodata & \nodata &    $..$  &    $..$  &  \nodata &   $..$  &  0 &  \\
                    MCM2005b77  &   30  &  15.01 & 12.71 & 11.63 & \nodata & 15.05 & 11.72 & 10.81 & 10.96 & 10.64 & \nodata & \nodata & \nodata & \nodata & \nodata & \nodata & \nodata &    $..$  &    $..$  &  \nodata &   $..$  &  0 &  \\
                   DBS2003-172  &    7  &   9.24 &  8.65 &  8.51 & 10.19 &  9.19 &  8.54 &  8.40 &  8.44 &  8.41 &  8.52 & \nodata & \nodata & \nodata &  12.48 &  11.77 &  10.81 &    2.21$\pm$   0.08 &    1.27$\pm$   0.07 &  11.09 &   0.97$\pm$  0.04 &  0 &  \\
                   DBS2003-172  &    8  &  14.32 & 10.42 &  8.63 & \nodata & 14.11 &  8.59 &  7.46 &  7.45 &  7.00 &  7.00 & \nodata &  7.86 &  7.47 & \nodata & \nodata & \nodata &    $..$  &    $..$  &  \nodata &   $..$  &  0 &  \\
                   DBS2003-172  &    9  &  10.00 &  9.36 &  9.07 & 11.42 &  9.91 &  9.09 &  8.94 &  9.04 &  8.95 &  8.82 & \nodata &  8.81 &  8.84 &  13.91 &  12.84 &  11.99 &   $-$1.20$\pm$   0.07 &   $-$2.72$\pm$   0.06 &  12.57 &   0.33$\pm$  0.04 &  0 &  \\
                   DBS2003-172  &   10  &  14.47 & 11.10 &  9.49 & \nodata & 14.18 &  9.54 & \nodata & \nodata & \nodata & \nodata & \nodata & \nodata & \nodata & \nodata & \nodata & \nodata &    $..$  &    $..$  &  \nodata &   $..$  &  0 &  \\
                   DBS2003-172  &   11$^*$  &   17.36 & 12.76 & 10.45 & \nodata & \nodata & 10.42 &  8.86 &  8.66 &  7.94 & \nodata & \nodata &  8.69 &  7.54 & \nodata & \nodata & \nodata &    $..$  &    $..$  &  \nodata &   $..$  &  0 &  \\

\hline
\end{tabular}
\begin{list}{}{}{}
\item[](a) For the observed late-type stars a few alias names are found in the SIMBAD database.
MCM2005b77-18 coincides with CD-50$-$10319 and MCM2005b77-22 with IRAS 16136$-$5020.
MCM2005b77-22 is listed as a possible carbon star by \citet{skiff14}.~ 
($^{**}$) Proper motions values are from GAIA DR2.
\end{list}
\end{sidewaystable*}

\subsection{ Estimates of extinction}

For early-type stars, estimates of reddenings (color excess) are obtained by assuming  intrinsic colors 
per spectral type, as listed in  \citet[][]{messineo11}. For late-type stars, intrinsic colors are taken from \citet{koornneef83}.
Reddenings in $J-$\Ks,  $H-$\Ks, and $J-H$ are converted into \Aks\ values using the
coefficients provided by \citet{messineo05} for an index $\alpha=-1.9$.
This  index is consistent with the recent determination by \citet{wang14}, which is based
on spectroscopic observations of K-type giants.
Tables  \ref{spectra.late} and \ref{exttable} list our  estimates of \Aks.
The sample mainly consists of ``naked" early-type stars, as inferred by the
consistent \Aks\ values  from multicolors and from the small values of $|Q1|$ and $|Q2|$
\citep{messineo12}. Star MCM2005b77 12 (\brg\ in emission) has an 8 \um\ excess.
Stars DBS2003-172 5A and 5B are embedded in a nebula.

\begin{table*}
\caption{ \label{exttable} Infrared color properties of early-type stars in the direction of 
in  MCM2005b72/DBS2003-157 (GRS G331.34$-$0.36), MCM2005b77, and DBS2003-172 (GRS G337.92$-$0.48).}
\begin{tabular}{@{\extracolsep{-.08in}}lrl    rr rrr rr l}
\hline
\hline
Field                            &       ID         &     Sp. Type          & $(J-K_{\mathrm s})_o$ &  $(H-K_{\mathrm s})_o$  &       
{\it A$_{\it K_{\rm s}}$ (JH)} &  {\it A$_{\it K_{\rm s}}$ (JK$_{\mathrm s}$) } &  {\it A$_{\it K_{\rm s}}$ (HK$_{\mathrm s}$) } & $Q1$ & $Q2$ & Comment\\   
\hline

                  G331.34$-$0.36   &   1   &                OBAF   &  $-$0.060     & $-$0.010     &  0.829     &  0.760     &  0.637     &                   0.191$\pm$ 0.199     &                   0.226$\pm$ 0.037     &                   \\
                  G331.34$-$0.36   &   2   &                B0-5   &  $-$0.130     & $-$0.030     &  0.554     &  0.524     &  0.471     &                   0.048$\pm$ 0.083     &                   0.695$\pm$ 0.060     &   FG$^a$                  \\
                  G331.34$-$0.36   &   3   &                O4-6   &  $-$0.180     & $-$0.040     &  1.315     &  1.242     &  1.112     &                   0.165$\pm$ 0.080     &                   0.253$\pm$ 0.041     &                   \\
                  G331.34$-$0.36   &   4   &                OBAF   &  $-$0.060     & $-$0.010     &  0.929     &  0.830     &  0.652     &                   0.292$\pm$ 0.084     &                               $..$     &                   \\
                  G331.34$-$0.36   &   5   &          O9-B0   &  $-$0.160     &  $-$0.040     &  1.096     &  1.047     &  0.960     &                   0.104$\pm$ 0.085     &                               $..$     &                   \\
                  G331.34$-$0.36   &   6   &           B0-3   &  $-$0.130     &  $-$0.030     &  0.956     &  0.918     &  0.851     &                   0.071$\pm$ 0.124     &                  $-$5.661$\pm$ 0.008     &                   \\
                  G331.34$-$0.36   &   7   &                OBAF   &  $-$0.060     & $-$0.010     &  1.686     &  1.656      &  1.602 &             0.053$\pm$ 0.927     &                               $..$     &                   \\
                  G331.34$-$0.36   &   8   &                OBAF   &  $-$0.060     & $-$0.010     &  0.333     &  0.317     &  0.289     &                   0.019$\pm$ 0.080     &                               $..$     &    FG$^a$               \\

                    MCM2005b77   &   1   &                B0-3   &  $-$0.080     & $-$0.040     &  0.838     &  0.851     &  0.875     &                  $-$0.021$\pm$ 0.099     &                   0.020$\pm$ 0.005     &         Member$^b$    \\
                    MCM2005b77   &   2   &                OBAF   &  $-$0.050     & $-$0.040     &  2.843     &  2.763     &  2.621     &                   0.302$\pm$ 0.080     &                   0.367$\pm$ 0.006     &               BG$^c$    \\
                    MCM2005b77   &   3   &                B0-5   &  $-$0.050     & $-$0.020     &  1.144     &  1.114     &  1.059     &                   0.098$\pm$ 0.091     &                  $-$0.206$\pm$ 0.010     &             BG$^c$    \\
                    MCM2005b77   &   4   &                B0-3   &  $-$0.080     & $-$0.040     &  0.789     &  0.793     &  0.800     &                   0.011$\pm$ 0.083     &                  $-$0.083$\pm$ 0.011     &         Member    \\
                    MCM2005b77   &   5   &                B0-3   &  $-$0.080     & $-$0.040     &  0.787     &  0.794     &  0.806     &                   0.002$\pm$ 0.084     &                  $-$0.312$\pm$ 0.004     &         Member    \\
                    MCM2005b77   &   6   &                B0-5   &  $-$0.050     & $-$0.020     &  0.980     &  0.933     &  0.850     &                   0.154$\pm$ 0.080     &                  $-$0.082$\pm$ 0.010     &         Member    \\
                    MCM2005b77   &   7   &                B0-5   &  $-$0.050     & $-$0.020     &  0.890     &  0.851     &  0.782     &                   0.127$\pm$ 0.078     &                   0.037$\pm$ 0.014     &         Member    \\
                    MCM2005b77   &   8   &                B0-5   &  $-$0.050     & $-$0.020     &  0.746     &  0.709     &  0.645     &                   0.120$\pm$ 0.085     &                   0.386$\pm$ 0.009     &                   \\
                    MCM2005b77   &   9   &                B0-5   &  $-$0.050     & $-$0.020     &  0.798     &  0.760     &  0.691     &                   0.127$\pm$ 0.139     &                  $-$0.330$\pm$ 0.005     &         Member    \\
                    MCM2005b77   &  10   &                B0-5   &  $-$0.050     & $-$0.020     &  0.874     &  0.867     &  0.854     &                   0.021$\pm$ 0.095     &                   0.631$\pm$ 0.014     &                  \\
                    MCM2005b77   &  11   &                B0-5   &  $-$0.050     & $-$0.020     &  0.959     &  0.917     &  0.841     &                   0.139$\pm$ 0.800     &                               $..$     &         Member    \\
                    MCM2005b77   &  12   &                  OB   &  $-$0.050     & $-$0.020     &  1.726     &  1.837     &  2.035     &                  $-$0.381$\pm$ 0.084     &                  $-$3.887$\pm$ 0.012     &                 \\

                   DBS2003-172   &   1   &                OBAF   &  $-$0.060     & $-$0.010     &  1.045     &  1.053     &  1.067     &                  $-$0.067$\pm$ 0.148     &                               $..$     &                   \\
                   DBS2003-172   &   2   &                OBAF   &  $-$0.060     & $-$0.010     &  0.894     &  0.836     &  0.732     &                   0.155$\pm$ 0.081     &                               $..$     &                   \\
                   DBS2003-172   &   3   &                  OB   &  $-$0.060     & $-$0.010     &  1.088     &  1.126     &  1.192     &                  $-$0.168$\pm$ 0.094     &                               $..$     &                   \\
                   DBS2003-172   &   4   &                OBAF   &  $-$0.060     & $-$0.010     &  0.889     &  0.824     &  0.708     &                   0.178$\pm$ 0.088     &                   0.099$\pm$ 0.008     &                   \\
                   DBS2003-172   &   5A   &                  OB   &  $-$0.060     & $-$0.010     &  2.142     &  1.897     &  1.459     &                   0.770$\pm$ 1.122     &                               $..$     &                   \\
                   DBS2003-172   &   5B   &                  OB   &  $-$0.060     & $-$0.010     &   $..$      &   $..$      &  3.575     &                               $..$     &                               $..$     &                   \\
                   DBS2003-172   &   6   &                OBAF   &  $-$0.060     & $-$0.010     &  0.375     &  0.353     &  0.314     &                   0.038$\pm$ 0.074     &                               $..$     &       FG$^a$             \\
                  
\hline
\end{tabular}
\begin{list}{}{}
\item ($^a$) FG= object in the background of the \HH\ region. ($^b$) Member= cluster member on the basis of the radial distance and extinction; 
($^c$) BG= object in the background of the \HH\ region.
\end{list}
\end{table*}

\section{Massive  members of clusters or star forming regions}
\label{clusteranalysis}
          
We  have detected several  massive stars   in the direction 
of each of the candidate clusters MCM2005b72/DBS2003-157 (GRS G331.34$-$0.36), 
MCM2005b77, and DBS2003-172 (GRS G337.92$-$0.48).
In this section, we analyze  the infrared properties of those stars, such as 
extinction and bolometric magnitudes,  \Mbol. We also estimate
their spectrophotometric distances, in order 
to investigate their association with the nearby  \HH\ regions.
 
Detected late-type stars are not discussed further, because
they appear to be giants that are unrelated to the star forming regions analyzed.  
As shown in  Table \ref{spectra.late}, for the assumed distances to the star forming regions,
the \Mbol\ for all but two of the detected late-type stars are $\ga -5.27$ mag, i.e.
typical of giants.
The  two bright stars are asymptotic giant branch stars (AGBs) with
strong  water absorption in their spectra
\citep[see also][]{messineo14a,messineo14}.

\subsection{S62/GRS G331.34$-$00.36}
\label{s62}
The candidate clusters MCM2005b72 \citep{mercer05}, DBS2003-157 \citep{dutra03}, 
and Cl063 \citep{borissova11} are projected onto the \HH\ region GRS G331.34$-$00.36.
The \HH\ emission extends over an area of $3\farcm8 \times 2\farcm0$, as
measured  by \citet{culverhouse11} at 100 and 150 GHz.
At this  location, a bubble (S62) is seen at mid-infrared wavelengths 
\citep{churchwell06,simpson12};
dust emission shapes and defines several subregions.  
The candidate cluster DBS2003-157 overlaps with a large portion of
GRS G331.34$-$00.36; it was  detected on 2MASS images as a region  
associated with nebular emission \citep{dutra03}.
The candidate cluster MCM2005b72 was detected as an overdensity of stellar datapoints 
 at 3.6 \um\ \citep{mercer05} and is  adjacent to 
DBS2003-157. 
The candidate cluster Cl063 corresponds to a bright clump of dust emission 
on the northeastern side of  GRS G331.34$-$00.36 associated with  IRAS 16089$-$5137, 
it was detected with VVV data \citep{borissova11}. 
These three candidate clusters are most probably subsets of the same young population 
associated with  GRS G331.34$-$00.36.

The \HH\ region indicates the presence of massive stars.
We assume a thermal emission from an optically thin plasma at a temperature of about 10,000 K
and neglect dust
contribution to the flux. 
Using the formula of  \citet{martinhernandez03}, which is equivalent to that of \citet{rubin68} 
when the recombination rate coefficients  are taken from \citet{storey95},
and using the \HH\ radio fluxes  at 4.85 GHz (19.169 Jy), 100 GHz ($7.05\pm0.18$ Jy), and 150 GHz  
($9.29\pm0.27$ Jy) listed in \citet{kuchar97} and \citet{culverhouse11}, 
we obtain log$_{10}$(N$_{\rm lyc}$[photons s$^{-1}$])= 49.25, 48.8, and 49.0, respectively.
 These values are consistent with that inferred from the flux at 22 GHz by \citet{pinheiro12}.
This amount can be generated by a single  O4-6 star.

The region is still an active site of star formation. 
Several molecular clumps appear in the 870 \um\ ATLASGAL map of S62 \citep{schuller09}, 
which is shown in Fig.\ \ref{maps}.  
Sixteen compact molecular clumps, that is, 12 protostellar  and four prestellar cores 
were identified in the direction of  S62 by \citet{elia17}.
 Highly reliable radial velocities are available for 9 clumps  
\citep[\Vlsr\ from $-65.0$ to $-66.4$ \kms,][]{urquhart18}. 
Details are provided in the Appendix.
In summary, all  clumps are likely at the close distance of about 4 kpc, 
with the exception of Hi-GAL \#46272 \citep[\Vlsr=-97.3 \kms, 5.8 kpc;][]{urquhart18} 
and \#46184 \citep[10.6 kpc,][]{elia17}, which are external  to  S62, as
seen in the GLIMPSE 8 \um\ map.
The blackbody temperatures of the clumps range from  11.9 to 30.2 K and, for a common distance of 3.9 kpc,
individual masses are from $\approx$ 30 to $\approx$ 600 \Msun,  by having excluded
Hi-GAL \#46272 and  \#46184.
 Thirteen of these 14 clumps, equivalent to 3700 \Msun,  are  above their critical mass threshold 
for virial equilibrium \citep{koenig17,bertoldi92}.

\subsubsection{Properties of the massive stars detected in S62/GRS G331.34$-$00.36}

We detected one O4-6 star (\#3) and one  O9-B0 star (\#5) in the direction of GRS G331.34$-$00.36 (S62).
 \citet{pinheiro12} identified the same two stars (\#3 and \#5) as the main ionizing sources.
The O4-6  star  has interstellar extinction  \Aks($JH$) = 1.3 mag, \Aks($J$\Ks)=1.2 mag,
\Aks($H$\Ks) = 1.1 mag, with an average value of $1.2$ mag and  $\sigma=0.1$ mag.
Assuming a dwarf class,  as inferred in \citet{pinheiro12}, and the \Mk\ listed by \citet{martins06}, we derive a
distance modulus,  DM, of  $12.66\pm0.78$ mag. If a giant class would be assumed, then DM=$13.30\pm0.63$ mag.
The  \HH\  region GRS G331.34$-$00.36 has a
velocity of  \Vlsr$=-64.8\pm2.9$ \kms\ \citep[][]{bronfman96}. 
Using  this  \Vlsr\ and the Galactic rotation curve from \citet{reid09},
we estimate a near-kinematic distance of $3.9\pm0.3$ kpc, 
DM=$12.98\pm0.18$ mag. 
We thus conclude that star \#3  is associated with S62.  
The \Mbol\ values derived using the near-kinematic distance are listed 
in Table \ref{abstable}. 

\begin{figure}
\begin{center}
\resizebox{1.0\hsize}{!}{\includegraphics[angle=0]{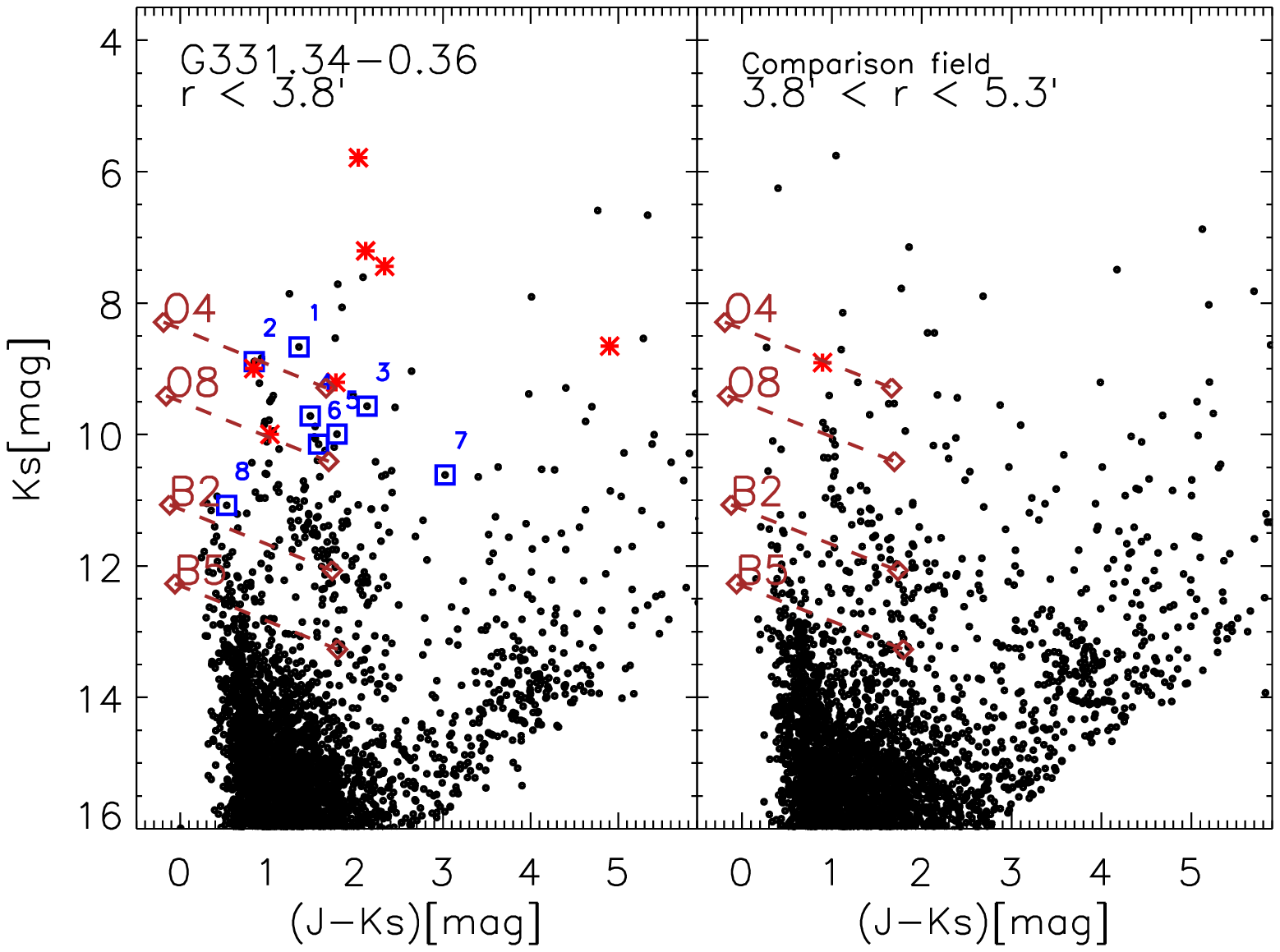}}
\resizebox{1.0\hsize}{!}{\includegraphics[angle=0]{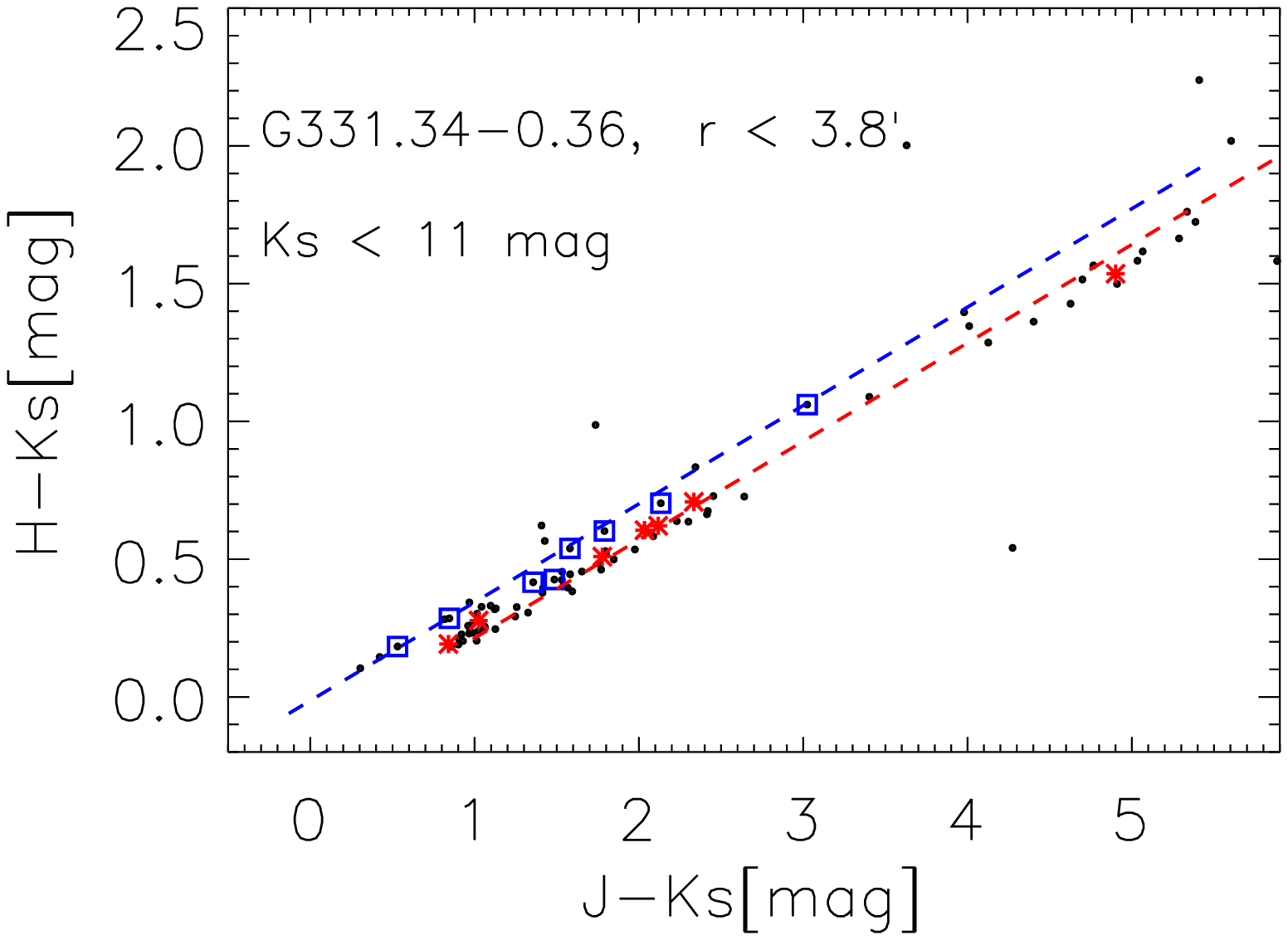}}
\caption{{ \it Upper panel:} \label{cmdg331.31} $J-$\Ks\ versus \Ks\ diagram with 2MASS and VVV 
(\Ks $> 11.5$ mag) sources detected in the direction of the star-forming region  
GRS G331.34$-$00.36. 
 Squares mark detected early stars, and asterisks mark detected late-type stars.
Labels and diamonds indicate the locations of O4, O8, B2, and B5 dwarfs  
for a distance of 3.9 kpc at zero extinction \citep[e.g.][]{martins05,bibby08,messineo11}. 
Dashed lines connect the zero extinction locations with
those at \Aks=1.0 mag.
 {\it Bottom left figure:} $J-$\Ks\ versus $H-$\Ks\ diagram of 2MASS datapoints
with \Ks$< 11$ mag, labels are  as for the top panels.  The two dashed lines
show the locus defined by an O9 (upper line) and an M1 (lower line) star with increasing \Aks. 
}   
\end{center}
\end{figure}

Star \#5  is 20\arcsec\ away from star \#3 and has 
\Aks=$1.02\pm0.05$ mag. Similar estimates of extinction  
suggest that star \#5  is physically associated with the bubble.
In the  color-magnitude diagram (CMD) in Fig.\ \ref{cmdg331.31}, 
these two massive stars appear among the brightest stars.  
Assuming an O9-9.5 type for star \#5,
we obtain a DM between $12.20\pm0.76$ mag (dwarf) and $13.37\pm0.65$ mag (giant).

Star \#2 (B0-5) is located on the northeastern side of the region, outside the bubble, 
at a significantly lower extinction of \Aks=0.5 mag. A star that is foreground  
to  the \HH\ region GRS G331.34$-$00.36 cannot be ruled out.
Assuming a dwarf, we derived DM= $10.72 \pm 1.27$ mag (1.4 kpc), assuming a giant,
we obtained DM=$11.81\pm1.43$ mag. 
Unfortunately, the spectra of stars \#1, \#4, \#6, \#7, and \#8 have poor signal-to-noise 
ratio and we can only conclude that they are not late-type stars.
 Star \#8 (\Aks$\approx$0.3 mag)  is likely to be foreground to the complex.
From Fig. \ref{cmdg331.31}, stars \#4 (\Aks$\approx$0.8 mag) and 
\#7 (\Aks$\approx$  1.6 mag)  are plausibly late-O stars related to S62.  For star 
\#6 (\Aks$\approx$  0.9 mag) \citet{pinheiro12} report an early B type,
which yields a DM=$12.66\pm1.43$ mag for a giant or $11.57\pm1.27$ mag for a dwarf.

 By assuming that the stars are at the nebular distance of 3.9 kpc, 
star \#3 (O4-6V) is able to produce log$_{10}$(N$_{\rm lyc}$) 
from 49.0 to 49.5 photons s$^{-1}$,
star \#5 (O9-9.5V) from 47.6 to 47.9  photons s$^{-1}$, and
star \#6 (B0-B3V) from 43.9 to 47.6  \citep{martins05,panagia73}.
The O4-6 star is the  main source of ionizing flux,
as it accounts for  the observed radio flux from this HII bubble (see above).

By inspecting the $H-$\Ks\ versus $J-$\Ks\ colors of stars brighter than \Ks=11 mag,
we find that   a few other  faint ionizing sources remain unobserved 
(\Ks$ < 10.5$ mag, see Fig.\ \ref{cmdg331.31}).

\subsection{MCM2005b77}
\label{clMCM2005b77}

MCM2005b77 was detected by \citet{mercer05}, with GLIMPSE and 2MASS data.
This stellar cluster shows a well-peaked surface density in the 2MASS \Ks\ image, 
 with a  half-light diameter of 98\arcsec, and is rich in early-type stars. 
Our spectroscopic study revealed 12 early-type stars in the direction of this candidate cluster.

\subsubsection{The color-magnitude diagram of MCM2005b77}
\begin{figure*}
\begin{center}
\resizebox{0.7\hsize}{!}{\includegraphics[angle=0]{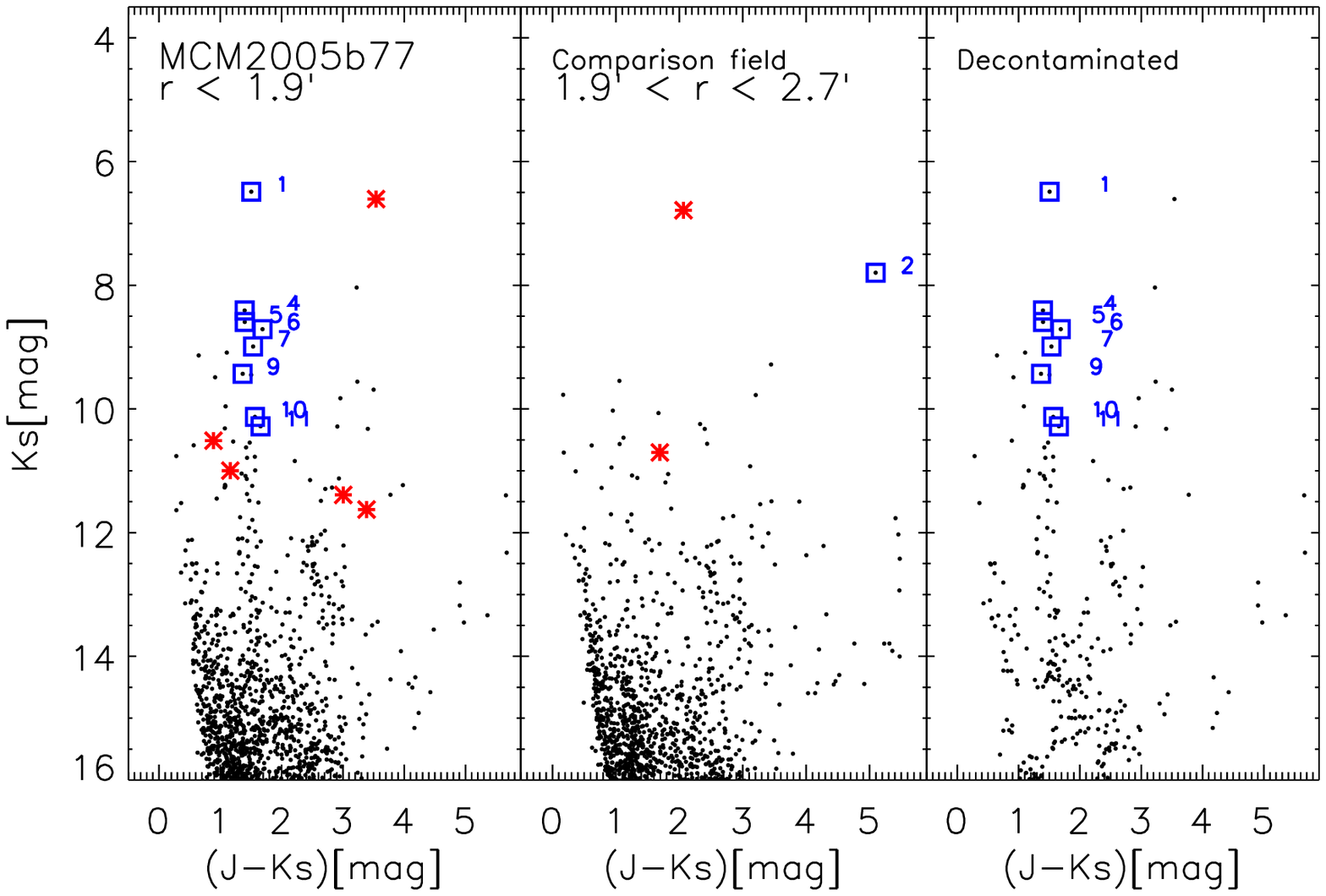}}
\end{center}
\begin{center}
\resizebox{0.49\hsize}{!}{\includegraphics[angle=0]{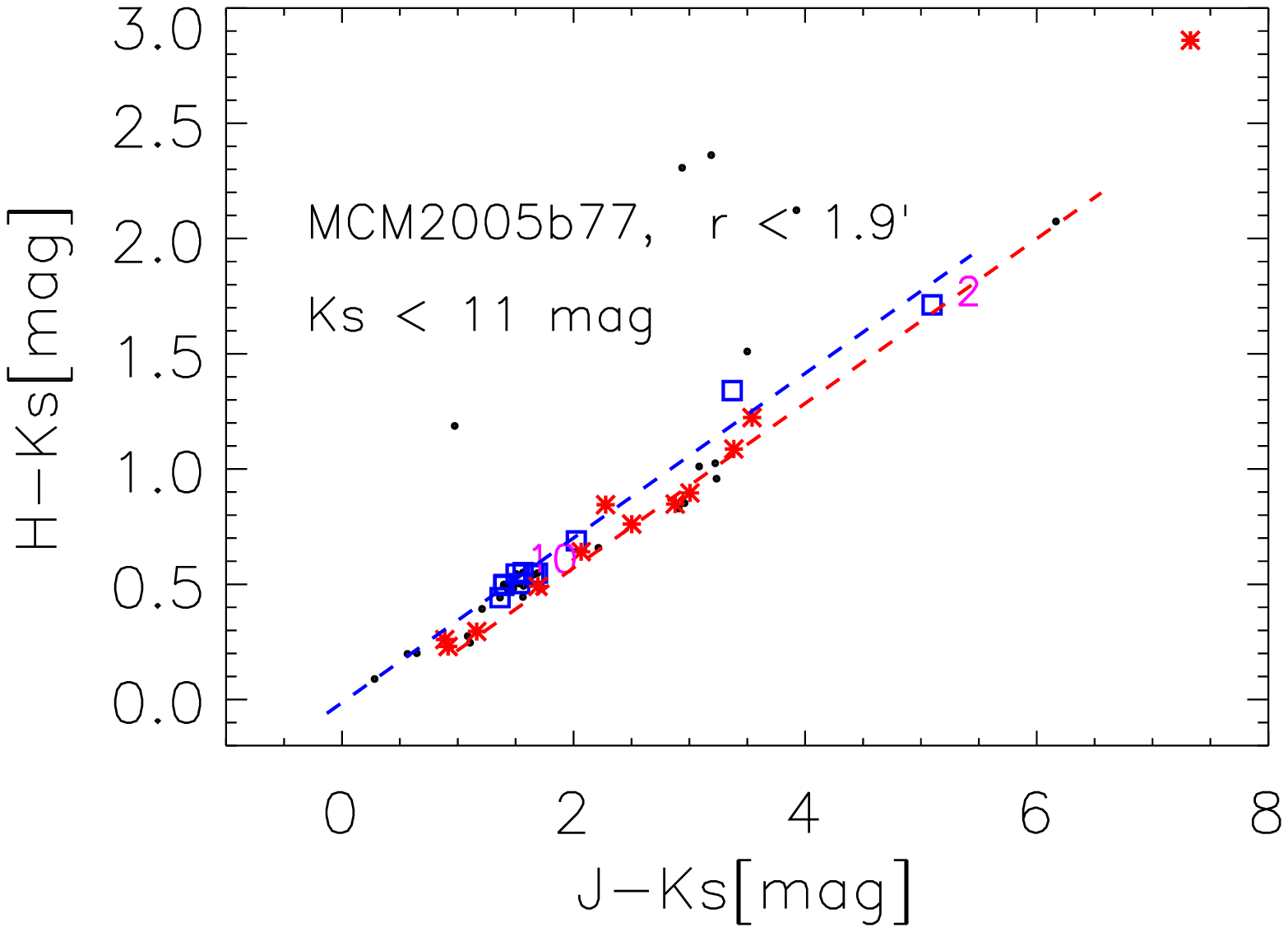}}
\resizebox{0.49\hsize}{!}{\includegraphics[angle=0]{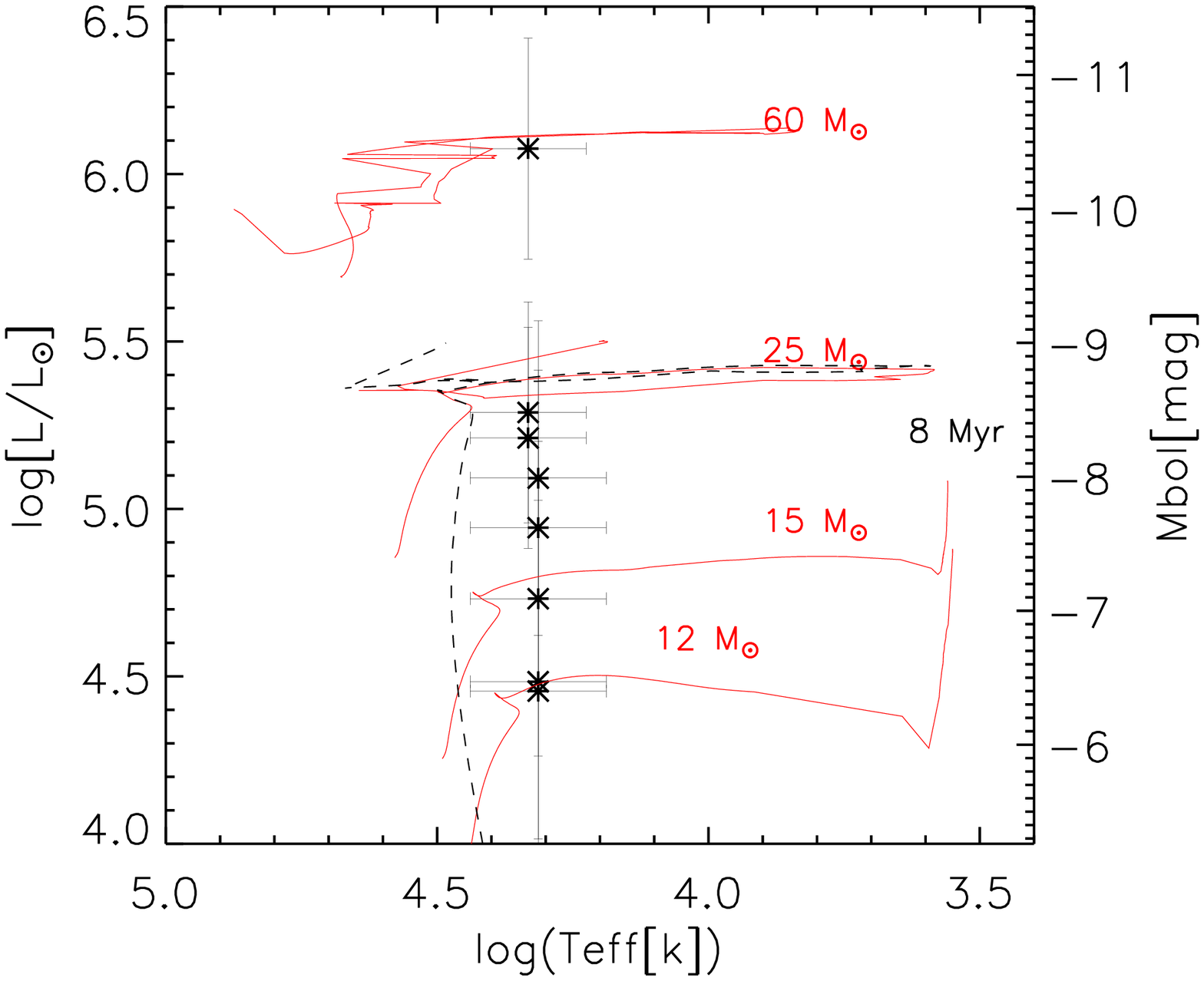}}
\end{center}
\caption{ \label{cmd77}  {\it Top Figures:} $J-$\Ks\ versus \Ks\ diagrams of MCM2005b77.
2MASS-VVV datapoints (small circles) in the direction of  the stellar cluster are shown 
in the left panel; datapoints of a comparison field of equal area
are shown in the middle panel; a  statistically
decontaminated   cluster diagram  is displayed in the right panel.
Detected early-type stars are indicated with squares and late-type stars with
asterisks. {\it Bottom left figure:} $J-$\Ks\ versus $H-$\Ks\ diagram of 2MASS datapoints
with \Ks$< 11$ mag, labels are  as for the top panels.  The two dashed lines
show the locus defined by an O9 (upper line) and an M1 (lower line) star with increasing \Aks. 
The locations of star \#10 and the outlier star \#2, which have poor  spectra, are shown in magenta.
In the {\it bottom right figure}, luminosities versus \Teff\ of 
detected early types in the cluster MCM2005b77 are shown. 
 Tracks (dashed lines) of rotating stars with masses of
12, 15, 25, and 60 \Msun, as well as an isochrone (continuous line) of 8 Myr extracted from 
the same  set of tracks, are overplotted  \citep{ekstrom12}. }
\end{figure*}

A $J-$\Ks\ versus \Ks\ diagram of 2MASS datapoints  in the direction of MCM2005b77 
is shown in Fig. \ref{cmd77}. The diagram shows a dwarf foreground sequence
at $J-$\Ks$\approx 0.5$ mag, a red clump sequence (not well defined) 
that crosses the CMD from \Ks$\approx 10$ mag  and $J-$\Ks$\approx 1$ mag  to 
\Ks$\approx 12$ mag  and $J-$\Ks$\approx 2.5$ mag, and  a tail of redder
giants with $J-$\Ks$\ga 2.5$  mag.  The cluster sequence, consisting of B-type stars, appears
above the red clump  sequence at \Ks$ \la 10.5$ mag  and $J-$\Ks$\approx 1.4$ mag.
On the CMD, it is difficult to identify this at lower magnitudes because of the high number 
of field clump stars. Late-type and early-type stars  separate out in the 
$J-$\Ks\ and $H-$\Ks\ diagram shown in Fig. \ref{cmd77}.
In the inner 1.9\arcmin, we count a total of 16 photometric 
early-type stars with (\Ks $ < 11$ mag), which include the 8 spectroscopically 
confirmed early types  (\#1, \#4, \#5, \#6, \#7, \#9, \#10, and \#11). 
Average proper motions in the right ascension and declination directions are
$\mu_{\alpha}$ and $\mu_{\delta}$, respectively, for all but star \#10:  $\mu_{\alpha}=-4.15\pm0.5$ mas/yr and
$\mu_{\delta}=-4.15\pm0.5$ mas/yr. Star \#10 is more than 3 $\sigma$ off, with 
$\mu_{\alpha}=-7.73\pm0.54$ mas/yr and $\mu_{\delta}=-7.14\pm0.32$ mas/yr. It could be unrelated to the cluster;
however, star \#10 (OBAF) is similar in colors and magnitudes to the star \#11 (B0-5), 
which, together with its location, indicates that it could be another B0-5 star
 with anomalous velocity. The average \Aks\ and standard deviation of the eight 
spectroscopic early-type stars are 0.91 mag and 0.08 mag, respectively. 
The small dispersion of \Aks\ values confirms
that these stars are physically at the same distance and are members
of the  stellar cluster MCM2005b77 (see Table \ref{exttable}).
Parallaxes from GAIA DR2 range from $-0.54$ to $1.0$ mas, and  
the average is $0.3\pm0.4$ mas.

There are four other spectroscopic early-type stars detected outside the cluster radius.
Stars \#2 (OBAF) and \#3 (B0-5), with \Aks\ values of 2.9 mag and 1.2 mag,
are probably unrelated background stars. Red $J-$\Ks\ and $H-$\Ks\ colors   
indicate that star \#2 (OBAF)  is a G-F star (see Fig. \ref{cmd77}).
Star \#12 (\brg\ in emission)  is still enshrouded in dust (\Aks = 1.8 mag).
Star \#8 (B0-5, \Kso=8.55 mag) has an \Aks\ similar to that of the cluster (0.8 mag) 
and derived magnitudes consistent with those of cluster members, but it
resides a few arcminutes  away from the cluster (see Fig. \ref{maps}).

\subsubsection{Spectrophotometric distance of MCM2005b77}

The \HH\ region G332.769-0.007 \citep{bronfman96} coincides with
IRAS 16137$-$5025 (MSX G332.7673$-$00.0069) and is   $\sim2$\arcmin\ away from  the cluster center.
With  \Vlsr$=-95.0\pm2.8$ \kms\ 
\citep[CS observations with a beam width of 50\arcsec,][]{bronfman96},
this \HH\ region is at a  near-kinematic distance of  $5.2\pm0.1$ kpc 
\citep[DM=$13.60\pm0.17$ mag,][]{reid09};
a far-kinematic distance would yield 9.7 kpc.
$^{13}$CO observations   confirm the main  component at $-95.6$ \kms, 
and a weaker components at $-65.3$ \kms\ \citep{urquhart07}.

Along the line of sight to MCM2005b77, the model  of Galactic dust distribution 
by \citet{drimmel03}  predicts an \Aks\ value of 0.8-0.9 mag 
(i.e., the extinction of MCM2005b77) for a  heliocentric 
distance of 5.2 kpc. We thus conclude that the cluster MCM2005b77 (\Aks=0.91 mag) 
is most probably at the same distance as the nearby IRAS 16137$-$5025.

To estimate a spectrophotometric distance to the cluster, we analyze the seven B-type members within 
the cluster radius, using the  intrinsic colors and absolute magnitudes per spectral type  tabulated
in \citet{messineo11}.
The spectra of stars \#1,  \#4, and \#5 have the \ion{He}{I} line at 2.058 \um\ in emission,
which is common among  early-B supergiants \citep[e.g.][]{davies12}. Their dereddened \Ks\ values, $(K_{\mathrm s})_{\mathrm o}$,
range from 5.65 mag to 7.81 mag. Star \#6 (B0-5), with  $(K_{\mathrm s})_{\mathrm o}=7.73$ mag, is
also assumed to be  a supergiant. These four stars yield an average  DM of $13.53\pm1.08$ mag
(see Table \ref{spdist77}), which is  compatible with the gaseous near-kinematic distance of the
nearby source IRAS 16137$-$5025.

Assuming a kinematic distance of $5.2\pm0.1$ kpc,  the \Mbol\ values of all detected cluster 
members range from $-6.4$ to $-10.5$ mag \citep[see Table \ref{abstable}; 
these values are typical of evolved B-type stars, class I and III,][]{martins06}.

\begin{table*}
\caption{\label{spdist77} Spectrophotometric distance to the cluster MCM2005b77.}
\begin{tabular}{llllllll}
\hline
\hline
CLUSTER-ID         & Sp. group&  $(K_{\mathrm s})_{\mathrm o}$ & $M_K^*$ & DM & Phot. type  & $M_K^{**}$ & DM(phot)$^{**}$ \\
                   &          &          [mag]                       &   [mag]     & [mag]    &             &  [mag]      & [mag]   \\
\hline
     MCM2005b77  1 &     B0-3 &  5.65 & $-$6.27 &  11.92 $\pm$   0.92 &                  B3/B4 & $-$6.70 & 12.35 \\
     MCM2005b77  4 &     B0-3 &  7.62 & $-$6.27 &  13.89 $\pm$   0.92 &                  B0 & $-$5.85 & 13.47\\
     MCM2005b77  5 &     B0-3 &  7.81 & $-$6.27 &  14.08 $\pm$   0.92 &                  B0 & $-$5.85 & 13.66 \\
     MCM2005b77  6 &     B0-5 &  7.73 & $-$6.49 &  14.22 $\pm$   1.14 &                  B0 & $-$5.85 & 13.58 \\
\hline
Average            &          &       &       & 13.53$\pm$1.08 &                        &  \\
\hline
\end{tabular}
\begin{list}{}{}
\item[{\bf Notes.}]  ($^*$)= $M_K$s are average values for supergiants with types from B0 to B3 (or B5).
~  ($^{**}$)=$M_K$ and DM(Phot) are for photometrically inferred types (Phot. type) \citep{bibby08}.
\end{list}
\end{table*}

\subsubsection{Discussion of MCM2005b77}

We found a concentration of eight  early-type stars at the location of MCM2005b77.
These   have similar interstellar extinctions and appear  along 
the same cluster sequence  in the CMD. 
Assuming a distance of 5.2 kpc, the  \Mbol\ values of the members 
range  from $-6.40$ mag to $-10.45$ mag and
the new rotating stellar  models from the Geneva group 
yield stellar masses from 12 to 60 \Msun\ \citep{ekstrom12}.
Within errors, all detected members except star \#1 are consistent with a coeval 
population of about 8 Myr \citep{ekstrom12}.
The maximum initial unexploded stellar mass  in such a population
has a mass of $\approx 28$ \Msun, while  masses of 18 \Msun\ are already evolving after the
main sequence.

We counted a number of 16 photometric early-type stars, down to \Ks=11 mag, which is the limit at which completeness can be assumed. 
Assuming a Salpeter mass function, a minimum mass of $\approx 350$ \Msun\ is derived and by extrapolating to 1 \Msun\ we infer a cluster mass of  about $3000\pm900$ \Msun.

Star \#1 has an \Aks\ consistent with those of fainter members and infrared colors
and spectral features consistent with a normal B star.
It is exceptionally bright ($M_K = -7.95$ mag, \Mbol=$-$10.45 mag)
for an 8 Myr old population.
The second brightest star has  $M_K = -6.52$ mag and a companion could
brighten it by only $-0.75$ mag. Assuming a single star, a mass of 60 \Msun\ is derived.

For a distance of 5.2 kpc, in the MCM2005b77 cluster, we counted four probable 
blue supergiants, one giant, and six dwarfs ( $< 18$ \Msun) \citep{martins05}. 
Assuming B0 types, we estimate log$_{10}$(N$_{\rm lyc}$) = 49.2 photons s$^{-1}$, 
assuming B1 stars,  47.4  photons s$^{-1}$, 
and assuming B2 stars,  46.83 photons s$^{-1}$  \citep{panagia73}.
The cluster may be  contributing ionizing photons for 
the nearby  \HH\  region IRAS 16137$-$5025 (MSX G332.7673$-$00.0069). 
The \HH\ region has a flux density of $1.42\pm0.15$ Jy at 150 GHz and
log$_{10}$(N$_{\rm lyc}$) $\approx$ 48.52 photons s$^{-1}$  are sufficient
to power it.  
In the direction of  IRAS 16137$-$5025, some molecular clumps are currently collapsing. 
 The ATLASGAL clumps have  integrated fluxes from 8.19 to 12.79 Jy \citep{urquhart14,wienen15}. 
 Four compact clumps are  detected with Hi-GAL data \citep[numbers 47117,47118,47127, and 17129;][]{elia17}
and classified as proto-stellar clumps (with mid-infrared detections).  
Estimated dust temperatures range from 12.09 to 20.39  K and,  
for an assumed distance  of 5.2 kpc, masses   
from 500 to 1500 \Msun, 
 make up a total mass of about 3800 \Msun\ enclosed within a circle of 2 pc radius (see the Appendix). 
For  Hi-GAL 47118 	(AGAL332.767-00.019)	and 47127	(AGAL332.774-00.009), 
similarly high masses  of 1340 and 1982 \Msun, respectively,
are estimated by \citet{heyer16}. 
 Radial velocities confirm near-kinematic distances and association with IRAS 16137$-$5025 
for the four Hi-GAL/ATLASGAL 
sources \citep[from $-94.8$ to $-95.6$ \kms;][]{urquhart18}.

\subsection{DBS2003-172}

DBS2003-172 is reported as an infrared candidate stellar cluster by \citet{dutra03}.
Deeper infrared photometry confirmed a rich population of young stars 
in a star-forming region \citep{borissova06}.

DBS2003-172  appears to be the probable source 
of ionizing radiation for GRS G337.92$-$00.48, 
a \HH\ region of $2\farcm2 \times 2\farcm0$  
\citep{culverhouse11}, as shown in Fig.  \ref{maps}.
This \HH\ region coincides with the bulk of 8 \um\ emission from 
the bubble S36 \citep{churchwell06}. In  the direction of this \HH\ region,
\citet{huang99} measured  three CO clouds at \Vlsr\ $-58.3$,$-40.5$, and $-33.4$ \kms, respectively,
which yield near-kinematic distances of 3.9, 3.0, 2.7 kpc 
\citep[or far-kinematic distances of 11.7, 12.5, and 12.9 kpc;][]{reid09}. The $-40.5$ \kms\ 
cloud is far stronger than the other two clouds.
\citet{walsh98}  detected a site of active star formation 
with three methanol masers, whose velocities ( \Vlsr= $-36.5$, $-38.0$, and $-38.9$ \kms)
indicate an association with  the  CO cloud at $-40.5$ \kms.
We thus  assume  a \Vlsr=$-40.5$ \kms\
and a kinematic distance of 3.0 kpc (near), or 12.5 kpc (far), for DBS2003-172.
 We assume the near distance.

Considering the whole \HH\ region, 
the radio flux densities measured at 100 GHz  and 
at 150 GHz  of GRS G337.92-00.48 \citep{culverhouse11} are  
$14.92\pm0.14$ Jy and $18.88\pm0.16$ Jy, and \citet{caswell87} report $12.80$ Jy at 5 GHz. Assuming thermal emission  and a temperature of 10,000 K (5600 K)
and using the formula by \citet{martinhernandez03}, we computed
log$_{10}$(N$_{\rm lyc}$)=48.74 (48.85), 48.93 (49.05), and  49.05 (49.17) 
photons s$^{-1}$ with the 5 GHz, 100 GHz, and 150 GHz flux densities respectively.
This is equivalent to the log$_{10}$(N$_{\rm lyc}$) emitted by a single O9-B0 supergiant,
or by a dozen B0 dwarfs \citep{panagia73}.

The  submillimeter emission detected by ATLASGAL nicely follows the  mid-infrared bubble. 
Seven prestellar and  10 protostellar  compact ($< 1$ pc) clumps  identified 
with Hi-GAL by \citet{elia17} are  in the direction of this bubble.
Their 870 \um\ fluxes range from 0.0 to 8.21 Jy \citep{elia17}.
Temperatures were estimated by fitting a grey 
blackbody and range from 10.96 to 31.13 K. 
Assuming a common distance of 3.0 kpc, 
rescaled masses range from 50 to 3500 \Msun, with two clumps above 1000 \Msun.

Elia et al.  provide distances of about 3 kpc for 12 out of 17 clumps (see the Appendix). 
The just-released  catalog of highly reliable 
velocities of ATLASGAL clumps \citep{urquhart18}  proposes close kinematic distances for 
seven Hi-GAL clumps (\Vlsr\ from $-$37.6 to $-$40.0 \kms) 
and only the  clumps 50264, 50313, and 50322 remain  at the far distance.
However, by looking at the bubble morphology and filaments,  
it appears possible to us   that they are all associated with bubble S36.

In the ATLASGAL map, the candidate stellar cluster DBS2003-172 lies between two
extended ATLASGAL clumps (AGAL 337.916$-$00.477 
and AGAL 337.922$-$00.456) of 74.83 and 86.41 Jy, respectively \citep{urquhart14}.
For a distance of 3.0 kpc, their angular sizes 
( $30^{\prime \prime} \times 17^{\prime \prime}$ and $45^{\prime \prime} \times 35^{\prime \prime}$ ) correspond 
to $1.5 \times 0.9$ and $2.1 \times 1.7$  pc$^2$. 
AGAL 337.916$-$00.477 is one of the mid-IR bright sources studied by \citet{koenig17},
a high-mass protostar with  a temperature of $34.4\pm1.8$ K and a mass of 1100 \Msun\ (for 3.0 kpc).
 With  a velocity of $-39.6$ \kms,
AGAL $337.916-00.477$ is associated with GRS G337.92$-$00.48 (S36) \citep{giannetti14,urquhart18}. 
The other extended clump AGAL337.922$-$00.456 breaks into two different Hi-GAL sources listed by Elia et al.
(\#50279 and \#50266).
AGAL337.922$-$00.456  is given at the far distance by \citet[\Vlsr=$-38.7$ \kms,][]{urquhart14,urquhart18}, but 
at the close distance by \citet[][]{kim17}; \Vlsr=$-43$ \kms.
The stellar cluster location and symmetry of the molecular structure around it suggest close distance.

\begin{figure}[h]
\resizebox{1.0\hsize}{!}{\includegraphics[angle=0]{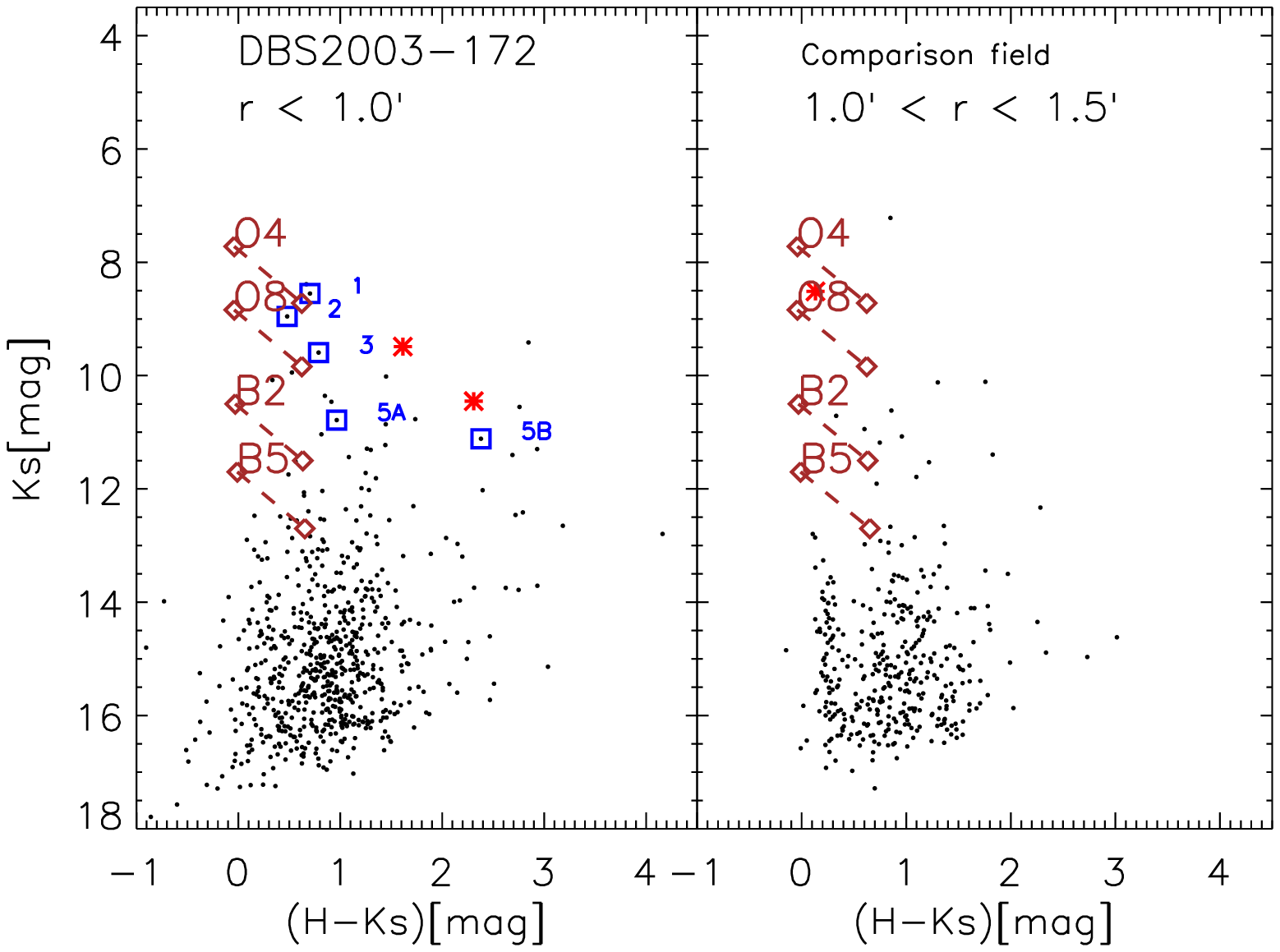}}
\resizebox{0.8\hsize}{!}{\includegraphics[angle=0]{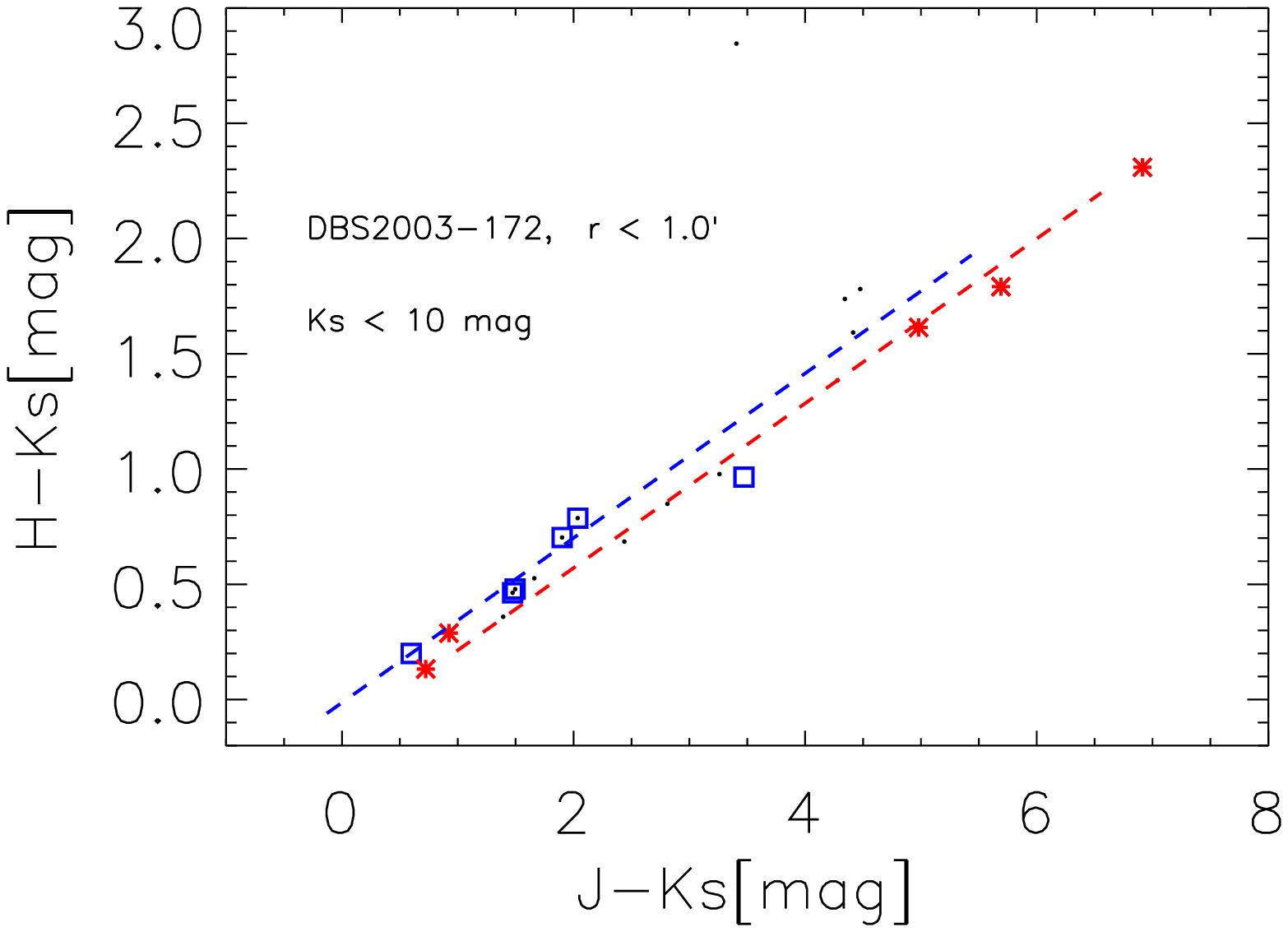}}
\caption{  {\it Top:}  \label{cmddbs2003172} combined VVV and 2MASS  \Ks\ versus $H-$\Ks\ 
diagram of DBS2003-172. 
Squares mark detected early stars, and asterisks mark detected late-type stars.
The locations of O4, O8, B2, and B5 dwarfs  (diamonds) at  \Aks=0.0 and 1.0 mag 
are shown in coral, as in Fig.\ \ref{cmdg331.31}; this time for a distance of 3.0 kpc. 
{\it Left panel:} data points within 1\arcmin. 
{\it Right panel:} comparison CMD with datapoints from an annular region of area
equal to that used in the right panel. 
 {\it Bottom:} $J-$\Ks\ versus $H-$\Ks\ diagram. Symbols are as in the upper panels.
The two dashed lines show the locus of an O9 (upper line) and an M1 (lower line) 
star with increasing \Aks.} 
\end{figure}

\subsubsection{Massive stars in DBS2003-172}
Six  stars were  spectroscopically detected as probable early types.
Five spectra present only  \brg\ in absorption, which does not allow 
for a precise classification.
Their $Q1$ parameters, which combine $H-$\Ks\  and $J-$\Ks\ colors, 
are consistent with those of early-type stars 
\citep[see Table  \ref{abstable}, and][]{messineo12}.
Indeed, in the \Ks\ versus $H-$\Ks\ diagram in 
Fig. \ref{cmddbs2003172}, a small overdensity
of stars is found in the range 8 mag $<$ \Ks $<$ 11 mag.
Stars \#1, \#2,  \#3, and  \#4 yield an average \Aks=1.0 mag with 
$\sigma=0.1$ mag and are likely to be members of the star forming region. 
Spectral types have not been assigned, since the spectra have poor signal-to-noise ratio
and only \brg\ is detected. 
With \Ks\ from 8.55 to 9.98 mag and  with detected \brg\ in absorption, which
is  the strongest absorption line in the spectrum of O-type stars, 
stars \#1, \#2, and \#4  have photometry consistent with those of  O-type stars.
Star \#3 (\Aks=1.1 mag) has a strong \brg\ line in emission and $JH$\Ks\ 
colors consistent with those reddened by interstellar extinction. 
Assuming a young dwarf, we estimate an O7-O8 type, which would be enough to make 
star \#3 alone the main contributor to the total  N$_{\rm lyc}$ necessary 
to excite this \HH\ region (see previous subsection). 
Star \# 6 is  foreground to the star-forming complex (\Aks=0.37 mag).
Four other possible O-type stars remain unobserved (as shown in the color-color plot of Fig. \ref{cmddbs2003172}).

Stars \#5A and \#5B are   part of the star forming complex  DBS2003-172, 
as deduced by their colors and morphological interaction with the ISM.
In VVV $J,H,$\Ks\ images, a comet-shaped nebula is seen   to surround them.
With \Aks\ of 1.9 and 3.6 mag, stars \#5A and \#5B   are  likely embedded 
in the  nebula. The infrared nebula coincides with an ultracompact 
\HH\ region  at RA=16h41m08.0s and Dec=$-$47d06m46s,  detected by \citet{walsh98}.  
The spectra of DBS2003-172-5  (A and B) display stellar and circumstellar \ion{He}{I} and 
\brg\ (see Fig. \ref{acqn3}). Owing to the presence of strong \ion{He}{I} and  \brg\ lines, 
they resemble typical spectra  of  ultracompact (UC) \HH\ regions
\citep[e.g.][]{doherty94} and of  \HH\ regions associated with  evolved massive stars: 
in particular, IRS13, the remains of a stellar cluster close to the Galactic center, 
and the M1-78 nebula, which also have  [FeIII] at 2.2178 \um\ \citep{blum95,eckart13,martin08}. 
Detections of  [FeIII] emission lines at 2.2178 \um\  have also been reported 
in the direction of planetary nebulae, but together with several \ion{H}{2} lines
\citep[][]{likkel04}.
The ratio of the flux densities of the \ion{He}{I} and \brg\ lines indicates
stars hotter than an O7 star \citep[e.g.][]{doherty94}. 
When this ratio is above unity and the star is a supergiant, it is typically
a transitional object (e.g., Ofpe/WN9 stars and LBVs) or a WR star
\citep[e.g.][]{blum95, najarro97,messineo09,morris96}. 
The  IRS13 E2 component (the brightest IR component of the IRS13 cluster) is
a  WN8  \citep{martins07}. 
The M1-78 nebula is most probably a combination of 
a \HH\ region and the ejecta from a central massive evolved star (O+WR)
\citep{martin08}.
The UC radio source (IRAS 16374$-$4701) detected at the location of stars \#5A and \#5B 
by \citet{walsh98} has  flux densities of 452 mJy and  396 mJy
at  8.64 GHz and  6.67 GHz (VIZIER), respectively.
For a  optically thin thermal nebula at 10,000 K and 3 kpc distant, we estimated 
log$_{10}$(N$_{\rm lyc}$)=47.3 photons s$^{-1}$, using the formula in \citet{martines03}.
A star of O7 type or earlier produces log$_{10}$(N$_{\rm lyc}) \ga 48.9$  
photons s$^{-1}$ for a dwarf class and  
$\ga 49.3$ photons s$^{-1}$ for a supergiant class.
On the other hand, if we could use the same formula for ten known WR stars
with available radio continuum measurements 
 \citep{leitherer97}, we would obtain an average log$_{10}$(N$_{\rm lyc}) =47.24\pm034$   photons s$^{-1}$. 
Assuming a distance of 3.0 kpc and a \BCK\ of $-4.3$ mag 
(typical of O7-O8 stars, hotter stars have a larger correction), 
with the \Aks\ values listed in Tables \ref{exttable} and 
\ref{abstable}, we inferred luminosities of 5.03 and 5.57 log($L/L_\odot$).
These values are consistent with those of evolved supergiants
\citep[see, for example, ][]{hamann06,martins07}.

\section{Bubbles and exciting stars}

We emphasize that our main goal is  to put at the disposal 
of the  astronomical community ionizing stars detected
in the studied \HH\ regions. These  spectroscopically observed 
stars are precious benchmarks for  stellar population studies.
In this section we comment on the relation of the identified stars with their surrounding bubbles.

Bubbles are prototype laboratories of current triggers and sequential star formation.
Given a molecular cloud and the first massive star forming within it, 
a \HH\ region forms and expands, creating a layer of compressed matter at its border, 
where molecular clumps condense, collapse, and interact.
Bubbles are very common: about $5000$ have already been identified, covering approximately 
10\% of the inner Galactic plane \citep{simpson12}.
There is ionized gas in about 90\% of the $\approx 600$
bubbles studied by \citet[][]{churchwell09} and \citet{deharveng10}, but 
only 13\% of them have been  associated with stellar clusters.
In the larger sample (1814)  analyzed by \citet{bufano18},  
60\% of bubbles  are associated with  \HH\ regions excited by young O, B stars, 
2\% by blue supergiants, and about 38\% are still of unknown origin. 
The underlying problem with classifying bubbles is the difficulty of detecting 
isolated massive stars and even  stellar clusters. 
A cluster is defined as an overdensity of  stars that are likely to be
physically close and of similar age; see, for example, MCM2005b77 (see Sect. \ref{clMCM2005b77}).
For details on detection algorithms and associated problematics, we remind the
reader of  the summary by \citet{schmeja11}.  
However, lists of candidate clusters from different surveys 
provide several clusters with different sizes and centroids for the same bubble.
 These candidate clusters  must be discussed in the context of 
the bubble morphology  and extinction pattern, with reference to benchmark stars.

\begin{figure}[h]
\begin{center}
\resizebox{0.8\hsize}{!}{\includegraphics[angle=0]{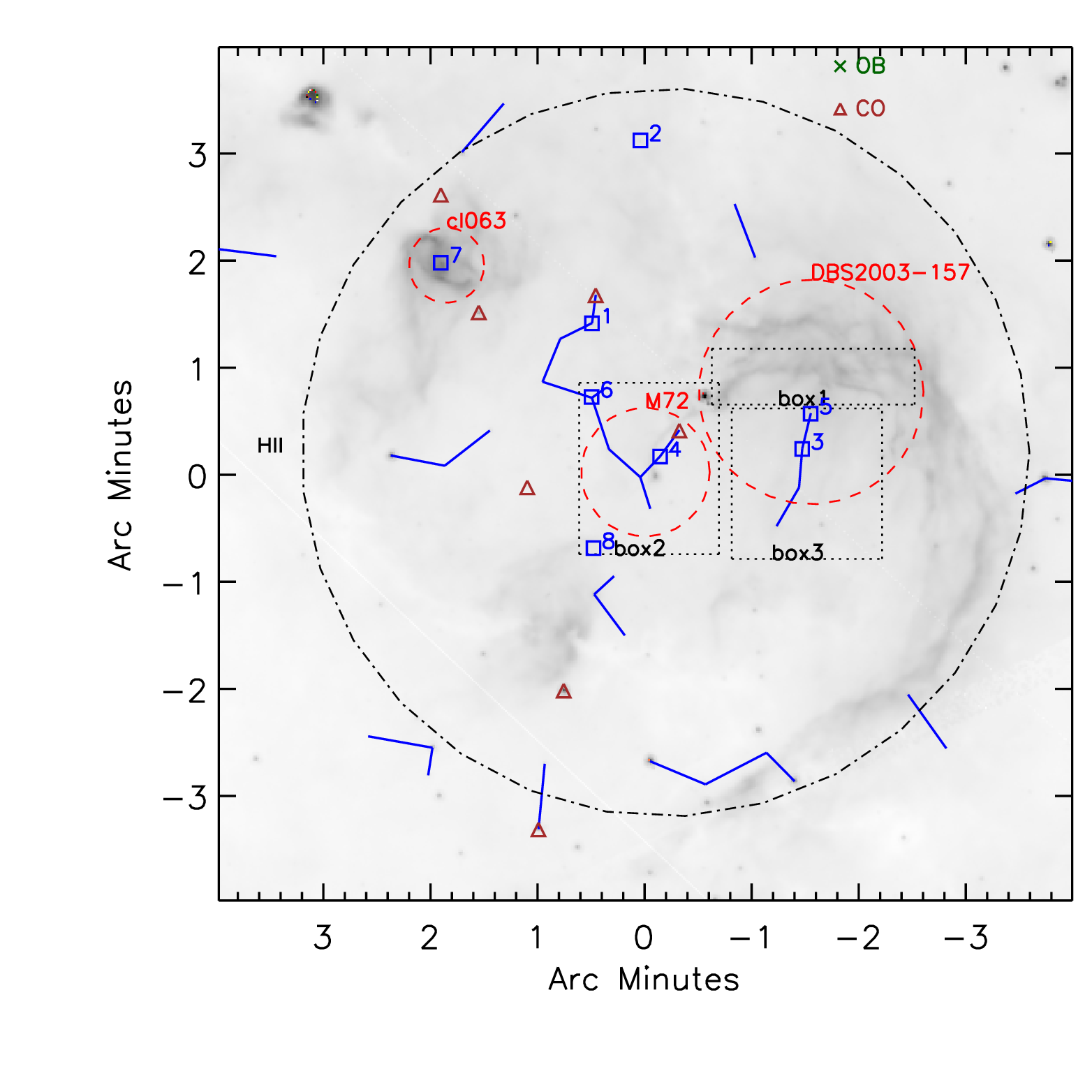}}
\end{center}
\caption{ \label{fig.detect} 
The gray scale shows a GLIMPSE 8.0 \um\  image of GRS G$331.31-00.34$. 
Red long-dashed circles mark the candidate clusters projected into this \HH\ region.
The \HH\ region is shown with a dotted-dashed black circle.
The three dotted boxes indicate regions plotted in the CMDs of Fig. \ref{boxes} (see text).
Observed stars are marked with blue squares (blue stars) and with 
red triangles (late-type stars).
The blue trees are the results of an MST search with a cutoff distance of 40\arcsec\ (0.8 pc).
} 
\end{figure}

\begin{figure}[h]
\begin{centering}
\resizebox{0.49\hsize}{!}{\includegraphics[angle=0]{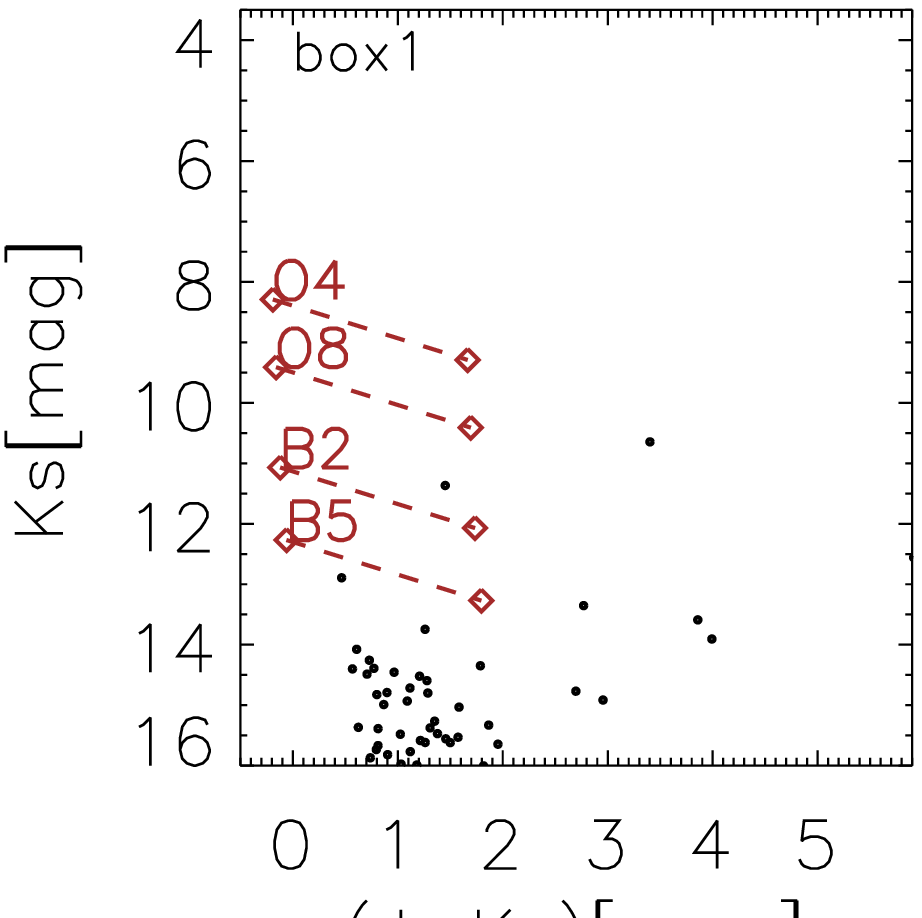}}
\resizebox{0.49\hsize}{!}{\includegraphics[angle=0]{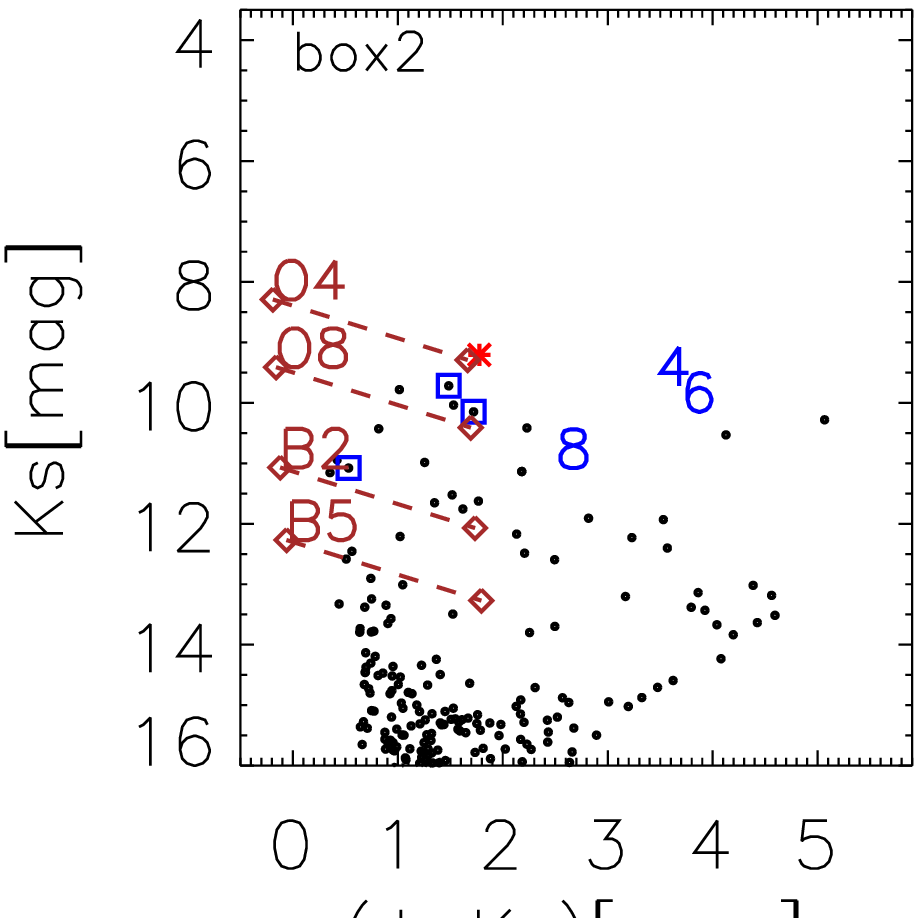}}
\resizebox{0.49\hsize}{!}{\includegraphics[angle=0]{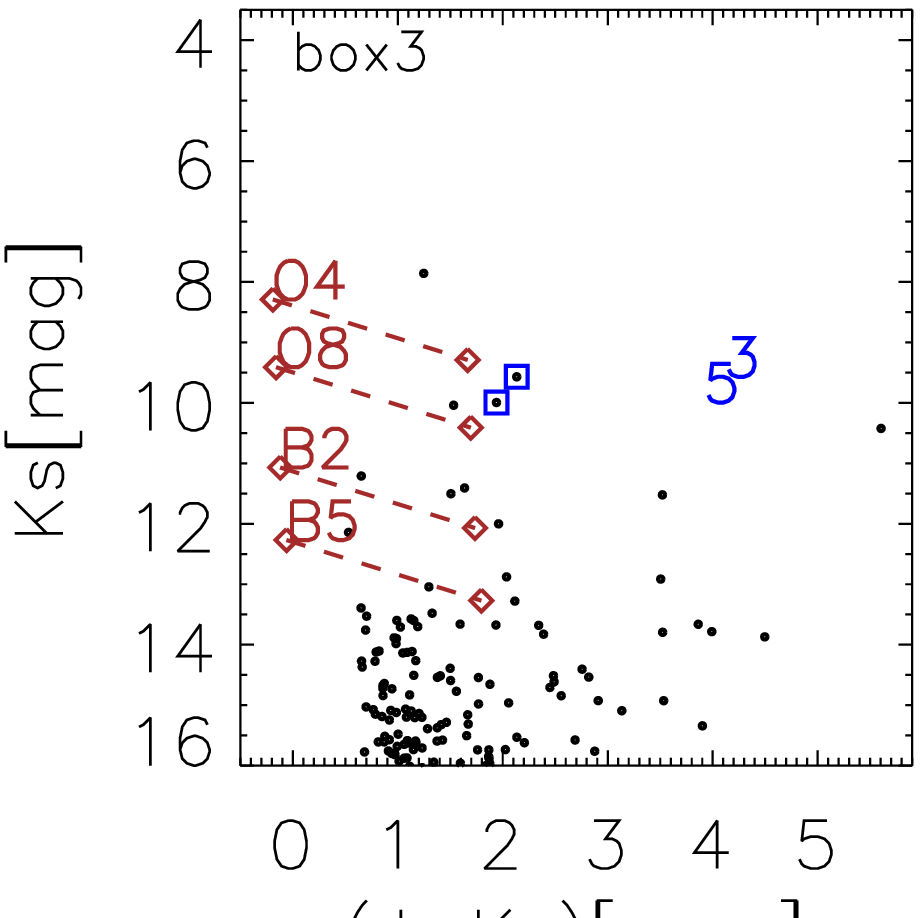}}
\end{centering}
\caption{ \label{boxes}
$J-$\Ks\ versus \Ks\ diagrams of the three  dotted-squared regions 
marked in Fig. \ref{fig.detect} (see text). 
Squares mark detected early stars, and asterisks mark detected late-type stars.
As in Fig.\ \ref{cmdg331.31}, diamonds indicate the locations of O4, O8, B2, and B5 dwarfs  
for a distance of 3.9 kpc at \Aks=0.0 and 1.0 mag.  
Dashed lines connect the values at \Aks=0 and those  at \Aks=1.0 mag.
} 
\end{figure}

The bubbles S36 and S62 were selected because they are \HH\ 
regions with candidate stellar clusters where the brightest stars have \Ks\ from 6 to 11 mag, 
suitable for a 4 m class telescope. The case of S62  is an exemplary one: 
S62 hosts several  candidate clusters, 
which were identified with different datasets and methods (see Sect. \ref{s62}).
The bubble has a diameter of about 8 pc at 3.9 kpc.
MCM2005b72 is a clump of bright stars (e.g., stars \#4, \#6, and  \#8) 
at mid- and near-infrared wavelengths, 
with a diameter of  72\arcsec\ (1.3 pc at 3.9 kpc), detected with an automated algorithm.   
DBS2003-172 has a diameter of 126\arcsec\ (2.3 pc at 3.9 kpc) and was visually detected 
and centered on   an arc of a bright 8 \um\ knot of S62. 
Cl063  is a bright 8 \um\ nebula located on the border of the bubble S62 and it coincides 
with the ATLASGAL clump  AGALG331.3873$-$0.360 (Hi-GAL 46282).   
We tried to redetect groups of bright GLIMPSE stars in S62 with the minimum spanning 
tree (MST) algorithm. The resulting maps  depend on the chosen cutoff distance 
and on the sample stars analyzed, but have the advantage of being independent 
of centroids \citep[e.g.,][]{wright14}. 
When considering stars bright in \Ks\ ($<11.$ mag) and [3.6] mag, i.e. OB stars at 3.9 kpc,
one can nicely reproduce the detection of MCM2005b72 with a cutoff distance from  
20\arcsec\ to 40\arcsec\ (0.4-0.8 pc at 3.9 kpc). 
A tree also appears at the location of the massive stars 
\#3 and \#5 (see Fig. \ref{fig.detect}). 
CMDs of stars detected by 2MASS and GLIMPSE 
in selected areas of Fig. \ref{fig.detect} are shown in Fig.\ \ref{boxes}.
MCM2005b72 (box2) is  located in the center of the bubble, 
so  it may  contain  massive stars from the first generation of star 
formation, such as stars \#4 and \#6
(late-O-type dwarfs, \Aks=0.8 and 0.9 mag, 
\Mk=$-$4.2 and $-$3.9 mag).  In box3, star \#5  has similar properties
-- an O9-B type (\Aks=1.1, \Mk=$-$4.1 mag). 
Star \#3 is the main ionizing star since it is a massive O4-6 (\Aks=1.3 mag),
of about 40 \Msun\ and  an age of 5-7 Myr \citep{ekstrom12}.

In conclusion, naked OB stars are not concentrated, 
as one would expect for a mass-segregated population,
but are detected inside the bubble in  two groups; while 
condensing molecular clumps are currently located on the  shell surrounding the bubble.
In the direction of Cl63 (AGALG331.3873$-$0.360), which is on the shell,  
we detected  another early-type, star \#7, which 
is likely to be contributing to the heating of the protostar  AGALG331.3873$-$0.360.   
These findings are in line with the current literature. 
Indeed, bound stellar clusters are a rarity and  in more 
than 90\% of the cases (infant mortality and cluster dissolution), we expect
associations of massive stars. For example, the massive  association  
Cygnus OB2 (about 7 pc in diameter) is at least 5-7 Myrs old, is not associated with gas, 
and is made of discrete clumps of stars 
\citep[1-2 pc in size,][]{wright14,wright16}. 
The authors conclude that Cygnus OB2  originated with 
discrete clumps, and there is no trace of dynamical evolution (e.g. mass segregation).
Similar evidence is provided for the younger W33 complex by 
\citet{messineo15},  5 pc in diameter, which is made up of several condensing molecular 
clumps and has sparsely distributed exciting stars (2-4 Myr).

\begin{table*}
\caption{ \label{abstable} Absolute properties of detected early-type stars in  MCM2005b72/DBS2003-157, MCM2005b77, and DBS2003-172.}
\begin{tabular}{l l r r r r r r r r}
\hline
\hline
ID             &      Sp &  $A_{K_{\mathrm s}}$ &${K_{\mathrm s}}_{\mathrm o}$ & $M_{K}$ &  $BC_{K_{\mathrm s}}$ & $DM$ & \Mbol  & log($L$) &\\
               &         &   [mag]              & [mag]                        & [mag]   &  [mag]                & [mag]&[mag]  & [$L_\odot$]&\\      
\hline
   G331.34$-$0.36   2 &      B0$-$5& 0.55 $\pm$ 0.03 &    8.34 $\pm$   0.04 &  $-$4.64 $\pm$   0.31 &  $-$3.12 $\pm$   0.66 &  12.98 $\pm$   0.31 &  $-$7.76 $\pm$    0.73 & 5.00$_{ 0.29}^{ 0.29}$ \\
   G331.34$-$0.36   3 &      O4$-$6& 1.32 $\pm$ 0.03 &    8.25 $\pm$   0.04 &  $-$4.72 $\pm$   0.31 &  $-$4.55 $\pm$   0.15 &  12.98 $\pm$   0.31 &  $-$9.27 $\pm$    0.35 & 5.61$_{ 0.14}^{ 0.14}$ \\
   G331.34$-$0.36   5 &    O9$-$B0 & 1.10 $\pm$ 0.02 &    8.90 $\pm$   0.04 &  $-$4.08 $\pm$   0.31 &  $-$3.85 $\pm$   0.13 &  12.98 $\pm$   0.31 &  $-$7.93 $\pm$    0.34 & 5.07$_{ 0.14}^{ 0.14}$ \\
   G331.34$-$0.36   6 &      B0$-$3& 0.96 $\pm$ 0.04 &    9.19 $\pm$   0.05 &  $-$3.79 $\pm$   0.31 &  $-$3.12 $\pm$   0.66 &  12.98 $\pm$   0.31 &  $-$6.91 $\pm$    0.73 & 4.66$_{ 0.29}^{ 0.29}$ \\
\hline
     MCM2005b77   1 &      B0$-$3& 0.84 $\pm$ 0.03 &    5.65 $\pm$   0.04 &  $-$7.95 $\pm$   0.17 &  $-$2.50 $\pm$   0.80 &  13.60 $\pm$   0.17 & $-$10.45 $\pm$    0.82 & 6.08$_{ 0.33}^{ 0.33}$ \\
     MCM2005b77   3 &      B0$-$5& 1.14 $\pm$ 0.03 &    7.08 $\pm$   0.04 &  $-$6.52 $\pm$   0.17 &  $-$2.12 $\pm$   1.17 &  13.60 $\pm$   0.17 &  $-$8.64 $\pm$    1.18 & 5.35$_{ 0.47}^{ 0.47}$ \\
     MCM2005b77   4 &      B0$-$3& 0.79 $\pm$ 0.03 &    7.62 $\pm$   0.04 &  $-$5.98 $\pm$   0.17 &  $-$2.50 $\pm$   0.80 &  13.60 $\pm$   0.17 &  $-$8.48 $\pm$    0.82 & 5.29$_{ 0.33}^{ 0.33}$ \\
     MCM2005b77   5 &      B0$-$3& 0.79 $\pm$ 0.03 &    7.81 $\pm$   0.04 &  $-$5.79 $\pm$   0.17 &  $-$2.50 $\pm$   0.80 &  13.60 $\pm$   0.17 &  $-$8.29 $\pm$    0.82 & 5.21$_{ 0.33}^{ 0.33}$ \\
     MCM2005b77   6 &      B0$-$5& 0.98 $\pm$ 0.03 &    7.73 $\pm$   0.04 &  $-$5.87 $\pm$   0.17 &  $-$2.12 $\pm$   1.17 &  13.60 $\pm$   0.17 &  $-$7.99 $\pm$    1.18 & 5.09$_{ 0.47}^{ 0.47}$ \\
     MCM2005b77   7 &      B0$-$5& 0.89 $\pm$ 0.03 &    8.10 $\pm$   0.03 &  $-$5.50 $\pm$   0.17 &  $-$2.12 $\pm$   1.17 &  13.60 $\pm$   0.17 &  $-$7.62 $\pm$    1.18 & 4.94$_{ 0.47}^{ 0.47}$ \\
     MCM2005b77   8 &      B0$-$5& 0.75 $\pm$ 0.03 &    8.55 $\pm$   0.04 &  $-$5.05 $\pm$   0.17 &  $-$2.12 $\pm$   1.17 &  13.60 $\pm$   0.17 &  $-$7.17 $\pm$    1.18 & 4.76$_{ 0.47}^{ 0.47}$ \\
     MCM2005b77   9 &      B0$-$5& 0.80 $\pm$ 0.04 &    8.63 $\pm$   0.06 &  $-$4.97 $\pm$   0.18 &  $-$2.12 $\pm$   1.17 &  13.60 $\pm$   0.17 &  $-$7.09 $\pm$    1.18 & 4.73$_{ 0.47}^{ 0.47}$ \\
     MCM2005b77  10 &      B0$-$5& 0.87 $\pm$ 0.03 &    9.26 $\pm$   0.04 &  $-$4.35 $\pm$   0.17 &  $-$2.12 $\pm$   1.17 &  13.60 $\pm$   0.17 &  $-$6.47 $\pm$    1.18 & 4.48$_{ 0.47}^{ 0.47}$ \\
     MCM2005b77  11 &      B0$-$5& 0.96 $\pm$ 0.05 &    9.32 $\pm$   0.80 &  $-$4.28 $\pm$   0.82 &  $-$2.12 $\pm$   1.17 &  13.60 $\pm$   0.17 &  $-$6.40 $\pm$    1.43 & 4.46$_{ 0.57}^{ 0.57}$ \\
     MCM2005b77  12 &        OB& 1.70 $\pm$ 0.03 &    9.26 $\pm$   0.04 &  $-$4.34 $\pm$   0.17 &  $-$2.87 $\pm$   1.21 &  13.60 $\pm$   0.17 &  $-$7.21 $\pm$    1.22 & 4.78$_{ 0.49}^{ 0.49}$ \\
\hline
    DBS2003-172   3 &        OB& 1.05 $\pm$ 0.03 &    8.55 $\pm$   0.04 &  $-$3.87 $\pm$   0.50 &  $-$3.27 $\pm$   0.86 &  12.41 $\pm$   0.50 &  $-$7.14 $\pm$    1.00 & 4.75$_{ 0.40}^{ 0.40}$ \\
    DBS2003-172   5A &        OB& 1.90 $\pm$ 0.02 &    8.89 $\pm$   0.04 &  $-$3.53 $\pm$   0.50 &  $-$4.30 $\pm$   1.00 &  12.41 $\pm$   0.50 &  $-$7.83 $\pm$    1.12 & 5.03$_{ 0.45}^{ 0.45}$ \\
    DBS2003-172   5B &        OB& 3.58 $\pm$ 0.28 &    7.54 $\pm$   0.31 &  $-$4.87 $\pm$   0.59 &  $-$4.30 $\pm$   1.00 &  12.41 $\pm$   0.50 &  $-$9.17 $\pm$    1.16 & 5.57$_{ 0.46}^{ 0.46}$ \\
\hline
\end{tabular}
\end{table*}

\section{Summary}
\label{summary}

We performed a spectroscopic survey of bright infrared stars
in the direction of GRS G331.34$-$00.36 (S62), GRS G337.92$-$00.48 (S36),
and MCM2005b77, and for the first time we detected
massive members in these regions.

\begin{itemize}
\item[a]
A total number of 27 early-type and 32 late-type stars were detected.

\item[b]
We  confirm one O4-6 star and one B0-5 star as the main sources of 
ionizing radiation of GRS G331.34$-$00.36 (S62),
at a kinematic distance of $3.9 \pm 0.3$ kpc. The stars have
\Aks\ of 1.3 mag and 1.0 mag. 
With N$_{\rm lyc}$=49 photons s$^{-1}$,
the massive O4-6 star can  alone account for most of 
the energy needed to maintain this \HH\ region.

\item[c] 
In GRS G337.92-00.48 (S36), at the  distance of about 3.0 kpc, 
we detected four photometric O-type stars at \Aks=1.0 mag
and two  massive emission line stars surrounded by a nebula. 
The newly discovered massive stars can account for the bulk of  
radio continuum emitted by this \HH\ region.

\item[d] We detected a cluster of B-type stars at the location of MCM2005b77.
The cluster has an average \Aks=$0.91\pm0.08$ mag and a spectrophotometric distance 
modulus of $13.53\pm1.08$ mag that agrees 
with the kinematic distance of the adjacent \HH\ IRAS 16137$-$5025.
We inferred an age of 8 Myr and a mass of about 3000 \Msun.

 \item[e] Star formation is currently ongoing in these bubbles, with new condensations
detected by ATLASGAL with masses from  30 to 3500 \Msun.

\end{itemize}

\begin{appendix}
\section{Charts of the targets}
In Figure \ref{charts}, \Ks\ charts from the VVV survey  are shown 
for the detected stars.
Charts are aligned with the Celestial axis: north is up, and east is to the left.
Each chart has a field of view of $1$\arcmin$\times1$\arcmin.

\begin{figure*}
\begin{centering}
\resizebox{0.8\hsize}{!}{\includegraphics[angle=0]{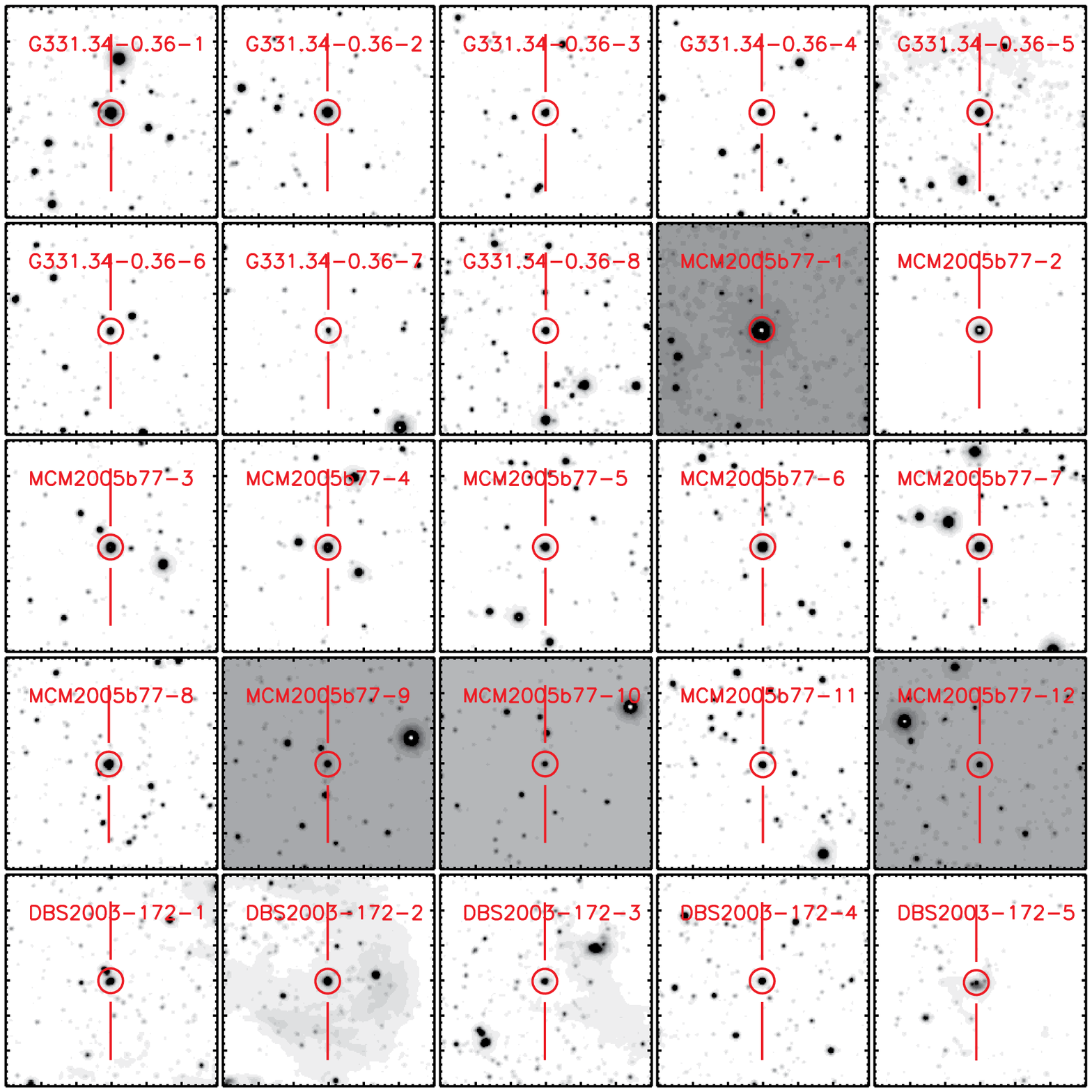}}
\end{centering}
\caption{ \label{charts} Charts of spectroscopically observed stars.
$K$ and \Ks\ images, $1\times1$ \arcmin\ large (see text); North is up and East to the Left.
Targets are marked and labeled as in Tables \ref{spectra.early} and \ref{spectra.late}. }
\end{figure*}

\addtocounter{figure}{-1}
\begin{figure*}
\begin{centering}
\resizebox{0.8\hsize}{!}{\includegraphics[angle=0]{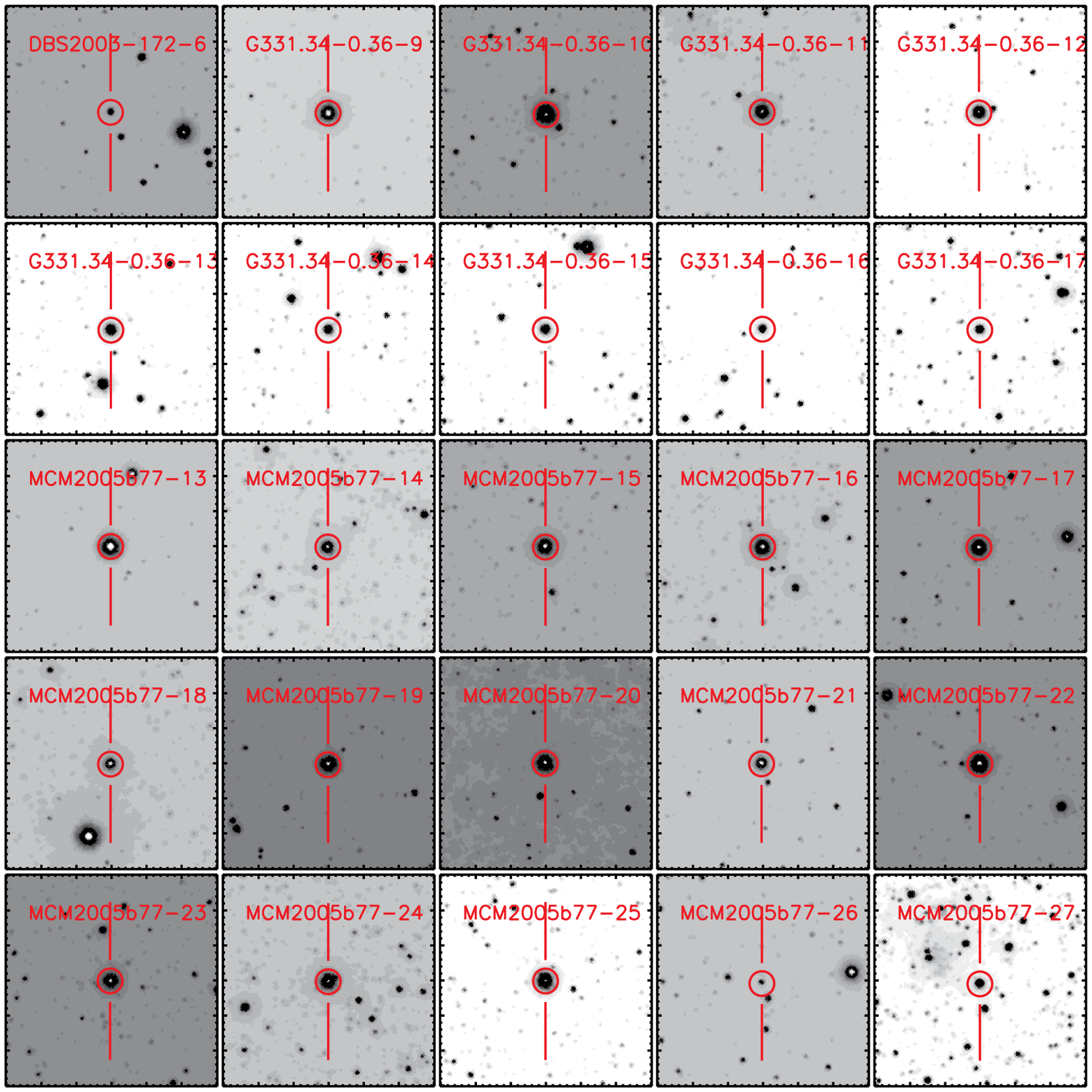}}
\end{centering}
\caption{ Continuation of Fig.\ \ref{charts}. }
\end{figure*}

\addtocounter{figure}{-1}
\begin{figure*}
\begin{centering}
\resizebox{0.8\hsize}{!}{\includegraphics[angle=0]{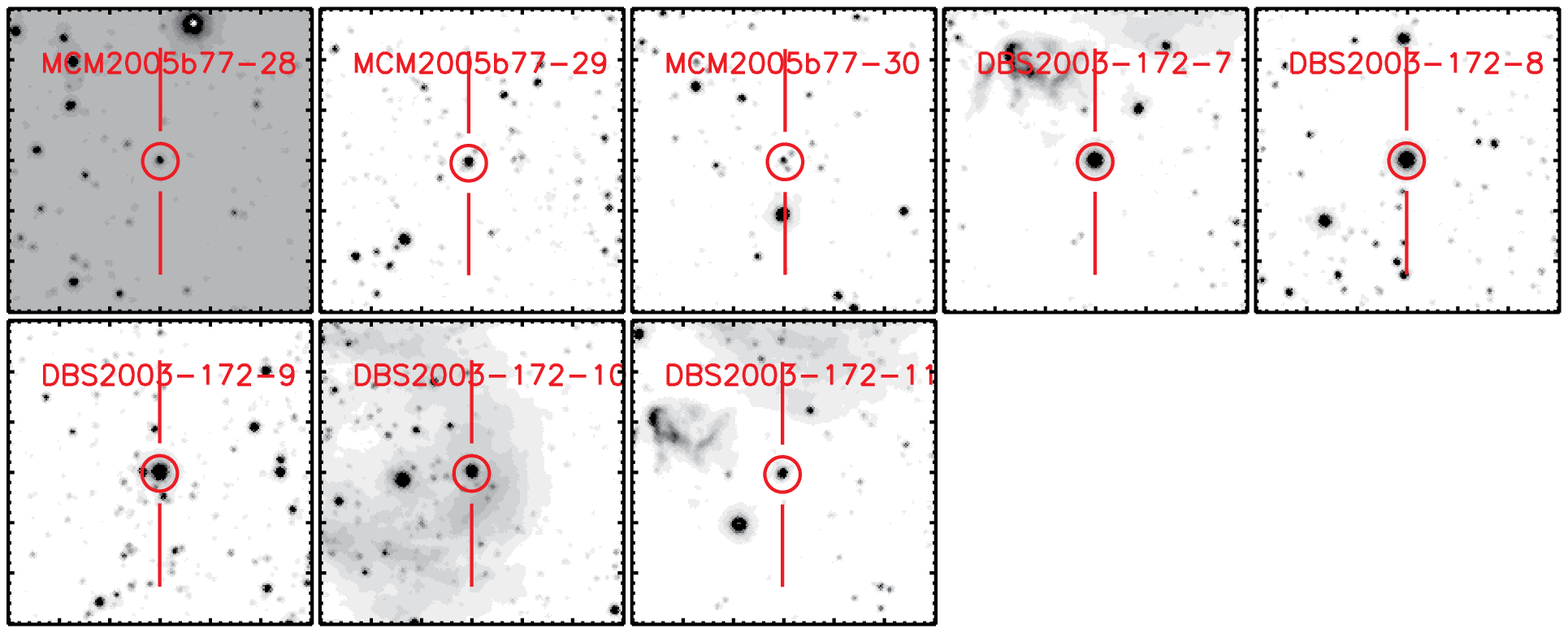}}
\end{centering}
\caption{Continuation of Fig.\ \ref{charts}. }
\end{figure*}

\begin{figure*}
\begin{centering}
\resizebox{0.49\hsize}{!}{\includegraphics[angle=0]{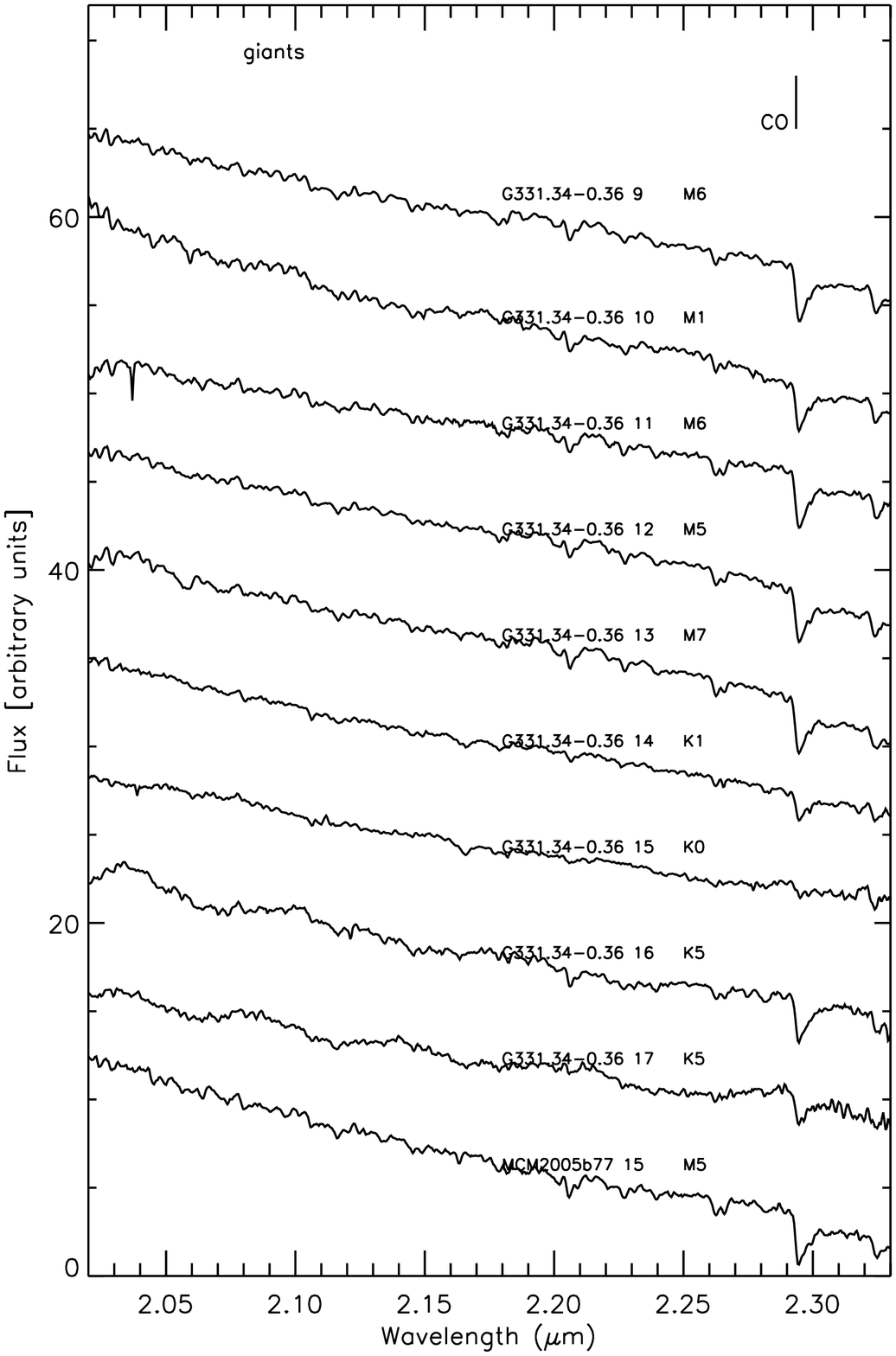}}
\resizebox{0.49\hsize}{!}{\includegraphics[angle=0]{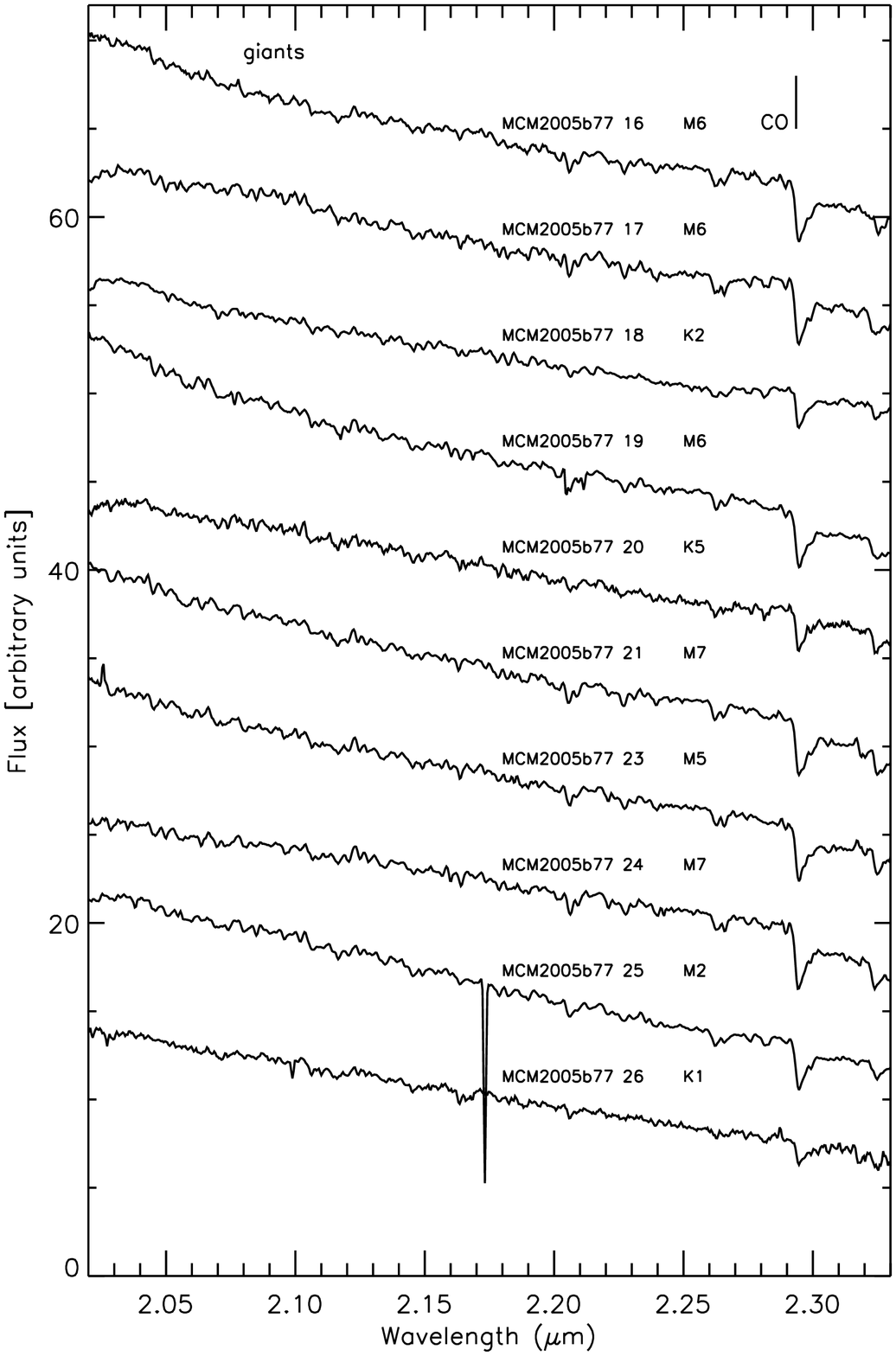}}
\end{centering}
\caption{ \label{spectralate} Spectra of late-type stars. }
\end{figure*}

\addtocounter{figure}{-1}
\begin{figure*}
\begin{centering}
\resizebox{0.49\hsize}{!}{\includegraphics[angle=0]{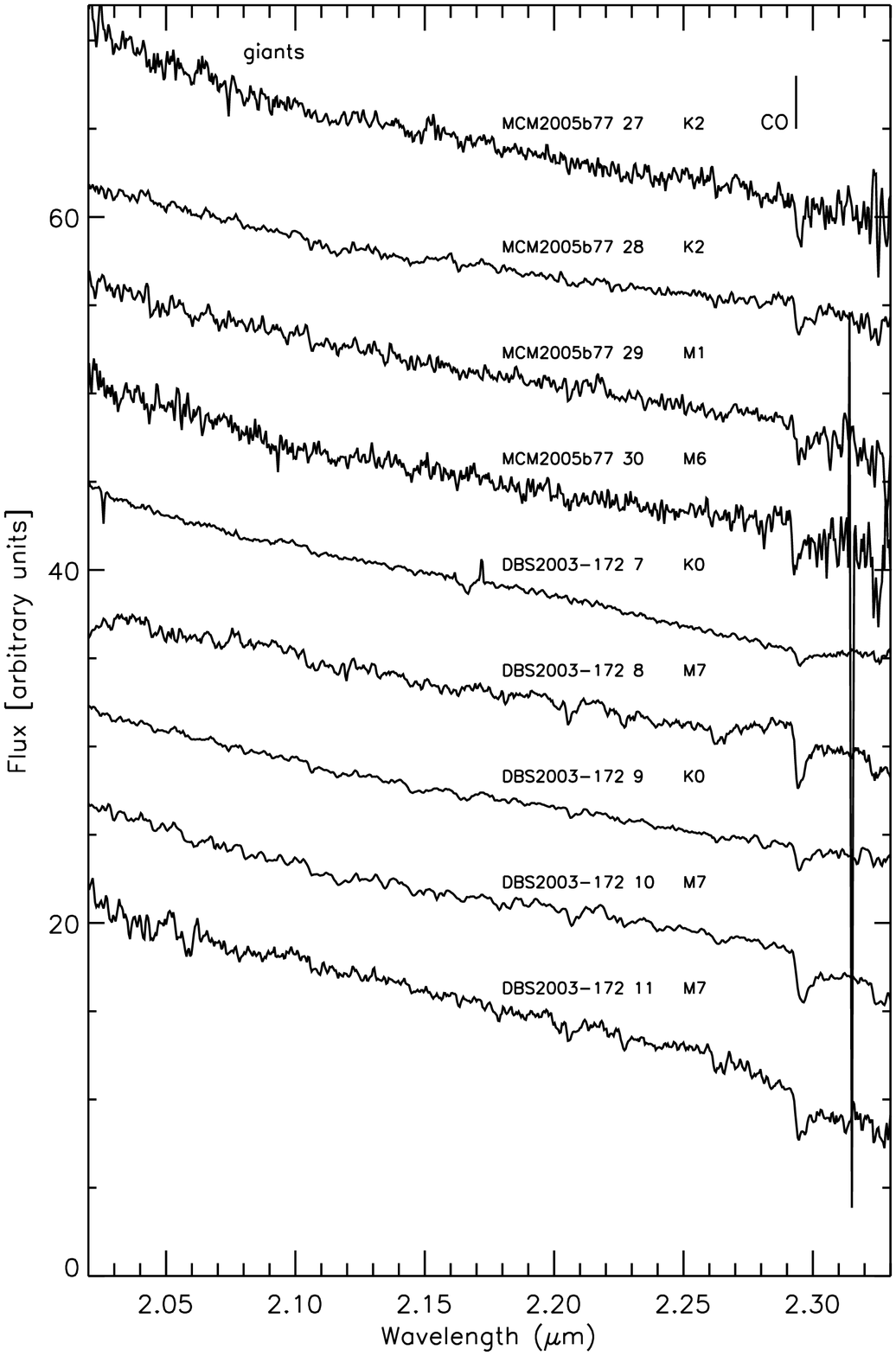}}
\resizebox{0.49\hsize}{!}{\includegraphics[angle=0]{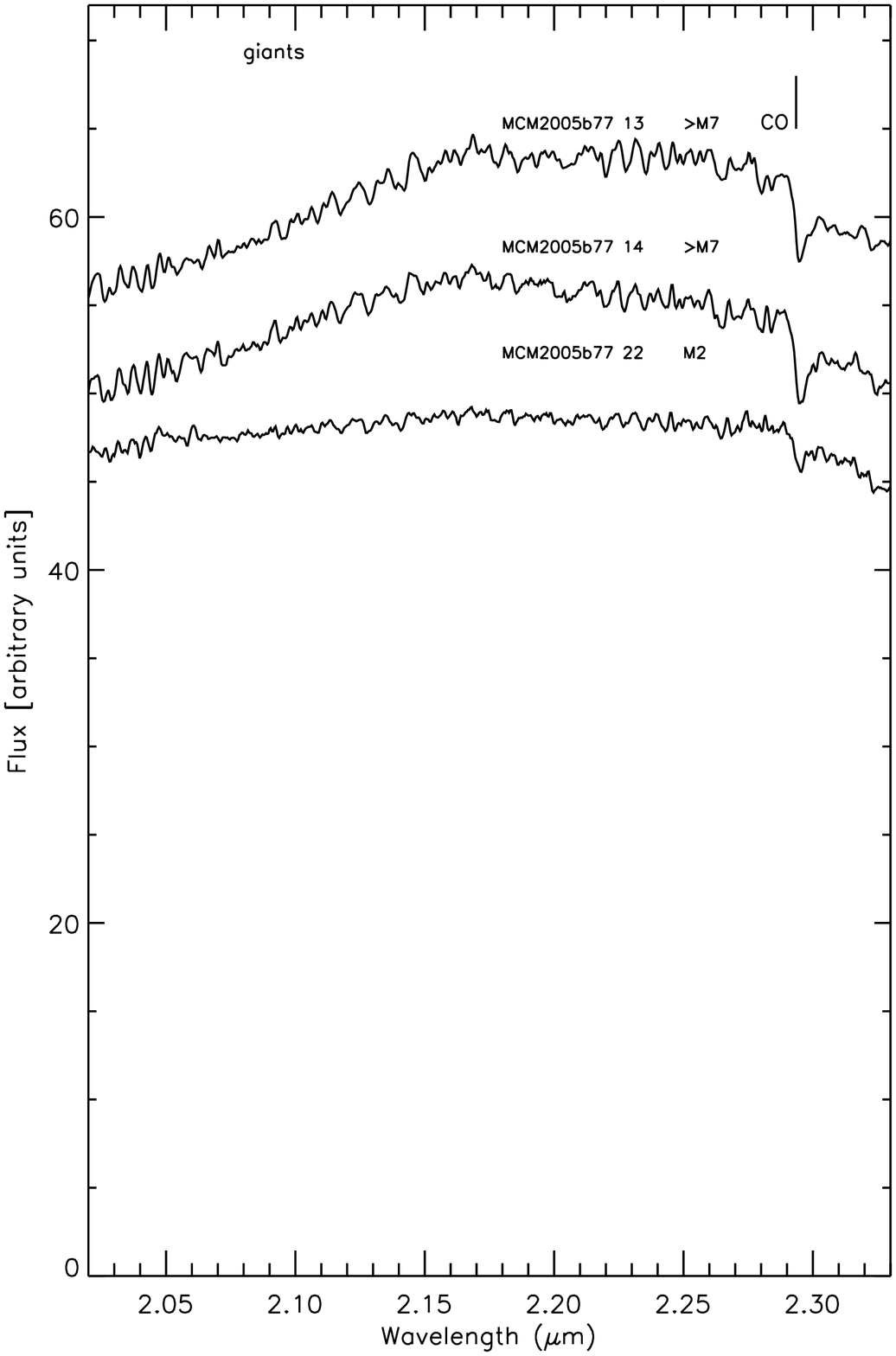}}
\end{centering}
\caption{ Continuation of Fig.\ \ref{spectralate}. }
\end{figure*}

\section{List of molecular clumps associated with the studied \HH\ regions}

In Table \ref{listclump} we list the  parameters of  Hi-GAL clumps 
in the direction of the studied \HH\ regions.
Mid-infrared  sources detected with 
the Hershel satellite  were matched with the ATLASGAL point sources to identify
protostar and prestellar objects by \citet{elia17}.
\citet{elia17} estimate distances with available $^{12}$CO and $^{13}$CO 
observations and the Galactic rotation curve.
 
Highly reliable radial velocities of $\approx 8000$ ATLASGAL clumps are now available from
line observations of  C$^{18}$O(2-1), C$^{17}$O(3-2), NH$_3$, and CS \citep{urquhart18}.
Distances have been revised with the catalog of radial velocities by \citet{urquhart18},
and masses  rescaled to the adopted distances.

\begin{table}
\caption{\label{listclump} List of Hi-GAL clumps associated with the studied \HH\ regions}
\begin{tabular}{lllllrrrrrrr}
\hline
\hline
     \HH   & ID&  Temp$^a$&    EV$^b$&  COND$^c$&  Mass$^a$        &  Dist$^d$ &      Vlsr$^e$   & sep$^f$     & a$^f$  & D(Urq)$^d$    & D(Elia)$^d$ \\ 
           &      &  [K] &      &      &  [M$_\odot$] &  [kpc]&    [\kms] & [\arcsec]& [\arcsec]& [kpc]  & [kpc] \\
\hline     
       S62 &  46184 &  11.20 &   1 &   1 &   9592 &     10.64 &      $..$ &      $..$ &      $..$ &      $..$ &     10.64  \\
       S62 &  46196 &  20.03 &   2 &   1 &    105 &      3.94 &    $-$65.50 &     13.53 &     16.00 &      4.00 &     10.64  \\
       S62 &  46198 &  23.26 &   2 &   1 &     76 &      3.94 &    $-$65.30 &      8.79 &     11.00 &      4.00 &     10.64  \\
       S62 &  46204 &  11.86 &   2 &   1 &    668 &      3.94 &    $-$66.40 &     11.04 &     16.00 &      4.00 &     10.64  \\
       S62 &  46210 &  28.58 &   2 &   1 &     30 &      3.94 &      $..$ &      $..$ &      $..$ &      $..$ &      4.22  \\
       S62 &  46219 &  15.83 &   2 &   1 &    569 &      3.94 &    $-$66.00 &      9.19 &     39.00 &      4.00 &      4.22  \\
       S62 &  46225 &  16.42 &   2 &   1 &    350 &      3.94 &    $-$65.90 &     25.77 &     32.00 &      4.00 &     10.64  \\
       S62 &  46235 &  16.65 &   2 &   1 &    146 &      3.94 &      $..$ &      $..$ &      $..$ &      $..$ &      4.22  \\
       S62 &  46238 &  22.89 &   2 &   1 &    529 &      3.94 &    $-$65.50 &      0.45 &     30.00 &      4.00 &      4.22  \\
       S62 &  46239 &  15.59 &   1 &   1 &    272 &      3.94 &    $-$65.00 &     11.79 &     21.00 &      3.90 &     10.64  \\
       S62 &  46246 &  27.87 &   2 &   0 &     76 &      3.94 &      $..$ &      $..$ &      $..$ &      $..$ &      4.22  \\
       S62 &  46248 &  13.34 &   2 &   1 &    376 &      3.94 &      $..$ &      $..$ &      $..$ &      $..$ &      4.22  \\
       S62 &  46272 &  10.84 &   2 &   1 &   2342 &      5.80 &    $-$97.30 &     12.12 &     22.00 &      5.80 &      4.22  \\
       S62 &  46282 &  30.19 &   2 &   1 &    158 &      3.94 &    $-$65.60 &      7.63 &     16.00 &      4.00 &      4.22  \\
       S62 &  91379 &  13.69 &   1 &   1 &    176 &      3.94 &      $..$ &      $..$ &      $..$ &      $..$ &      4.22  \\
       S62 &  91393 &  15.83 &   1 &   1 &    201 &      3.94 &      $..$ &      $..$ &      $..$ &      $..$ &      4.22  \\
IRAS 16137$-$5025 &  47117 &  20.39 &   2 &   1 &    582 &      5.20 &    $-$95.60 &      4.15 &     21.00 &      5.60 &     10.96  \\
IRAS 16137$-$5025 &  47118 &  12.09 &   2 &   1 &   1167 &      5.20 &    $-$95.30 &      9.87 &     18.00 &      5.60 &      3.84  \\
IRAS 16137$-$5025 &  47127 &  15.11 &   2 &   1 &   1488 &      5.20 &    $-$95.00 &      8.85 &     16.00 &      5.60 &      3.84  \\
IRAS 16137$-$5025 &  47129 &  14.05 &   2 &   1 &    520 &      5.20 &    $-$94.80 &      6.29 &     20.00 &      5.60 &      3.84  \\
       S36 &  50256 &  14.52 &   1 &   1 &    309 &      3.04 &    $-$40.00 &      6.25 &     38.00 &      3.00 &      3.30  \\
       S36 &  50259 &  10.96 &   2 &   1 &   3553 &      3.04 &    $-$40.00 &     32.84 &     38.00 &      3.00 &      3.30  \\
       S36 &  50261 &  30.90 &   2 &   1 &     73 &      3.04 &      $..$ &      $..$ &      $..$ &      $..$ &      3.13  \\
       S36 &  50264 &  11.31 &   1 &   1 &   5925 &     12.68 &      $..$ &      $..$ &      $..$ &      $..$ &     12.68  \\
       S36 &  50266$^*$ &  31.13 &   2 &   1 &   3394 &     12.60? &    $-$38.70 &     33.56 &     45.00 &     12.60 &      3.13  \\
       S36 &  50269 &  13.63 &   1 &   1 &    327 &      3.04 &    $-$39.50 &     53.48 &     93.00 &      2.90 &      3.57  \\
       S36 &  50271 &  18.91 &   2 &   1 &    350 &      3.04 &      $..$ &      $..$ &      $..$ &      $..$ &      3.13  \\
       S36 &  50279$^*$ &  23.83 &   2 &   1 &  19733 &     12.60? &    $-$38.70 &     26.32 &     45.00 &     12.60 &      3.13  \\
       S36 &  50285 &  17.25 &   2 &   1 &    846 &      3.04 &    $-$38.00 &      2.91 &     39.00 &      2.80 &     12.68  \\
       S36 &  50287 &  15.83 &   2 &   0 &    122 &      3.04 &    $-$39.50 &     71.62 &     93.00 &      2.90 &      3.57  \\
       S36 &  50305 &  17.43 &   2 &   1 &     50 &      3.04 &      $..$ &      $..$ &      $..$ &      $..$ &      3.13  \\
       S36 &  50313 &  15.88 &   2 &   1 &   3087 &     12.93 &      $..$ &      $..$ &      $..$ &      $..$ &     12.93  \\
       S36 &  50322 &  13.16 &   2 &   1 &   3930 &     12.93 &      $..$ &      $..$ &      $..$ &      $..$ &     12.93  \\
       S36 &  93621 &  18.26 &   1 &   1 &     84 &      3.04 &    $-$39.50 &     13.31 &     93.00 &      2.90 &      3.13  \\
       S36 &  93623 &  18.20 &   1 &   1 &    137 &      3.04 &    $-$37.60 &      6.72 &     12.00 &      2.80 &     12.68  \\
       S36 &  93634 &  17.96 &   1 &   1 &     61 &      3.04 &      $..$ &      $..$ &      $..$ &      $..$ &      3.57  \\
       S36 &  93636 &  13.58 &   1 &   1 &    106 &      3.04 &      $..$ &      $..$ &      $..$ &      $..$ &      3.57  \\
       \hline
\end{tabular}
\begin{list}{}
\item Notes:
   $(^a)$ Temperatures (Temp) and masses (Mass) are taken from \citet{elia17}; masses are rescaled to the assumed distances.   \\    
   $(^b)$  EV=1 (prestellar) and EV=2 (protostellar), as in \citet{elia17}. \\    
   $(^c)$ COND=1 if(mass $> 580 \times ({\mathrm diameter}/2)^{1.33}$). The clump is likely collapsing \citep{koenig17}. \\    
   $(^d)$ Dist= assumed distance ; D(Urq)= distance assumed in the work of \citet{urquhart18}; D(Elia)=distances from the work of  \citet{elia17}.\\
   $(^e)$ \Vlsr\ data are taken from \citet{urquhart18}.\\
   $(^f)$ Distance (sep) to the ATLASGAL clump centroid  by  \citet{urquhart18}; the clump  has a semi major axis = a.\\
   $(^*)$  \citet{kim17} argue for a close distance  based on  the $H$ radio-recombination line, \Vlsr=$-43$ \kms.
\end{list}
\end{table}
\end{appendix}


\begin{thebibliography}{97}
\expandafter\ifx\csname natexlab\endcsname\relax\def\natexlab#1{#1}\fi

\bibitem[{{Benjamin} {et~al.}(2003){Benjamin}, {Churchwell}, {Babler}, {Bania},
  {Clemens}, {Cohen}, {Dickey}, {Indebetouw}, {Jackson}, {Kobulnicky},
  {Lazarian}, {Marston}, {Mathis}, {Meade}, {Seager}, {Stolovy}, {Watson},
  {Whitney}, {Wolff}, \& {Wolfire}}]{benjamin03}
{Benjamin}, R.~A., {Churchwell}, E., {Babler}, B.~L., {et~al.} 2003, \pasp,
  115, 953

\bibitem[{{Bertoldi} \& {McKee}(1992)}]{bertoldi92}
{Bertoldi}, F. \& {McKee}, C.~F. 1992, \apj, 395, 140

\bibitem[{{Bibby} {et~al.}(2008){Bibby}, {Crowther}, {Furness}, \&
  {Clark}}]{bibby08}
{Bibby}, J.~L., {Crowther}, P.~A., {Furness}, J.~P., \& {Clark}, J.~S. 2008,
  \mnras, 386, L23

\bibitem[{{Blum} {et~al.}(1995){Blum}, {Depoy}, \& {Sellgren}}]{blum95}
{Blum}, R.~D., {Depoy}, D.~L., \& {Sellgren}, K. 1995, \apj, 441, 603

\bibitem[{{Blum} {et~al.}(2003){Blum}, {Ram{\'{\i}}rez}, {Sellgren}, \&
  {Olsen}}]{blum03}
{Blum}, R.~D., {Ram{\'{\i}}rez}, S.~V., {Sellgren}, K., \& {Olsen}, K. 2003,
  \apj, 597, 323

\bibitem[{{Borissova} {et~al.}(2011){Borissova}, {Bonatto}, {Kurtev}, {Clarke},
  {Pe{\~n}aloza}, {Sale}, {Minniti}, {Alonso-Garc{\'{\i}}a}, {Artigau},
  {Barb{\'a}}, {Bica}, {Baume}, {Catelan}, {Chen{\`e}}, {Dias}, {Folkes},
  {Froebrich}, {Geisler}, {de Grijs}, {Hanson}, {Hempel}, {Ivanov}, {Kumar},
  {Lucas}, {Mauro}, {Moni Bidin}, {Rejkuba}, {Saito}, {Tamura}, \&
  {Toledo}}]{borissova11}
{Borissova}, J., {Bonatto}, C., {Kurtev}, R., {et~al.} 2011, \aap, 532, A131

\bibitem[{{Borissova} {et~al.}(2006){Borissova}, {Ivanov}, {Minniti}, \&
  {Geisler}}]{borissova06}
{Borissova}, J., {Ivanov}, V.~D., {Minniti}, D., \& {Geisler}, D. 2006, \aap,
  455, 923

\bibitem[{{Borissova} {et~al.}(2005){Borissova}, {Ivanov}, {Minniti},
  {Geisler}, \& {Stephens}}]{borissova05}
{Borissova}, J., {Ivanov}, V.~D., {Minniti}, D., {Geisler}, D., \& {Stephens},
  A.~W. 2005, \aap, 435, 95

\bibitem[{{Bronfman} {et~al.}(1996){Bronfman}, {Nyman}, \& {May}}]{bronfman96}
{Bronfman}, L., {Nyman}, L.-A., \& {May}, J. 1996, \aaps, 115, 81

\bibitem[{{Bufano} {et~al.}(2018){Bufano}, {Leto}, {Carey}, {Umana}, {Buemi},
  {Ingallinera}, {Bulpitt}, {Cavallaro}, {Riggi}, {Trigilio}, \&
  {Molinari}}]{bufano18}
{Bufano}, F., {Leto}, P., {Carey}, D., {et~al.} 2018, \mnras, 473, 3671

\bibitem[{{Camargo} {et~al.}(2012){Camargo}, {Bonatto}, \& {Bica}}]{camargo14}
{Camargo}, D., {Bonatto}, C., \& {Bica}, E. 2012, \mnras, 423, 1940

\bibitem[{{Caswell} \& {Haynes}(1987)}]{caswell87}
{Caswell}, J.~L. \& {Haynes}, R.~F. 1987, \aap, 171, 261

\bibitem[{{Churchwell} {et~al.}(2009){Churchwell}, {Babler}, {Meade},
  {Whitney}, {Benjamin}, {Indebetouw}, {Cyganowski}, {Robitaille}, {Povich},
  {Watson}, \& {Bracker}}]{churchwell09}
{Churchwell}, E., {Babler}, B.~L., {Meade}, M.~R., {et~al.} 2009, \pasp, 121,
  213

\bibitem[{{Churchwell} {et~al.}(2006){Churchwell}, {Povich}, {Allen}, {Taylor},
  {Meade}, {Babler}, {Indebetouw}, {Watson}, {Whitney}, {Wolfire}, {Bania},
  {Benjamin}, {Clemens}, {Cohen}, {Cyganowski}, {Jackson}, {Kobulnicky},
  {Mathis}, {Mercer}, {Stolovy}, {Uzpen}, {Watson}, \& {Wolff}}]{churchwell06}
{Churchwell}, E., {Povich}, M.~S., {Allen}, D., {et~al.} 2006, \apj, 649, 759

\bibitem[{{Culverhouse} {et~al.}(2011){Culverhouse}, {Ade}, {Bock}, {Bowden},
  {Brown}, {Cahill}, {Castro}, {Church}, {Friedman}, {Ganga}, {Gear}, {Gupta},
  {Hinderks}, {Kovac}, {Lange}, {Leitch}, {Melhuish}, {Memari}, {Murphy},
  {Orlando}, {Pryke}, {Schwarz}, {O'Sullivan}, {Piccirillo}, {Rajguru},
  {Rusholme}, {Taylor}, {Thompson}, {Turner}, {Wu}, {Zemcov}, \& {QUaD
  Collaboration}}]{culverhouse11}
{Culverhouse}, T., {Ade}, P., {Bock}, J., {et~al.} 2011, \apjs, 195, 8

\bibitem[{{Davies} {et~al.}(2012){Davies}, {Clark}, {Trombley}, {Figer},
  {Najarro}, {Crowther}, {Kudritzki}, {Thompson}, {Urquhart}, \&
  {Hindson}}]{davies12}
{Davies}, B., {Clark}, J.~S., {Trombley}, C., {et~al.} 2012, \mnras, 419, 1871

\bibitem[{{Deharveng} {et~al.}(2010){Deharveng}, {Schuller}, {Anderson},
  {Zavagno}, {Wyrowski}, {Menten}, {Bronfman}, {Testi}, {Walmsley}, \&
  {Wienen}}]{deharveng10}
{Deharveng}, L., {Schuller}, F., {Anderson}, L.~D., {et~al.} 2010, \aap, 523,
  A6

\bibitem[{{Doherty} {et~al.}(1994){Doherty}, {Puxley}, {Doyon}, \&
  {Brand}}]{doherty94}
{Doherty}, R.~M., {Puxley}, P., {Doyon}, R., \& {Brand}, P.~W.~J.~L. 1994,
  \mnras, 266, 497

\bibitem[{{Drimmel} {et~al.}(2003){Drimmel}, {Cabrera-Lavers}, \&
  {L{\'o}pez-Corredoira}}]{drimmel03}
{Drimmel}, R., {Cabrera-Lavers}, A., \& {L{\'o}pez-Corredoira}, M. 2003, \aap,
  409, 205

\bibitem[{{Dutra} \& {Bica}(2001)}]{dutra01}
{Dutra}, C.~M. \& {Bica}, E. 2001, \aap, 376, 434

\bibitem[{{Dutra} {et~al.}(2003){Dutra}, {Bica}, {Soares}, \&
  {Barbuy}}]{dutra03}
{Dutra}, C.~M., {Bica}, E., {Soares}, J., \& {Barbuy}, B. 2003, \aap, 400, 533

\bibitem[{{Eckart} {et~al.}(2013){Eckart}, {Mu{\v z}i{\'c}}, {Yazici}, {Sabha},
  {Shahzamanian}, {Witzel}, {Moser}, {Garcia-Marin}, {Valencia-S.}, {Jalali},
  {Bremer}, {Straubmeier}, {Rauch}, {Buchholz}, {Kunneriath}, \&
  {Moultaka}}]{eckart13}
{Eckart}, A., {Mu{\v z}i{\'c}}, K., {Yazici}, S., {et~al.} 2013, \aap, 551, A18

\bibitem[{{Egan} {et~al.}(2003){Egan}, {Price}, \& {Kraemer}}]{egan03}
{Egan}, M.~P., {Price}, S.~D., \& {Kraemer}, K.~E. 2003, in Bulletin of the
  American Astronomical Society, Vol.~35, American Astronomical Society Meeting
  Abstracts, 1301

\bibitem[{{Ekstr{\"o}m} {et~al.}(2012){Ekstr{\"o}m}, {Georgy}, {Eggenberger},
  {Meynet}, {Mowlavi}, {Wyttenbach}, {Granada}, {Decressin}, {Hirschi},
  {Frischknecht}, {Charbonnel}, \& {Maeder}}]{ekstrom12}
{Ekstr{\"o}m}, S., {Georgy}, C., {Eggenberger}, P., {et~al.} 2012, \aap, 537,
  A146

\bibitem[{{Elia} {et~al.}(2017){Elia}, {Molinari}, {Schisano}, {Pestalozzi},
  {Pezzuto}, {Merello}, {Noriega-Crespo}, {Moore}, {Russeil}, {Mottram},
  {Paladini}, {Strafella}, {Benedettini}, {Bernard}, {Di Giorgio}, {Eden},
  {Fukui}, {Plume}, {Bally}, {Martin}, {Ragan}, {Jaffa}, {Motte}, {Olmi},
  {Schneider}, {Testi}, {Wyrowski}, {Zavagno}, {Calzoletti}, {Faustini},
  {Natoli}, {Palmeirim}, {Piacentini}, {Piazzo}, {Pilbratt}, {Polychroni},
  {Baldeschi}, {Beltr{\'a}n}, {Billot}, {Cambr{\'e}sy}, {Cesaroni},
  {Garc{\'{\i}}a-Lario}, {Hoare}, {Huang}, {Joncas}, {Liu}, {Maiolo}, {Marsh},
  {Maruccia}, {M{\`e}ge}, {Peretto}, {Rygl}, {Schilke}, {Thompson},
  {Traficante}, {Umana}, {Veneziani}, {Ward-Thompson}, {Whitworth}, {Arab},
  {Bandieramonte}, {Becciani}, {Brescia}, {Buemi}, {Bufano}, {Butora},
  {Cavuoti}, {Costa}, {Fiorellino}, {Hajnal}, {Hayakawa}, {Kacsuk}, {Leto}, {Li
  Causi}, {Marchili}, {Martinavarro-Armengol}, {Mercurio}, {Molinaro},
  {Riccio}, {Sano}, {Sciacca}, {Tachihara}, {Torii}, {Trigilio}, {Vitello}, \&
  {Yamamoto}}]{elia17}
{Elia}, D., {Molinari}, S., {Schisano}, E., {et~al.} 2017, \mnras, 471, 100

\bibitem[{{Epchtein} {et~al.}(1999){Epchtein}, {Deul}, {Derriere},
  {Borsenberger}, {Egret}, {Simon}, {Alard}, {Bal{\'a}zs}, {de Batz}, {Cioni},
  {Copet}, {Dennefeld}, {Forveille}, {Fouqu{\'e}}, {Garz{\'o}n}, {Habing},
  {Holl}, {Hron}, {Kimeswenger}, {Lacombe}, {Le Bertre}, {Loup}, {Mamon},
  {Omont}, {Paturel}, {Persi}, {Robin}, {Rouan}, {Tiph{\`e}ne}, {Vauglin}, \&
  {Wagner}}]{epchtein99}
{Epchtein}, N., {Deul}, E., {Derriere}, S., {et~al.} 1999, \aap, 349, 236

\bibitem[{{Figer} {et~al.}(2006){Figer}, {MacKenty}, {Robberto}, {Smith},
  {Najarro}, {Kudritzki}, \& {Herrero}}]{figer06}
{Figer}, D.~F., {MacKenty}, J.~W., {Robberto}, M., {et~al.} 2006, \apj, 643,
  1166

\bibitem[{{Froebrich} {et~al.}(2007){Froebrich}, {Scholz}, \&
  {Raftery}}]{froebrich07}
{Froebrich}, D., {Scholz}, A., \& {Raftery}, C.~L. 2007, \mnras, 374, 399

\bibitem[{{Gaia Collaboration}(2018)}]{gaia18}
{Gaia Collaboration}. 2018, \aap, 1804.09365

\bibitem[{{Geballe} {et~al.}(2000){Geballe}, {Najarro}, \& {Figer}}]{geballe00}
{Geballe}, T.~R., {Najarro}, F., \& {Figer}, D.~F. 2000, \apjl, 530, L97

\bibitem[{{Georgelin} \& {Georgelin}(1976)}]{georgelin76}
{Georgelin}, Y.~M. \& {Georgelin}, Y.~P. 1976, \aap, 49, 57

\bibitem[{{Giannetti} {et~al.}(2014){Giannetti}, {Wyrowski}, {Brand},
  {Csengeri}, {Fontani}, {Walmsley}, {Nguyen Luong}, {Beuther}, {Schuller},
  {G{\"u}sten}, \& {Menten}}]{giannetti14}
{Giannetti}, A., {Wyrowski}, F., {Brand}, J., {et~al.} 2014, \aap, 570, A65

\bibitem[{{Glushkova} {et~al.}(2010){Glushkova}, {Koposov}, {Zolotukhin},
  {Beletsky}, {Vlasov}, \& {Leonova}}]{glushkova10}
{Glushkova}, E.~V., {Koposov}, S.~E., {Zolotukhin}, I.~Y., {et~al.} 2010,
  Astronomy Letters, 36, 75

\bibitem[{{Gonzalez} {et~al.}(2012){Gonzalez}, {Rejkuba}, {Zoccali}, {Valenti},
  {Minniti}, {Schultheis}, {Tobar}, \& {Chen}}]{gonzalez12}
{Gonzalez}, O.~A., {Rejkuba}, M., {Zoccali}, M., {et~al.} 2012, \aap, 543, A13

\bibitem[{{Hamann} {et~al.}(2006){Hamann}, {Gr{\"a}fener}, \&
  {Liermann}}]{hamann06}
{Hamann}, W.-R., {Gr{\"a}fener}, G., \& {Liermann}, A. 2006, \aap, 457, 1015

\bibitem[{{Hanson} {et~al.}(1996){Hanson}, {Conti}, \& {Rieke}}]{hanson96}
{Hanson}, M.~M., {Conti}, P.~S., \& {Rieke}, M.~J. 1996, \apjs, 107, 281

\bibitem[{{Hanson} {et~al.}(2005){Hanson}, {Kudritzki}, {Kenworthy}, {Puls}, \&
  {Tokunaga}}]{hanson05}
{Hanson}, M.~M., {Kudritzki}, R.-P., {Kenworthy}, M.~A., {Puls}, J., \&
  {Tokunaga}, A.~T. 2005, \apjs, 161, 154

\bibitem[{{Heyer} {et~al.}(2016){Heyer}, {Gutermuth}, {Urquhart}, {Csengeri},
  {Wienen}, {Leurini}, {Menten}, \& {Wyrowski}}]{heyer16}
{Heyer}, M., {Gutermuth}, R., {Urquhart}, J.~S., {et~al.} 2016, \aap, 588, A29

\bibitem[{{Huang} {et~al.}(1999){Huang}, {Bania}, {Bolatto}, {Chamberlin},
  {Ingalls}, {Jackson}, {Lane}, {Stark}, {Wilson}, \& {Wright}}]{huang99}
{Huang}, M., {Bania}, T.~M., {Bolatto}, A., {et~al.} 1999, \apj, 517, 282

\bibitem[{{Kim} {et~al.}(2017){Kim}, {Wyrowski}, {Urquhart}, {Menten}, \&
  {Csengeri}}]{kim17}
{Kim}, W.-J., {Wyrowski}, F., {Urquhart}, J.~S., {Menten}, K.~M., \&
  {Csengeri}, T. 2017, \aap, 602, A37

\bibitem[{{Kleinmann} \& {Hall}(1986)}]{kleinmann86}
{Kleinmann}, S.~G. \& {Hall}, D.~N.~B. 1986, \apjs, 62, 501

\bibitem[{{K{\"o}nig} {et~al.}(2017){K{\"o}nig}, {Urquhart}, {Csengeri},
  {Leurini}, {Wyrowski}, {Giannetti}, {Wienen}, {Pillai}, {Kauffmann},
  {Menten}, \& {Schuller}}]{koenig17}
{K{\"o}nig}, C., {Urquhart}, J.~S., {Csengeri}, T., {et~al.} 2017, \aap, 599,
  A139

\bibitem[{{Koornneef}(1983)}]{koornneef83}
{Koornneef}, J. 1983, \aap, 128, 84

\bibitem[{{Kuchar} \& {Clark}(1997)}]{kuchar97}
{Kuchar}, T.~A. \& {Clark}, F.~O. 1997, \apj, 488, 224

\bibitem[{{Leitherer} {et~al.}(1997){Leitherer}, {Chapman}, \&
  {Koribalski}}]{leitherer97}
{Leitherer}, C., {Chapman}, J.~M., \& {Koribalski}, B. 1997, \apj, 481, 898

\bibitem[{{Likkel} {et~al.}(2004){Likkel}, {Dinerstein}, {Lester}, {Bruch}, \&
  {Bartig}}]{likkel04}
{Likkel}, L., {Dinerstein}, H.~L., {Lester}, D., {Bruch}, J., \& {Bartig}, K.
  2004, in Astronomical Society of the Pacific Conference Series, Vol. 313,
  Asymmetrical Planetary Nebulae III: Winds, Structure and the Thunderbird, ed.
  M.~{Meixner}, J.~H. {Kastner}, B.~{Balick}, \& N.~{Soker}, 351

\bibitem[{{Mart{\'{\i}}n-Hern{\'a}ndez}
  {et~al.}(2008){Mart{\'{\i}}n-Hern{\'a}ndez}, {Esteban}, {Mesa-Delgado},
  {Bik}, \& {Puga}}]{martin08}
{Mart{\'{\i}}n-Hern{\'a}ndez}, N.~L., {Esteban}, C., {Mesa-Delgado}, A., {Bik},
  A., \& {Puga}, E. 2008, \aap, 482, 215

\bibitem[{{Mart{\'{\i}}n-Hern{\'a}ndez}
  {et~al.}(2003{\natexlab{a}}){Mart{\'{\i}}n-Hern{\'a}ndez}, {van der Hulst},
  \& {Tielens}}]{martinhernandez03}
{Mart{\'{\i}}n-Hern{\'a}ndez}, N.~L., {van der Hulst}, J.~M., \& {Tielens},
  A.~G.~G.~M. 2003{\natexlab{a}}, \aap, 407, 957

\bibitem[{{Mart{\'{\i}}n-Hern{\'a}ndez}
  {et~al.}(2003{\natexlab{b}}){Mart{\'{\i}}n-Hern{\'a}ndez}, {van der Hulst},
  \& {Tielens}}]{martines03}
{Mart{\'{\i}}n-Hern{\'a}ndez}, N.~L., {van der Hulst}, J.~M., \& {Tielens},
  A.~G.~G.~M. 2003{\natexlab{b}}, \aap, 407, 957

\bibitem[{{Martins} {et~al.}(2007){Martins}, {Genzel}, {Hillier}, {Eisenhauer},
  {Paumard}, {Gillessen}, {Ott}, \& {Trippe}}]{martins07}
{Martins}, F., {Genzel}, R., {Hillier}, D.~J., {et~al.} 2007, \aap, 468, 233

\bibitem[{{Martins} \& {Plez}(2006)}]{martins06}
{Martins}, F. \& {Plez}, B. 2006, \aap, 457, 637

\bibitem[{{Martins} {et~al.}(2005){Martins}, {Schaerer}, \&
  {Hillier}}]{martins05}
{Martins}, F., {Schaerer}, D., \& {Hillier}, D.~J. 2005, \aap, 436, 1049

\bibitem[{{Mercer} {et~al.}(2005){Mercer}, {Clemens}, {Meade}, {Babler},
  {Indebetouw}, {Whitney}, {Watson}, {Wolfire}, {Wolff}, {Bania}, {Benjamin},
  {Cohen}, {Dickey}, {Jackson}, {Kobulnicky}, {Mathis}, {Stauffer}, {Stolovy},
  {Uzpen}, \& {Churchwell}}]{mercer05}
{Mercer}, E.~P., {Clemens}, D.~P., {Meade}, M.~R., {et~al.} 2005, \apj, 635,
  560

\bibitem[{{Messineo} {et~al.}(2015){Messineo}, {Clark}, {Figer}, {Kudritzki},
  {Najarro}, {Rich}, {Menten}, {Ivanov}, {Valenti}, {Trombley}, {Chen}, \&
  {Davies}}]{messineo15}
{Messineo}, M., {Clark}, J.~S., {Figer}, D.~F., {et~al.} 2015, \apj, 805, 110

\bibitem[{{Messineo} {et~al.}(2011){Messineo}, {Davies}, {Figer}, {Kudritzki},
  {Valenti}, {Trombley}, {Najarro}, \& {Rich}}]{messineo11}
{Messineo}, M., {Davies}, B., {Figer}, D.~F., {et~al.} 2011, \apj, 733, 41

\bibitem[{{Messineo} {et~al.}(2009){Messineo}, {Davies}, {Ivanov}, {Figer},
  {Schuller}, {Habing}, {Menten}, \& {Petr-Gotzens}}]{messineo09}
{Messineo}, M., {Davies}, B., {Ivanov}, V.~D., {et~al.} 2009, \apj, 697, 701

\bibitem[{{Messineo} {et~al.}(2010){Messineo}, {Figer}, {Davies}, {Kudritzki},
  {Rich}, {MacKenty}, \& {Trombley}}]{messineo10}
{Messineo}, M., {Figer}, D.~F., {Davies}, B., {et~al.} 2010, \apj, 708, 1241

\bibitem[{{Messineo} {et~al.}(2005){Messineo}, {Habing}, {Menten}, {Omont},
  {Sjouwerman}, \& {Bertoldi}}]{messineo05}
{Messineo}, M., {Habing}, H.~J., {Menten}, K.~M., {et~al.} 2005, \aap, 435, 575

\bibitem[{{Messineo} {et~al.}(2012){Messineo}, {Menten}, {Churchwell}, \&
  {Habing}}]{messineo12}
{Messineo}, M., {Menten}, K.~M., {Churchwell}, E., \& {Habing}, H. 2012, \aap,
  537, A10

\bibitem[{{Messineo} {et~al.}(2014{\natexlab{a}}){Messineo}, {Menten}, {Figer},
  {Davies}, {Clark}, {Ivanov}, {Kudritzki}, {Rich}, {MacKenty}, \&
  {Trombley}}]{messineo14a}
{Messineo}, M., {Menten}, K.~M., {Figer}, D.~F., {et~al.} 2014{\natexlab{a}},
  \aap, 569, A20

\bibitem[{{Messineo} {et~al.}(2014{\natexlab{b}}){Messineo}, {Zhu}, {Ivanov},
  {Figer}, {Davies}, {Menten}, {Kudritzki}, \& {Chen}}]{messineo14}
{Messineo}, M., {Zhu}, Q., {Ivanov}, V.~D., {et~al.} 2014{\natexlab{b}}, \aap,
  571, A43

\bibitem[{{Messineo} {et~al.}(2017){Messineo}, {Zhu}, {Menten}, {Ivanov},
  {Figer}, {Kudritzki}, \& {Chen}}]{messineo17}
{Messineo}, M., {Zhu}, Q., {Menten}, K.~M., {et~al.} 2017, \apj, 836, 65

\bibitem[{{Morris} {et~al.}(1996){Morris}, {Eenens}, {Hanson}, {Conti}, \&
  {Blum}}]{morris96}
{Morris}, P.~W., {Eenens}, P.~R.~J., {Hanson}, M.~M., {Conti}, P.~S., \&
  {Blum}, R.~D. 1996, \apj, 470, 597

\bibitem[{{Nahar} \& {Pradhan}(1996)}]{nahar96}
{Nahar}, S.~N. \& {Pradhan}, A.~K. 1996, \aaps, 119, 509

\bibitem[{{Najarro} {et~al.}(1997){Najarro}, {Krabbe}, {Genzel}, {Lutz},
  {Kudritzki}, \& {Hillier}}]{najarro97}
{Najarro}, F., {Krabbe}, A., {Genzel}, R., {et~al.} 1997, \aap, 325, 700

\bibitem[{{Panagia}(1973)}]{panagia73}
{Panagia}, N. 1973, \aj, 78, 929

\bibitem[{{Pinheiro} {et~al.}(2012){Pinheiro}, {Abraham}, {Copetti}, {Ortiz},
  {Falceta-Gon{\c c}alves}, \& {Roman-Lopes}}]{pinheiro12}
{Pinheiro}, M.~C., {Abraham}, Z., {Copetti}, M.~V.~F., {et~al.} 2012, \mnras,
  423, 2425

\bibitem[{{Povich} {et~al.}(2008){Povich}, {Benjamin}, {Whitney}, {Babler},
  {Indebetouw}, {Meade}, \& {Churchwell}}]{povich08}
{Povich}, M.~S., {Benjamin}, R.~A., {Whitney}, B.~A., {et~al.} 2008, \apj, 689,
  242

\bibitem[{{Price} {et~al.}(2001){Price}, {Egan}, {Carey}, {Mizuno}, \&
  {Kuchar}}]{price01}
{Price}, S.~D., {Egan}, M.~P., {Carey}, S.~J., {Mizuno}, D.~R., \& {Kuchar},
  T.~A. 2001, \aj, 121, 2819


\bibitem[{{Ram{\'{\i}}rez} {et~al.}(2000){Ram{\'{\i}}rez}, {Stephens},
  {Frogel}, \& {DePoy}}]{ramirez00}
{Ram{\'{\i}}rez}, S.~V., {Stephens}, A.~W., {Frogel}, J.~A., \& {DePoy}, D.~L.
  2000, \aj, 120, 833

\bibitem[{{Reid} {et~al.}(2009){Reid}, {Menten}, {Zheng}, {Brunthaler},
  {Moscadelli}, {Xu}, {Zhang}, {Sato}, {Honma}, {Hirota}, {Hachisuka}, {Choi},
  {Moellenbrock}, \& {Bartkiewicz}}]{reid09}
{Reid}, M.~J., {Menten}, K.~M., {Zheng}, X.~W., {et~al.} 2009, \apj, 700, 137

\bibitem[{{Richards} {et~al.}(2012){Richards}, {Lang}, {Trombley}, \&
  {Figer}}]{richards12}
{Richards}, E.~E., {Lang}, C.~C., {Trombley}, C., \& {Figer}, D.~F. 2012, \aj,
  144, 89

\bibitem[{{Robitaille} {et~al.}(2008){Robitaille}, {Meade}, {Babler},
  {Whitney}, {Johnston}, {Indebetouw}, {Cohen}, {Povich}, {Sewilo}, {Benjamin},
  \& {Churchwell}}]{robitalle08}
{Robitaille}, T.~P., {Meade}, M.~R., {Babler}, B.~L., {et~al.} 2008, \aj, 136,
  2413

\bibitem[{{Rubin}(1968)}]{rubin68}
{Rubin}, R.~H. 1968, \apj, 154, 391

\bibitem[{{Russeil}(2003)}]{russeil03}
{Russeil}, D. 2003, \aap, 397, 133

\bibitem[{{Schmeja}(2011)}]{schmeja11}
{Schmeja}, S. 2011, Astronomische Nachrichten, 332, 172

\bibitem[{{Schuller} {et~al.}(2009){Schuller}, {Menten}, {Contreras},
  {Wyrowski}, {Schilke}, {Bronfman}, {Henning}, {Walmsley}, {Beuther},
  {Bontemps}, {Cesaroni}, {Deharveng}, {Garay}, {Herpin}, {Lefloch}, {Linz},
  {Mardones}, {Minier}, {Molinari}, {Motte}, {Nyman}, {Reveret}, {Risacher},
  {Russeil}, {Schneider}, {Testi}, {Troost}, {Vasyunina}, {Wienen}, {Zavagno},
  {Kovacs}, {Kreysa}, {Siringo}, \& {Wei{\ss}}}]{schuller09}
{Schuller}, F., {Menten}, K.~M., {Contreras}, Y., {et~al.} 2009, \aap, 504, 415

\bibitem[{{Sidorin} {et~al.}(2014){Sidorin}, {Douglas}, {Palou{\v s}},
  {W{\"u}nsch}, \& {Ehlerov{\'a}}}]{sidorin14}
{Sidorin}, V., {Douglas}, K.~A., {Palou{\v s}}, J., {W{\"u}nsch}, R., \&
  {Ehlerov{\'a}}, S. 2014, \aap, 565, A6

\bibitem[{{Simpson} {et~al.}(2012){Simpson}, {Povich}, {Kendrew}, {Lintott},
  {Bressert}, {Arvidsson}, {Cyganowski}, {Maddison}, {Schawinski}, {Sherman},
  {Smith}, \& {Wolf-Chase}}]{simpson12}
{Simpson}, R.~J., {Povich}, M.~S., {Kendrew}, S., {et~al.} 2012, \mnras, 424,
  2442

\bibitem[{{Skiff}(2014)}]{skiff14}
{Skiff}, B.~A. 2014, VizieR Online Data Catalog, 1, 2023

\bibitem[{{Skrutskie} {et~al.}(2006){Skrutskie}, {Cutri}, {Stiening},
  {Weinberg}, {Schneider}, {Carpenter}, {Beichman}, {Capps}, {Chester},
  {Elias}, {Huchra}, {Liebert}, {Lonsdale}, {Monet}, {Price}, {Seitzer},
  {Jarrett}, {Kirkpatrick}, {Gizis}, {Howard}, {Evans}, {Fowler}, {Fullmer},
  {Hurt}, {Light}, {Kopan}, {Marsh}, {McCallon}, {Tam}, {Van Dyk}, \&
  {Wheelock}}]{skrutskie06}
{Skrutskie}, M.~F., {Cutri}, R.~M., {Stiening}, R., {et~al.} 2006, \aj, 131,
  1163

\bibitem[{{Solin} {et~al.}(2012){Solin}, {Ukkonen}, \& {Haikala}}]{solin12}
{Solin}, O., {Ukkonen}, E., \& {Haikala}, L. 2012, \aap, 542, A3

\bibitem[{{Soto} {et~al.}(2013){Soto}, {Barb{\'a}}, {Gunthardt}, {Minniti},
  {Lucas}, {Majaess}, {Irwin}, {Emerson}, {Gonzalez-Solares}, {Hempel},
  {Saito}, {Gurovich}, {Roman-Lopes}, {Moni-Bidin}, {Santucho}, {Borissova},
  {Kurtev}, {Toledo}, {Geisler}, {Dominguez}, \& {Beamin}}]{soto13}
{Soto}, M., {Barb{\'a}}, R., {Gunthardt}, G., {et~al.} 2013, \aap, 552, A101

\bibitem[{{Stetson}(1987)}]{stetson87}
{Stetson}, P.~B. 1987, \pasp, 99, 191

\bibitem[{{Storey} \& {Hummer}(1995)}]{storey95}
{Storey}, P.~J. \& {Hummer}, D.~G. 1995, \mnras, 272, 41

\bibitem[{{Tanner} {et~al.}(2006){Tanner}, {Figer}, {Najarro}, {Kudritzki},
  {Gilmore}, {Morris}, {Becklin}, {McLean}, {Gilbert}, {Graham}, {Larkin},
  {Levenson}, \& {Teplitz}}]{tanner05}
{Tanner}, A., {Figer}, D.~F., {Najarro}, F., {et~al.} 2006, \apj, 641, 891

\bibitem[{{Urquhart} {et~al.}(2007){Urquhart}, {Busfield}, {Hoare}, {Lumsden},
  {Oudmaijer}, {Moore}, {Gibb}, {Purcell}, {Burton}, \&
  {Marechal}}]{urquhart07}
{Urquhart}, J.~S., {Busfield}, A.~L., {Hoare}, M.~G., {et~al.} 2007, \aap, 474,
  891

\bibitem[{{Urquhart} {et~al.}(2018){Urquhart}, {K{\"o}nig}, {Giannetti},
  {Leurini}, {Moore}, {Eden}, {Pillai}, {Thompson}, {Braiding}, {Burton},
  {Csengeri}, {Dempsey}, {Figura}, {Froebrich}, {Menten}, {Schuller}, {Smith},
  \& {Wyrowski}}]{urquhart18}
{Urquhart}, J.~S., {K{\"o}nig}, C., {Giannetti}, A., {et~al.} 2018, \mnras,
  473, 1059

\bibitem[{{Urquhart} {et~al.}(2014){Urquhart}, {Moore}, {Csengeri}, {Wyrowski},
  {Schuller}, {Hoare}, {Lumsden}, {Mottram}, {Thompson}, {Menten}, {Walmsley},
  {Bronfman}, {Pfalzner}, {K{\"o}nig}, \& {Wienen}}]{urquhart14}
{Urquhart}, J.~S., {Moore}, T.~J.~T., {Csengeri}, T., {et~al.} 2014, \mnras,
  443, 1555

\bibitem[{{Walsh} {et~al.}(1998){Walsh}, {Burton}, {Hyland}, \&
  {Robinson}}]{walsh98}
{Walsh}, A.~J., {Burton}, M.~G., {Hyland}, A.~R., \& {Robinson}, G. 1998,
  \mnras, 301, 640

\bibitem[{{Wang} \& {Jiang}(2014)}]{wang14}
{Wang}, S. \& {Jiang}, B.~W. 2014, ApJ, 788, 12

\bibitem[{{Watson} {et~al.}(2008){Watson}, {Povich}, {Churchwell}, {Babler},
  {Chunev}, {Hoare}, {Indebetouw}, {Meade}, {Robitaille}, \&
  {Whitney}}]{watson08}
{Watson}, C., {Povich}, M.~S., {Churchwell}, E.~B., {et~al.} 2008, \apj, 681,
  1341

\bibitem[{{Watson} {et~al.}(2009){Watson}, {Schr{\"o}der}, {Fyfe}, {Page},
  {Lamer}, {Mateos}, {Pye}, {Sakano}, {Rosen}, {Ballet}, {Barcons}, {Barret},
  {Boller}, {Brunner}, {Brusa}, {Caccianiga}, {Carrera}, {Ceballos}, {Della
  Ceca}, {Denby}, {Denkinson}, {Dupuy}, {Farrell}, {Fraschetti}, {Freyberg},
  {Guillout}, {Hambaryan}, {Maccacaro}, {Mathiesen}, {McMahon}, {Michel},
  {Motch}, {Osborne}, {Page}, {Pakull}, {Pietsch}, {Saxton}, {Schwope},
  {Severgnini}, {Simpson}, {Sironi}, {Stewart}, {Stewart}, {Stobbart}, {Tedds},
  {Warwick}, {Webb}, {West}, {Worrall}, \& {Yuan}}]{watson09}
{Watson}, M.~G., {Schr{\"o}der}, A.~C., {Fyfe}, D., {et~al.} 2009, \aap, 493,
  339

\bibitem[{{Wienen} {et~al.}(2015){Wienen}, {Wyrowski}, {Menten}, {Urquhart},
  {Csengeri}, {Walmsley}, {Bontemps}, {Russeil}, {Bronfman}, {Koribalski}, \&
  {Schuller}}]{wienen15}
{Wienen}, M., {Wyrowski}, F., {Menten}, K.~M., {et~al.} 2015, \aap, 579, A91

\bibitem[{{Wright} {et~al.}(2010){Wright}, {Eisenhardt}, {Mainzer}, {Ressler},
  {Cutri}, {Jarrett}, {Kirkpatrick}, {Padgett}, {McMillan}, {Skrutskie},
  {Stanford}, {Cohen}, {Walker}, {Mather}, {Leisawitz}, {Gautier}, {McLean},
  {Benford}, {Lonsdale}, {Blain}, {Mendez}, {Irace}, {Duval}, {Liu}, {Royer},
  {Heinrichsen}, {Howard}, {Shannon}, {Kendall}, {Walsh}, {Larsen}, {Cardon},
  {Schick}, {Schwalm}, {Abid}, {Fabinsky}, {Naes}, \& {Tsai}}]{wright10}
{Wright}, E.~L., {Eisenhardt}, P.~R.~M., {Mainzer}, A.~K., {et~al.} 2010, \aj,
  140, 1868

\bibitem[{{Wright} {et~al.}(2016){Wright}, {Bouy}, {Drew}, {Sarro}, {Bertin},
  {Cuillandre}, \& {Barrado}}]{wright16}
{Wright}, N.~J., {Bouy}, H., {Drew}, J.~E., {et~al.} 2016, \mnras, 460, 2593

\bibitem[{{Wright} {et~al.}(2014){Wright}, {Parker}, {Goodwin}, \&
  {Drake}}]{wright14}
{Wright}, N.~J., {Parker}, R.~J., {Goodwin}, S.~P., \& {Drake}, J.~J. 2014,
  \mnras, 438, 639

\bibitem[{{Zacharias} {et~al.}(2005){Zacharias}, {Monet}, {Levine}, {Urban},
  {Gaume}, \& {Wycoff}}]{zacharias05}
{Zacharias}, N., {Monet}, D.~G., {Levine}, S.~E., {et~al.} 2005, VizieR Online
  Data Catalog, 1297

\end{thebibliography}

\begin{acknowledgements}

This publication makes use of data products from the Two Micron All Sky Survey, which is a joint project 
of the University of Massachusetts and the Infrared Processing and Analysis Center/California Institute 
of Technology, funded by the National Aeronautics and Space Administration and the National Science Foundation.
This work is based  on observations made with the Spitzer Space Telescope, 
which is operated by the Jet Propulsion Laboratory, California Institute of Technology under a contract with NASA.
DENIS is  a joint effort of several institutes mostly located in Europe. It has
    been supported mainly by the French Institut National des
    Sciences de l'Univers, CNRS, and French Education Ministry, the
    European Southern Observatory, the State of Baden-Wuerttemberg, and
    the European Commission under networks of the SCIENCE and Human
    Capital and Mobility programs, the Landessternwarte, Heidelberg, and
    Institut d'Astrophysique de Paris.
This research made use of data products from the
Midcourse Space Experiment, the processing of which was funded by the Ballistic Missile Defense Organization with additional
support from the NASA office of Space Science. This publication makes use of data products from
WISE, which is a joint project of the University of California, Los
Angeles, and the Jet Propulsion Laboratory/California Institute of Technology, funded by the National Aeronautics and
Space Administration. 
We gratefully acknowledge use of data from the ESO Public Survey program 
ID 179.B-2002 taken with the VISTA telescope, and data products from the Cambridge Astronomical 
Survey Unit.
 It is a pleasure to thank Dave Monet, who forwarded a copy of the
    NOMAD catalog to CDS.
    This work has made use of data from the European Space Agency (ESA) mission Gaia 
(https://www.cosmos.esa.int/gaia), processed by the Gaia Data Processing and Analysis Consortium (DPAC, https://www.cosmos.esa.int/web/gaia/dpac/consortium). Funding for the DPAC has been provided by national institutions, in particular the institutions participating in the Gaia Multilateral Agreement.

This research has made use of the SIMBAD database, 
operated at CDS, Strasbourg, France. This research made use of Montage, funded by the National Aeronautics and Space Administration's 
Earth Science Technology Office, Computational Technnologies Project, under Cooperative 
Agreement Number NCC5-626 between NASA and the California Institute of Technology. 
The code is maintained by the NASA/IPAC Infrared Science Archive.
This work was partially funded by the ERC Advanced Investigator Grant GLOSTAR (247078).
This work was partially supported by  the Fundamental Research 
Funds for the Central Universities in China, and USTC
grant KY2030000054.
A special thank you is  for the great support offered by  the European Southern Observatory.
We thank Dr. Michelle Doherty for her careful reading of our manuscript,  Dr. Ernesto Martins
for his algorithms to solve for the minimum spanning three paths.
We thank the  anonymous referee for his constructive comments and for having
suggested that we  comment on the candidate cluster multiplicity. 

\end{acknowledgements}

\end{document}